\documentclass[prd,12pt,reprint, twocolumn,noshowpacs,nofootinbib,amsmath,amssymb,superscriptaddress,preprintnumbers]{revtex4-1}
\usepackage{graphicx,color}
\usepackage{tikz-feynman} 
\usepackage{caption}
\usepackage{amssymb,amsmath,mathrsfs}
\usepackage{cancel}
\usepackage{subfig}
\usepackage{hyperref}
\usepackage{verbatim,slashed}
\usepackage{xspace}
\usepackage{afterpage}

\newcommand{\beq}{\begin{equation}}
\newcommand{\eeq}{\end{equation}}
\newcommand{\be}{\begin{eqnarray}}
\newcommand{\ee}{\end{eqnarray}}

\newcommand{\bi}{\begin{itemize}}
\newcommand{\ei}{\end{itemize}}
\newcommand{\benum}{\begin{enumerate}}
\newcommand{\eenum}{\end{enumerate}}

\definecolor{greenp1}{rgb}{0.8, 0, 0}

\definecolor{bluep1}{rgb}{0, 0, 0.8}

\definecolor{magentap1}{HTML}{B16286}

\definecolor{orange1}{rgb}{1, 0.5, 0}

\def\lsim{\mathrel{\rlap{\lower4pt\hbox{\hskip1pt$\sim$}}
    \raise1pt\hbox{$<$}}}
\def\gsim{\mathrel{\rlap{\lower4pt\hbox{\hskip1pt$\sim$}}
    \raise1pt\hbox{$>$}}}

\begin{document}
\title{Oscillation Baryogenesis via Ultraviolet Dark Matter Freeze-In}
\author{Tian Dong}
\affiliation{Harvey Mudd College, Claremont, CA 91711, USA}
\affiliation{Perimeter Institute for Theoretical Physics,  Waterloo, ON N2L 2Y5, Canada}
\author{Conor M. Floyd}
\affiliation{Harvey Mudd College,  Claremont, CA 91711, USA}
\author{Antonia Hekster}
\affiliation{Harvey Mudd College,  Claremont, CA 91711, USA}
\author{Derek J.~Li}
\affiliation{Harvey Mudd College,  Claremont, CA 91711, USA}
\affiliation{Stanford Institute for Theoretical Physics, Stanford University, Stanford, CA 94305, USA}
\author{Brian Shuve}
\affiliation{Harvey Mudd College,  Claremont, CA 91711, USA}
\author{David Tucker-Smith}
\affiliation{Department of Physics, Williams College, Williamstown, MA 01267, USA}

\date{\today}

\begin{abstract}
We investigate baryogenesis from dark matter oscillations in the ultraviolet freeze-in regime. We find that the mechanism can simultaneously accommodate the observed abundances of baryons and dark matter for dark matter masses in the 10 keV to MeV range, provided the reheat temperature lies between the temperature of the electroweak phase transition and  $\sim 10$~TeV. The mechanism predicts observable consequences due to the presence of a light dark matter component that is relativistic during structure formation, and X-ray bounds on decaying dark matter can set strong constraints depending on the operator mediating dark matter production.

\end{abstract}
\maketitle

\section{Introduction}\label{sec:intro}

The nature of dark matter (DM) and the origin of the baryon asymmetry are two of the 
biggest outstanding questions in particle physics. Both necessitate new particles or
interactions that are out of equilibrium at some point in the Universe's history,
and it is natural to consider the scenario in which there is a common departure from
equilibrium that is responsible for establishing both abundances. Several such 
scenarios have been studied in the literature, such as asymmetric DM \cite{Nussinov:1985xr,Kaplan:1991ah,Barr:1990ca,Barr:1991qn,Dodelson:1991iv,Fujii:2002aj,Kitano:2004sv,Farrar:2005zd,Gudnason:2006ug,Kitano:2008tk,Kaplan:2009ag}
and baryogenesis via freeze-out of symmetric DM \cite{Cui:2011ab,Bernal:2012gv,Kumar:2013uca,Bernal:2013bga,Baldes:2014gca}, both of which assume that DM was at some point in equilibrium with Standard Model (SM)
particles.

It is, however, possible that particles involved in resolving the questions of DM and baryogenesis
were \emph{never} in equilibrium throughout  cosmic history. Such species are 
produced via freeze-in \cite{McDonald:2001vt,Choi:2005vq,Kusenko:2006rh,Petraki:2007gq,Hall:2009bx,Bernal:2017kxu}, where the particles' interaction rates are not large enough
to equilibrate them with the SM plasma. Freeze-in models have
become increasingly popular in recent years because they occur generically
in hidden-sector scenarios and they avoid increasingly stringent constraints from
laboratory and astrophysical searches for thermal DM. Unfortunately, this same feature
of freeze-in can also make it difficult to test, 
as the tiny couplings needed
to keep species out of equilibrium can also render signal rates negligible in experiments \cite{Bernal:2017kxu}.

If freeze-in simultaneously accounts for both baryogenesis and the DM abundance,
however, then the prospects for experimental discovery of the model can be 
considerably improved. In the recently proposed mechanism of freeze-in baryogenesis 
via DM oscillations \cite{Shuve:2020evk,Berman:2022oht}, DM must possess a coupling to SM particles in order to generate an 
asymmetry; this in turn necessitates the existence of new, electromagnetically charged states that mediate interactions
between DM and the SM. When DM is produced
from the on-shell decay of the charged states, the charged-particle masses must lie around the
weak scale and can be the target of collider searches. Furthermore, the masses of DM particles  affect
the timescale of DM oscillations, such that simultaneously accounting for the baryon asymmetry and DM abundance
restricts the DM to a relatively narrow mass range ($\sim\mathrm{keV}-\mathrm{MeV}$).
The freeze-in coupling can also not be too small because the baryon asymmetry would
otherwise be insufficient to account for the observed value. These properties are also found in the Akhmedov-Rubakov-Smirnov (ARS) mechanism for leptogenesis
via right-handed neutrino (RHNs) oscillations \cite{Akhmedov:1998qx,Asaka:2005pn}, which works with GeV-scale RHNs that
have couplings potentially large enough to allow experimental detection of the RHNs \cite{Gorbunov:2007ak}.

All prior studies of the cosmology and phenomenology of freeze-in 
baryogenesis via DM oscillations have been done within the framework of infrared (IR) 
freeze-in, in which the charged mediator between DM and the SM has a mass below the reheat
temperature of the Universe. In this scenario, DM production rates grow relative to the Hubble expansion
rate, $H$, down to energies comparable to the electroweak scale. Because $H\ll T$ for temperatures $T$ well below the Planck scale, IR freeze-in models typically feature very small couplings ($\lesssim10^{-7}$) to prevent
total equilibration of DM. 

There exists another limit:~ultraviolet (UV) freeze-in \cite{Moroi:1993mb,Hall:2009bx,Cheung:2011nn,Elahi:2014fsa}. In this case,
DM is produced by scattering via a virtual mediator at energies well
below the mediator mass. UV freeze-in occurs in cosmologies with a reheat temperature
that is sufficiently low as to exponentially suppress the on-shell production of the mediator.
 The couplings between the SM and DM
are now non-renormalizable, and the DM interaction rate falls relative to Hubble over
time. Consequently, the feebleness of the DM-SM interaction can be explained by
 hierarchies between the reheat temperature and the mass of the mediator rather than
by small dimensionless couplings; this is analogous to the reason for the feebleness of the weak interactions
at energies well below the electroweak scale.

We  undertake the first comprehensive study of baryogenesis from DM oscillations
 in the UV limit.
We find that it is possible to simultaneously account for DM and baryogenesis within UV freeze-in
provided the reheat temperature falls within the approximate range 130 GeV--10 TeV. The parameter space is much more 
restricted than for IR freeze-in, with the lightest DM state necessarily lying well below a mass of 1 keV. The model faces
stringent constraints from structure formation and X-ray line searches such that several implementations of UV freeze-in are already ruled out,
 and we expect that future improvements
in these observables should further test interesting parameter space for UV freeze-in baryogenesis. In contrast with IR freeze-in, however, 
colliders are no longer a promising avenue for discovering the new states associated with DM and baryogenesis, at least in the scenarios we consider here.

Baryogenesis through DM freeze-in can also occur through an entirely different mechanism, in which the baryon asymmetry and DM abundance are generated simultaneously through feeble $CP$-violating decays of or scattering induced by a heavy parent particle, rather than through DM oscillations and rescattering. This has been studied for both IR and UV freeze-in \cite{Unwin:2014poa,Goudelis:2021qla,Goudelis:2022bls,Asadi:2025vli}.

The code used to produce all numerical results from this paper can be found in Ref.~\cite{dong_2025_16786347}.

\section{Mechanisms for Freeze-In Baryogenesis via DM Oscillations}\label{sec:mechanisms}

\subsection{Review of IR Freeze-In}

We can obtain the UV limit of a freeze-in model by first considering IR freeze-in and then integrating out the heavy mediators. We therefore consider  a freeze-in leptogenesis model along the lines of Refs.~\cite{Shuve:2020evk,Berman:2022oht}. The model consists of two Majorana singlet fermion DM candidates, $\chi_1$ and $\chi_2$, with masses $M_1$ and $M_2$, respectively; and, one or more scalars, $\Phi_a$, where $a$ is an index running over the number of scalars. The scalars carry SM gauge quantum numbers that allow it to have renormalizable couplings to DM and either quarks or leptons. \\

\noindent {\bf DM interaction term:} 
Consider, for concreteness, the case in which the $\Phi_a$ have the same SM gauge charges as the SM Higgs doublet.   In terms of Weyl spinors, the DM interaction Lagrangian is
\be\label{eq:IR_lag}
\mathcal{L}_{\rm DM} &=& -F_{\alpha I}^a\,\Phi_a L_\alpha \chi_I + \mathrm{h.c.},
\ee
where $L_\alpha$ is the SM lepton doublet,  $\alpha$ is a SM lepton flavor index, $I$ is a DM flavor index in the mass eigenbasis, and $F^a_{\alpha I}$ is the coupling. This interaction  is sufficient to account for DM and baryogenesis through IR freeze-in.  We consider other, equally viable, choices for the SM quantum numbers of $\Phi_a$ when we make the transition to the UV freeze-in regime in Sec.~\ref{sec:IRtoUV}.

The coupling in Eq.~\ref{eq:IR_lag} is consistent with an exact $\mathrm{U}(1)_{B-L}$ symmetry where every $\Phi_a$ has $B-L$ charge of $+1$. It  is also consistent with a $Z_2$ symmetry where $\Phi_a$ and $\chi_I$ are odd and all SM fields are even. Such a $Z_2$ symmetry makes DM exactly stable.

  The couplings $F_{\alpha I}^a$ are sufficiently small that DM does not reach equilibrium, although $\Phi_a$ are nearly always in equilibrium as a result of  SM gauge interactions. We assume that the $\Phi_a$ do not acquire vacuum expectation values (VEVs).

If the reheat temperature of the universe exceeds the masses of the $\Phi_a$ particles, then DM is produced through  ${\Phi^*_a}\to  L_\alpha \chi$ and the conjugate process. Because DM is typically out of equilibrium, it is produced in interaction eigenstates that then undergo coherent propagation and oscillation\footnote{It is generally true for the parameters of interest to us that the DM mass splitting is sufficiently small that they are produced coherently.}.  To generate an asymmetry, higher-order processes are needed:~subsequent inverse decays $L_\beta\chi\to{\Phi_b^*}$ lead to net processes like ${\Phi^*_a} L_\beta\to{\Phi^*_b}  L_\alpha$. A combination of phases from oscillation and $CP$-violating couplings can lead to a difference in rate compared to the corresponding antiparticle process, ${\Phi_a} \bar L_\beta\to{\Phi_b} \bar L_\alpha$, at $\mathcal{O}(F^4)$. 

When there exists only a single scalar, this $CP$-violating difference leads to asymmetries in individual lepton flavors but not overall lepton number\footnote{The Majorana masses for $\chi$ can, in principle, produce a total lepton number at this order in perturbation theory, but such rates are suppressed by an additional factor of $M_\chi^2/T^2$ and are negligible for DM masses consistent with this mechanism, $M_\chi\sim100$~keV.}. A net lepton asymmetry originates only after washout erodes the asymmetries in SM lepton flavors at different rates, and thus the total asymmetry is $\mathcal{O}(F^6)$. 

Alternatively, when there exist multiple scalars,  
kinematic differences in $\Phi_a$ (inverse) decay between the time of DM production and annihilation 
can lead
to a total lepton asymmetry at $\mathcal{O}(F^4)$, which is a significant enhancement in the weak-washout limit\footnote{In the case where $\Phi_a$ couples to quarks instead of leptons, thermal mass effects from the large top quark Yukawa coupling can also lead to an enhanced asymmetry.}. 

In both scenarios, a portion of the total lepton asymmetry is  transferred to baryons via sphalerons, which decouple at a temperature $T_{\rm ew}\approx 132$~GeV during the electroweak phase transition \cite{DOnofrio:2014rug}. After sphaleron decoupling, the baryon asymmetry is locked in and protected from any subsequent dynamics that might otherwise wash out the asymmetry.

The baryon asymmetry is maximized when $\chi$ oscillations occur at a  rate comparable to the Hubble expansion rate for $T\approx M_{\Phi_1}$, where $\Phi_1$ is the lightest mediator species. If oscillations are much slower than this rate, then an insufficient time-evolution phase accrues by the time that DM decouples completely from the SM and  the resulting lepton asymmetry is negligible. Conversely, if oscillations are much faster than this rate, then they time-average to zero over much of the time of DM production, limiting the size of the asymmetry. The  optimal DM oscillation time typically corresponds to DM mass splittings on the order of $M_2^2-M_1^2\sim(100\,\,\mathrm{keV})^2$ for $M_{\Phi_1}\sim$~TeV. To avoid overproducing the DM energy density, both $\chi_1$ and $\chi_2$ mass eigenstates typically need to have masses at or below a few hundred keV.

 As mentioned earlier, the freeze-in coupling in Eq.~\eqref{eq:IR_lag} preserves an exact $\mathrm{U}(1)_{B-L}$ symmetry. Any lepton asymmetry produced from this interaction is exactly counterbalanced by an asymmetry in $\Phi_a$. If all $\Phi_a$ decay prior to the electroweak phase transition, the asymmetries sum to zero and baryogenesis fails. If, instead, at least some population of $\Phi_a$ persists until after sphalerons decouple at $T_{\rm ew}$, the part of the $B-L$ asymmetry stored in baryons is protected and baryogenesis can be successful. Viable baryogenesis therefore requires either:~$M_{\Phi_1}\lesssim$~TeV such that there is a non-negligible thermal abundance of $\Phi_1$ at the time of sphaleron decoupling, or $c\tau_{\Phi_1}\gtrsim H(T_{\rm ew})^{-1}\sim1$ cm so that a population of $\Phi_1$ survives even if it is much heavier than the electroweak scale.  \\

\noindent {\bf Explicit breaking of $\mathrm{U}(1)_{B-L}$:}~In addition to the minimal coupling in Eq.~\eqref{eq:IR_lag}, there might exist other interactions between $\Phi_a$ and SM fields that violate $\mathrm{U}(1)_{B-L}$,
such as 
\be\label{eq:IR_lag2}
\mathcal{L}_{\Phi-\mathrm{SM}} &=&  - \lambda^a_{\alpha \beta} \Phi_{a}^* Q_\alpha d_\beta^{\rm c}
+ \mathrm{h.c.}
\ee
Here, $Q_\alpha$ is the quark doublet and $d_\beta^{\rm c}$ is the singlet down-type antiquark. This operator can significantly impact  the asymmetry generated in the IR scenario.  If the $\Phi_a \rightarrow Qd^{\rm c}$ decay induced by this operator predominates, then the lepton asymmetry produced by DM scattering is not washed out by $\Phi_a$ decays and persists even after all $\Phi_a$ have disappeared. Viable parameter space with $M_{\Phi_1}\gg$~TeV consequently opens up even for short lifetimes, $c\tau_{\Phi_1}\ll H(T_{\rm ew})^{-1}$.

The explicit breaking of $\mathrm{U}(1)_{B-L}$ also breaks the $Z_2$ symmetry stabilizing DM. However, the same parametrics that keep  DM out of equilibrium also suppress its decay rate.
As a result, the $\chi_I$ are still metastable DM candidates with lifetimes longer than the age of the universe. Potential observational consequences of this scenario include indirect detection signatures associated with photons produced in DM decay, and flavor-violating signals induced by the $\lambda$ couplings. 

The particular $\mathrm{U}(1)_{B-L}$-breaking coupling of Eq.~\eqref{eq:IR_lag2} radiatively generates $\Phi-H$ mixing, which in turn leads to DM decay. These radiative corrections are quadratically divergent, suggesting that some fine tuning is needed to bring the DM decay rate below current constraints. Different $\Phi$ gauge charges permit different $\mathrm{U}(1)_{B-L}$-breaking operators, however, leading to a variety of predictions for the DM decay rate. We survey the full range of operators that can break $\mathrm{U}(1)_{B-L}$ and induce DM decay in Sec.~\ref{sec:IRtoUV}, and we provide a parametric estimate of DM lifetime there for one operator. We perform a more quantitative study in Sec.~\ref{sec:xray}.

Because $\mathrm{U}(1)_{B-L}$ is restored when $\lambda_{\alpha \beta} = 0$, it is technically natural for the $\lambda$ couplings to vanish.  Both $\mathrm{U}(1)_{B-L}$-preserving and $\mathrm{U}(1)_{B-L}$-violating scenarios are worth considering, and IR freeze-in leptogenesis is viable in both cases.  Conversely, we are about to see that $\mathrm{U}(1)_{B-L}$ must be explicitly broken for UV freeze-in leptogenesis to be successful.\\

\noindent More details on baryogenesis from IR freeze-in via $\Phi_a$ decays can be found in Refs.~\cite{Shuve:2020evk,Berman:2022oht}.

\subsection{From IR to UV using Effective Field Theory}\label{sec:IRtoUV}

All previous work on freeze-in leptogenesis via DM oscillations  assumed that DM is produced  through the decay of on-shell scalars $\Phi_a$, 
in which case most DM particles are produced at $T\sim M_{\Phi_a}$.
However, it is possible that the reheat temperature is well below the masses of all the $\Phi_a$, in which case DM is produced exclusively through $2\to2$ scattering mediated by virtual scalars. This is known as UV freeze-in because 
most DM production occurs 
around the time of reheating. 
In some ways, UV freeze-in is more theoretically appealing than IR freeze-in because the feeble DM-SM interaction can be naturally explained even with $\mathcal{O}(1)$ couplings provided $T_{\rm RH}\ll M_{\Phi_a}$. By contrast, IR freeze-in always requires tiny dimensionless couplings between the dark and visible sectors.
\\

\noindent {\bf A benchmark UV scenario:}~When there exists a large hierarchy between the reheat temperature and the mediator masses, the DM couplings are best described by an effective field theory (EFT) in which the scalar mediators have been integrated out. We refer to this as the ``fully UV" scenario. If we start from the two  interactions of Eqs.~(\ref{eq:IR_lag}) and (\ref{eq:IR_lag2}), 
the   operators responsible for DM scattering in the EFT are
\small
\be\label{eq:EFT}
\mathcal{L}_{\rm EFT} &=& C_{\alpha \beta IJ}(L_\alpha \chi_I)(\chi^\dagger_J L^\dagger_\beta) \\
\nonumber
&&
+\left[ D_{\alpha I}(L_\alpha \chi_I)(Qd^{\rm c}) 
+\mathrm{h.c.}
\right],
\ee
\normalsize
where  we assume for simplicity that only a single quark flavor couples: $\lambda^a_{\alpha \beta} \rightarrow \lambda^a$ in Eq.~(\ref{eq:IR_lag2}).
At leading order, the EFT coefficients are related to the IR couplings via
\small
\be
C_{\alpha\beta IJ} &=& \sum_a \frac{F^a_{\alpha I}{F^{a*}_{\beta J}}}{M_{\Phi_a}^2}, \label{eq:coupling:C}\\
D_{\alpha I} &=& \sum_a \frac{F^a_{\alpha I}\lambda^{a}}{M_{\Phi_a}^2}.
\ee
\normalsize
The advantage of working in the EFT is that we only need to consider the physically relevant  combinations of couplings and masses $C_{\alpha \beta IJ}$ and $D_{\alpha I}$.
Note that the first operator is self-adjoint and thus $C_{\alpha\beta IJ} = C^*_{\beta\alpha JI}$.

If $D_{\alpha I}=
0$, SM $B-L$ is exactly conserved in the EFT and no total SM $B-L$ asymmetry can be generated. It is therefore necessary for SM $B-L$ (and the associated DM-stabilizing $Z_2$ symmetry) to be explicitly broken, with 
$D_{\alpha I}$ 
non-zero.

In general, both of the EFT couplings in Eq.~(\ref{eq:EFT}) should be considered when computing the DM abundance and baryon asymmetry. However, since the $C$ couplings are products of two freeze-in couplings whereas the $D$ couplings have only one, it is possible for the $C$ couplings to be sufficiently subdominant that we can neglect them. The baryon asymmetry to DM abundance is also maximized in this limit, since the $C$ couplings contribute to the DM abundance but do not directly lead to a lepton asymmetry, making it harder to realize baryogenesis without overproducing DM. 
In what follows, our quantitative studies mainly focus on a benchmark scenario in which $D_{\alpha I}(L_\alpha \chi_I)(Qd^{\rm c})$ is the only important DM interaction term in the EFT.\\

\noindent {\bf Leptogenesis in the UV scenario:}~We now discuss how the asymmetry is produced, comparing and contrasting with IR leptogenesis. At $\mathcal{O}(D^4)$, processes such as $L_\alpha Q \to \chi_I^\dagger {d^{\rm c}}^\dagger$ followed by $\chi_I^\dagger {d^{\rm c}}^\dagger\to L_\beta Q$ can produce $CP$ asymmetries in individual lepton flavors, but not an overall lepton asymmetry. This is similar to the single-scalar IR scenario. At $\mathcal{O}(D^6)$, a total lepton asymmetry can be produced from flavor differential washout in a manner analogous to IR freeze-in leptogenesis mechanisms such as ARS leptogenesis \cite{Akhmedov:1998qx,Asaka:2005pn}. 

In UV leptogenesis, 
most DM interactions occur at high temperatures,  $T\sim T_\text{RH}$.
This is in contrast with IR leptogenesis, for which most DM is produced when $T\sim M_{\Phi_1}$. It is therefore optimal for the oscillation time to be matched to the expansion rate at reheating, $H(T_{\rm RH})$, which implies
\be\label{eq:osc-RH}
\frac{M_0 \Delta M^2}{6 T_{\rm RH}^3} \sim 1,
\ee
where $M_0$ is defined so that $H(T) \approx 1.66\sqrt{g_*}T^2/M_{\rm Pl} \equiv T^2/M_0$, and $\Delta M^2\equiv M_2^2-M_1^2$ is the difference of squared DM masses. A somewhat larger DM mass is preferred in UV leptogenesis relative to the IR scenario so that oscillations occur at early times, $T\sim T_{\rm RH}$. This larger DM mass scale can lead to overproduction of DM, and consequently the viable parameter space for UV leptogenesis is more constrained than for IR leptogenesis. We will show that UV leptogenesis necessitates relatively low reheat temperatures, $T_{\rm RH}\sim$~TeV.

As is common in the UV freeze-in literature, we assume that the universe reheats instantaneously to a temperature $T_{\rm RH}$, which we take to be a free parameter. Realistic models of reheating take into account the finite time over which reheating occurs, and there are typically periods prior to the end of reheating when the temperature is even higher than $T_{\rm RH}$. We expect, however, that any asymmetry and DM produced during reheating are suppressed by subsequent entropy injection during the final period of reheating. Previous studies  found that DM production during reheating is negligible when DM production is mediated by dimension-6 operators like the ones we study \cite{Garcia:2017tuj}, although larger deviations are possible for non-standard cosmologies \cite{Bernal:2019mhf}. 

For leptogenesis, it is not possible to make such a universal statement because there exists an additional time scale in the problem, namely that of DM oscillations.  DM production during reheating could conceivably still contribute  to the asymmetry if the DM oscillation rate is comparable to the expansion rate prior to the end of reheating. In this case, DM oscillations are too rapid after reheating and the asymmetry produced during the radiation-dominated epoch is negligible. This effect is most relevant for larger DM masses, for which we expect that the parameter space for leptogenesis would lead to an overproduction of DM and is not viable. Therefore the instantaneous reheat approximation should allow an adequate estimate of the successful parameter space for simultaneously accounting for DM and leptogenesis. We intend to  study the effects of realistic reheating on oscillation leptogenesis in future work.\\

\noindent {\bf Additional operators and alternative models:}~Eq.~(\ref{eq:IR_lag2}) includes just one possible $U(1)_{B-L}$-violating interaction involving $\Phi_a$ and SM fermions.  More generally, we might start instead with
\be\label{eq:IR_lag3}
\mathcal{L}_{\Phi-\mathrm{SM}} &=&  - \lambda^a_{\alpha \beta} \Phi_{a}^* Q_\alpha d_\beta^{\rm c}-\lambda_{\alpha\beta}'^a\Phi_a^* L_\alpha e_\beta^{\rm c}\\
\nonumber
& &-\lambda_{\alpha\beta}''^a\Phi_a Q_\alpha u_\beta^{\rm c} 
+\mathrm{h.c.}
\ee
In a simplified scenario in which only the $\lambda'^a$ couplings are significant, integrating out the $\Phi_a$ scalars produces a purely leptonic DM interaction term in the EFT, 
\small
\be\label{eq:EFT_2}
\mathcal{L}_{\rm EFT} &=& E_{\alpha\beta\gamma I}(L_\alpha\chi_I)(L_\beta e_\gamma^{\rm c})+\mathrm{h.c.},
\ee
\normalsize
with
\small
\be
E_{\alpha\beta\gamma I} &=& \sum_a \frac{F^a_{\alpha I}\lambda_{\beta\gamma}'^{a}}{M_{\Phi_a}^2}.
\ee
\normalsize
at leading order.   In Sec.~\ref{sec:UV}, we study baryogenesis and DM production both via the semileptonic $D$ couplings of Eq.~\ref{eq:EFT} and via the fully leptonic $E$ couplings of Eq.~\ref{eq:EFT_2}.  We find quantitatively similar results for  these two lepton-number-violating (LNV) operators\footnote{Integrating out $\Phi_a$ also generates operators involving only SM fields. While such operators can potentially affect the baryon asymmetry by mediating flavor-changing interactions among SM particles, they do not do so  for the benchmarks we consider in Sec.~\ref{sec:UV}.}. 

\begin{table}[]
    \centering
    \begin{tabular}{||c||c||c||}
    \hline\hline
 $\Phi_a$ charges & BNV & LNV \\
      \hline
           \hline
       $({\mathbf 3}, {\mathbf 2}, 1/6)$  & $-$ & $(Q\chi)(Ld^{\rm c} )$ \\
        \hline
              $({\mathbf 3}, {\mathbf 1}, 2/3)$  & $(u^{\rm c} \chi)(d^{\rm c}  d^{\rm c} )$ & $-$\\
         \hline
                $({\mathbf 3}, {\mathbf 2}, -1/3)$  & $(d^{\rm c}  \chi)(u^{\rm c}  d^{\rm c} )$,
                $(d^{\rm c}  \chi)(Q^\dagger  Q^\dagger)$ & $(d^{\rm c}  \chi)(QL)$, $(d^{\rm c}  \chi)({u^{\rm c}}^\dagger {e^{\rm c}}^\dagger)$ \\
         \hline
                $({\mathbf 2}, {\mathbf 1}, 1/2)$  & $-$ & $(L \chi)(Qd^{\rm c})$, $(L \chi)(Q^\dagger {u^{\rm c}}^\dagger)$,  \\
                 &  & $(L \chi)(Le^{\rm c})$ \\
         \hline
                $({\mathbf 1}, {\mathbf 1}, 1)$  & $-$ & $(e^{\rm c}  \chi)(LL)$ \\
         \hline
         \hline
    \end{tabular}
    \caption{Baryon-number violating (BNV) and lepton-number violating (LNV) dimension-six DM EFT operators that can be generated by integrating out scalars $\Phi_a$ in various representations of  $SU(3)_c\times SU(2)_w\times U(1)_y$, chosen to allow a  $\Phi_a^{(*)}$--$\chi_I$--SM-fermion Yukawa coupling.  All flavor indices are suppressed.     
    }
    \label{tab:optab}
\end{table}

Expanding our scope further, we can also consider $\Phi_a$ with various quantum numbers that allow renormalizable couplings of these scalars to $\chi_I$ and SM fermions.  Integrating out the $\Phi_a$ scalars generates the dimension-six DM EFT operators of Table \ref{tab:optab}. We continue to ignore operators with two $\chi$ fields, which affect DM production without contributing to baryogenesis.
Each EFT operator violates either lepton number or baryon number.  Proton-decay constraints on the baryon-number-violating (BNV) operators are generally too stringent for those operators to be relevant for baryogenesis in our scenario:~the reason is that viable baryogenesis requires $\chi$ to be much lighter than the proton mass such that proton decay modes like $p\to\pi^+\chi$ cannot be kinematically forbidden.  While we do not attempt a comprehensive study of the cosmology of various LNV operators, we expect that the results we obtain for the $(L \chi)(Qd^{\rm c})$ and $(L \chi)(L e^{\rm c})$ operators in Sec.~\ref{sec:UV} are reasonably representative, and that all of the LNV operators are therefore viable for DM and baryogenesis.

Regardless of which LNV operator we consider, the neutrino-portal operator $\chi L H$ is radiatively generated at some level.  After electroweak symmetry breaking, this operator leads to DM decay to a photon and a neutrino or antineutrino, in potential conflict with photon-line searches for decaying DM. For the  LNV operators in Table \ref{tab:optab} that involve the SM lepton doublets, the operator is generically induced at one-loop via a diagram obtained by connecting the appropriate SM fermion lines to the SM Higgs via a SM Yukawa interaction.  The associated contributions to the neutrino-portal coupling are suppressed by a loop factor, the DM and $U(1)_{B-L}$-violating couplings, and a SM Yukawa coupling.  In Sec.~\ref{sec:xray} we will see that parameter values that are viable for baryogenesis and DM require additional suppressions or cancellations  to be consistent with line searches.  

The flavor structures of the EFT operators do allow for such additional suppressions.  Suppose, for example, that the fully leptonic couplings $E_{\alpha \beta \gamma I}$ of Eq.~(\ref{eq:EFT_2}) are non-zero only when  $\gamma \neq \alpha$ and $\gamma \neq \beta$ are both satisfied.  This corresponds to there being no  overlap between the flavors of left-handed (LH) leptons and the flavors of right-handed (RH) leptons involved in  the beyond-SM couplings.   In this case, no neutrino-portal coupling is generated radiatively at any order in perturbation theory, in the approximation where we neglect SM neutrino masses.  The $L\chi H$ coupling induced by the semileptonic operator  $(L \chi)(Qd^{\rm c})$ can also be suppressed for particular flavor structures, although in this case simply making the EFT operator off-diagonal in quark flavor space (in the basis in which the down-type Yukawa couplings are diagonal) still leaves two-loop contributions  involving  two additional up-type-quark Yukawa couplings and a CKM mixing.     

In Sec.~\ref{sec:xray}, we provide more details of the DM lifetime in various scenarios. Here, we provide a parametric estimate in the case where only $\lambda'\neq0$ and the only flavor of lepton that shows up in both LH and RH fields in Eq.~\eqref{eq:EFT_2} is the electron. In this case, the DM lifetime is approximately
\small
\be
\tau_{\chi_2\to \gamma\nu} \sim 2\times10^{28}\,\,\mathrm{s}\left(\frac{100\,\,\mathrm{keV}}{M_2}\right)^3\left(\frac{M_\Phi}{1\,\,\mathrm{TeV}}\right)^4\left(\frac{10^{-9}}{F\lambda'}\right)^2.
\ee
\normalsize
While the age of the Universe is only $\sim10^{18}\,\,\mathrm{s}$, there are much stronger mass-dependent constraints on DM lifetime from X-ray lines ($\sim10^{28}\,\,\mathrm{s}$ for DM at the 100 keV scale). UV leptogenesis only depends on the quantity $F\lambda'/M_\Phi^2$; we see that this quantity should be $\sim10^{-30}\,\,\mathrm{GeV}^{-4}$, which corresponds to $M_\Phi\sim3\times10^7\,\,\mathrm{GeV}$ if $F,\lambda'\sim1$, with lower scalar masses giving the same decay rate if $F,\lambda'$ are smaller than 1.

The one-loop contributions to the neutrino-portal coupling induced in the EFT by $(L \chi)(Qd^{\rm c})$ can be understood as a consequence of one-loop $H-\Phi_a$ mixing in the UV theory, which only arises when the $\Phi_a$ have the same SM gauge charges as the SM Higgs.  Integrating out $\Phi_a$ scalars that transform instead as $\sim({\mathbf 3}, {\mathbf 1}, -1/3)$ under the SM gauge group
can generate a $(\chi d^{\rm c})({u^{\rm c}}^\dagger {e^{\rm c}}^\dagger)$ LNV operator in the EFT.  The neutrino-portal coupling is induced by this operator only at two loops, and it is suppressed by charged-lepton, up-type-quark, and down-type-quark Yukawa couplings, in addition to DM and $U(1)_{B-L}$-violating couplings.  We estimate that with particular flavor choices, the contributions to the neutrino-portal coupling can be rendered below current limits.

\subsection{Mixed UV-IR Scenario}

It is possible that there exists a hierarchy of scalar masses such that $M_{\Phi_1}\lesssim T_{\rm RH}\ll M_{\Phi_2}$. In this case, our theory consists of the EFT of Eq.~\eqref{eq:EFT}, along with the coupling of an on-shell scalar as in Eq.~\eqref{eq:IR_lag}. We refer to this as the ``mixed UV-IR'' scenario.

In this case, the results for DM and baryogenesis are very similar to the case where there exist two on-shell scalars with $M_{\Phi_1}\ll M_{\Phi_2}$:~a primordial abundance of DM (whether produced from virtual $\Phi_2$-mediated scattering or from on-shell $\Phi_2$ decays) undergoes inverse-decay at much later times into $\Phi_1$, leading to an asymmetry at $\mathcal{O}(F^4)$. When there exists a hierarchy of scales in the model such that a primordial abundance of DM is produced at temperatures $T\gg M_{\Phi_1}$, it doesn't matter much whether DM is produced by on-shell $\Phi_2$ decays or $2\to2$ scatterings mediated by virtual $\Phi_2$ exchange. In fact, for a fixed primordial DM abundance, the only difference between the two scenarios is the DM spectrum, which is hotter for production in the EFT relative to on-shell decays.

We do not undertake a full study of the mixed UV-IR scenario because of the similarity to earlier results for the two-scalar case from Ref.~\cite{Berman:2022oht}. However, the mixed UV-IR scenario does provide a controlled testing ground to better understand the effects of the DM spectrum on the parameter space of leptogenesis. We return to this study in Sec.~\ref{sec:mixed}.

\section{Fully UV Baryogenesis}\label{sec:UV}

\subsection{Quantum Kinetic Equations} \label{sec:QKE}

 The DM quantum state is represented by a pair of density matrices, $f_\chi$ and $f_{\bar\chi}\equiv\bar f_\chi$, whose evolution is determined by a set of quantum kinetic equations (QKEs) for each comoving DM momentum mode $y\equiv |\vec{k}|/T$ \cite{Akhmedov:1998qx,Asaka:2005pn,Hambye:2017elz}. For the parameters of interest to us, DM is always highly relativistic and so we approximate $E_\chi \approx |\vec{k}|$ everywhere except in the QKE term inducing oscillations. Oscillations are included by keeping the leading mass-dependent correction to the DM dispersion relation. The QKE for $f_\chi$ has the form
\be\label{eq:QKE_general}
\frac{df_\chi}{dt} &=& -i\left[ E_\chi, f_\chi \right]\nonumber\\
&&{}-\frac{1}{2}\sum_\alpha\left(\left\{\frac{\Gamma_\alpha^>}{2E_\chi},f_\chi\right\}-\left\{\frac{\Gamma_\alpha^<}{2E_\chi},1-f_\chi\right\}\right),
\ee
where $\Gamma^>_\alpha/2E_\chi$ and $\Gamma^<_\alpha/2E_\chi$ are the DM absorption and emission rates, respectively, for collisions involving lepton flavor $\alpha$ and DM comoving momentum $y$. For our main analysis, we consider DM production via the semileptonic operator in Eq.~\eqref{eq:EFT}. The DM absorption and emission rates are independent of $f_\chi$ but depend on the SM chemical potentials $\mu_{L_\alpha}$, $\mu_Q$, and $\mu_{d^{\rm c}}$. The QKE for $\bar f_\chi$ is found by taking $D\to D^*$ and $\mu\to-\mu$ relative to the QKE for $f_\chi$\footnote{This prescription  depends on the convention for how the $\bar\chi$ density matrix is defined, with several conventions appearing in the literature and not always clearly articulated. For the $D$ and $E$ EFT couplings (in Eqs.~\eqref{eq:EFT} and \eqref{eq:EFT_2}, respectively), for which each collision involves at most one DM particle, the choice of convention is irrelevant. When multiple DM states can interact in a single collision, such as for the $C$ EFT coupling in Eq.~\eqref{eq:EFT}, it is important to implement the QKEs consistently. We address this further in Appendix \ref{app:comment-on-chibar}. }. The energy $E_\chi$ in the commutator includes both tree-level masses and thermal corrections; since the thermal masses are aligned with the interaction eigenstates, they tend to suppress oscillations when they are the dominant contributions to the $\chi$ masses.

During the epoch of leptogenesis, there exists a generalized lepton number with $L(\chi) = -1$. This symmetry is broken explicitly by the $\chi$ Majorana masses, but helicity-flipping rates leading to lepton-number violation are suppressed by ${M_{\chi}^2}/{T^2}\ll1$ relative to the lepton-number-conserving rates and are irrelevant for $T>T_{\rm ew}$. This means that  the lepton asymmetry in flavor $\alpha$ is equal to a $\chi$ asymmetry produced in scatterings involving flavor $\alpha$. This allows us to derive evolution equations for the anomaly-free SM charge asymmetries $X_\alpha \equiv \Delta B/3-\Delta L_\alpha$ by using the $\chi$ and $\bar\chi$ QKEs to determine the production rate of an asymmetry between $\chi$ and $\bar\chi$ due to scattering with leptons of flavor $\alpha$. Note that we define $X_\alpha$ to include only the asymmetries in SM fields. We provide further details on the derivation of the QKEs in Appendix \ref{app:QKE}, including full expressions for the QKEs that we solve in our analysis.

\begin{figure}[t]
          \includegraphics[width=0.48\textwidth]{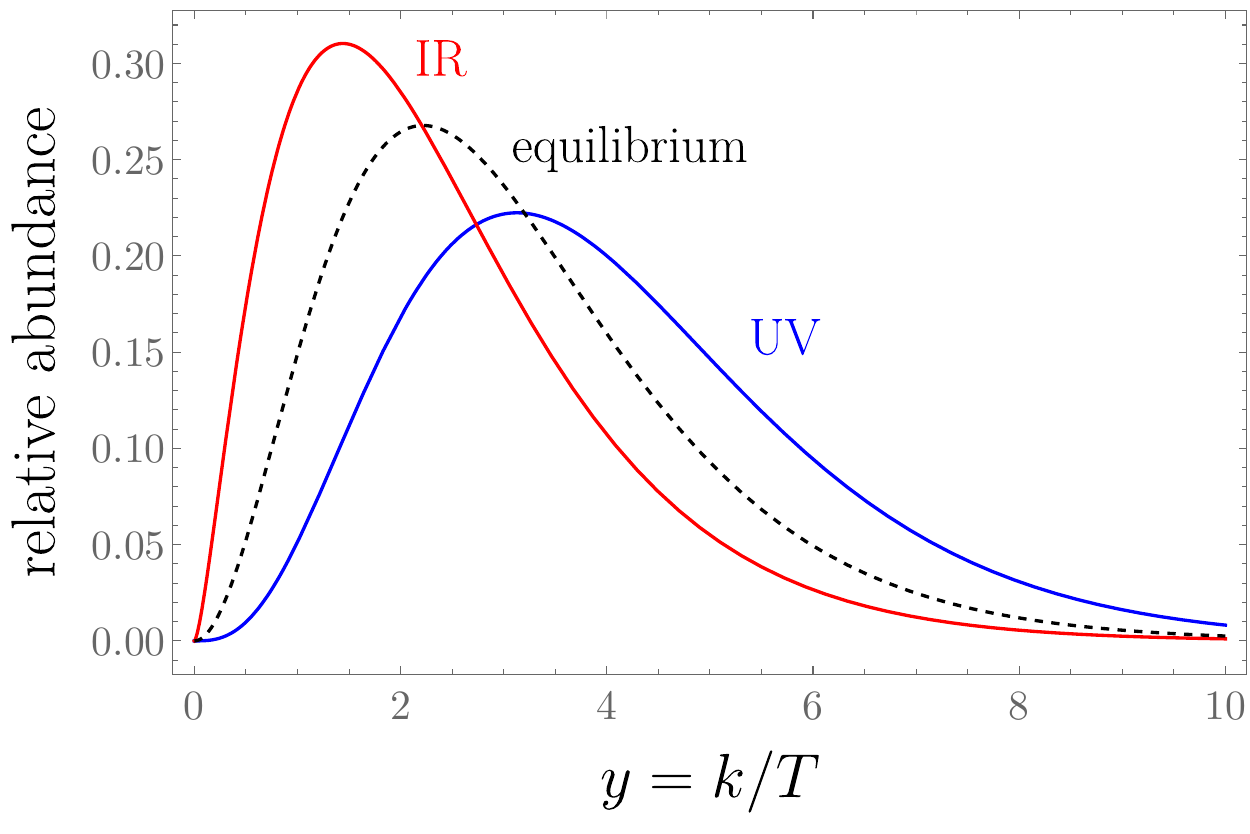}
\caption{
Comoving momentum spectrum of DM particles produced from (red) on-shell decay of $\Phi_a$; (blue) $2\to2$ scattering in the EFT with $\Phi_a$ integrated out; (dashed) Fermi-Dirac distribution. 
}
\label{fig:DM-spectrum}
\end{figure}

The QKEs in Eq.~\eqref{eq:QKE_general} are challenging to solve because each DM comoving momentum mode $y$ has its own differential equation, reflecting its own oscillation and equilibration time scales. All of these equations must be separately integrated to determine the values of $f_\chi$ and $\bar f_\chi$ for each momentum mode. If the DM spectrum is not too far from a Fermi-Dirac distribution, however, then we can simplify the QKEs by assuming that DM has a thermal spectrum, $f_\chi \propto f_{\rm Fermi-Dirac}$. We  then integrate over all momentum modes and  solve the resulting momentum-integrated QKEs,  significantly reducing the number of separate equations to solve. We refer to this as making a ``thermal ansatz'' for the DM spectrum. 

To motivate the use of a thermal ansatz, we show in Fig.~\ref{fig:DM-spectrum} the DM spectrum from UV production in the EFT, as well as from IR production via on-shell $\Phi_a$ decays~\cite{Berman:2022oht}, comparing both to a thermal distribution. The UV spectrum is given in Eq.~\eqref{eq:DM-spectrum}. The peaks differ by only an $\mathcal{O}(1)$ number from the Fermi-Dirac distribution, although the tails differ by larger factors. Based on the results of previous studies \cite{Ghiglieri:2017csp,Ghiglieri:2018wbs,Shuve:2020evk,Berman:2022oht}, we expect that the thermal ansatz induces at most $\mathcal{O}(1)$ shifts in the final DM abundances and baryon asymmetries obtained from the QKEs, apart from corners of parameter space that are sensitive to the tails of the spectrum. For the results in Sec.~\ref{sec:UV_full_results}, we employ the thermal ansatz to solve the QKEs, whereas we re-visit the validity of this approximation and the dependence of leptogenesis on the DM spectrum in Secs.~\ref{sec:UV_pert} and \ref{sec:mixed}.

Throughout, we normalize the DM abundance to the entropy density, $Y_\chi \equiv n_\chi/s$, and we use the dimensionless time variable $z\equiv T_{\rm ew}/T$ in solving our equations and reporting solutions.

\subsection{Results and Viable Parameters} \label{sec:UV_full_results}

We now determine the viable parameter space for baryogenesis and DM in the fully UV scenario. We use the semileptonic coupling benchmark in Appendix \ref{app:coupling_benchmark}, which provides the maximum asymmetry in the weak-washout limit relative to the DM abundance. In this benchmark, DM is only coupled to two lepton flavors because that is the minimum needed for an asymmetry to be generated. We expect the two-flavor benchmark to capture all of the relevant physics of leptogenesis while reducing the number of free parameters to consider. This allows us to more clearly understand the physics underlying the coupling structures that lead to successful leptogenesis.  

There are five free parameters:
\begin{itemize}

\item $T_{\rm RH}$, the reheat temperature (equivalently,  $z_{\rm RH}\equiv T_{\rm EW}/T_{\rm RH}$);
\item $M_1$, the mass of the lighter DM eigenstate;
\item $M_2$, the mass of the heavier DM eigenstate;
\item $\mathrm{Tr}D^\dagger D$, which parametrizes the overall (squared) coupling strength of leptons to DM, summed over both DM flavors;
\item $\theta$, an angle whose tangent parametrizes the  strength of the mediator coupling to $\chi_2$ relative to $\chi_1$: $\tan\theta \equiv \sqrt{(D^\dagger D)_{22}/(D^\dagger D)_{11}}$.

\end{itemize}

\begin{figure}[t]
          \includegraphics[width=3in]{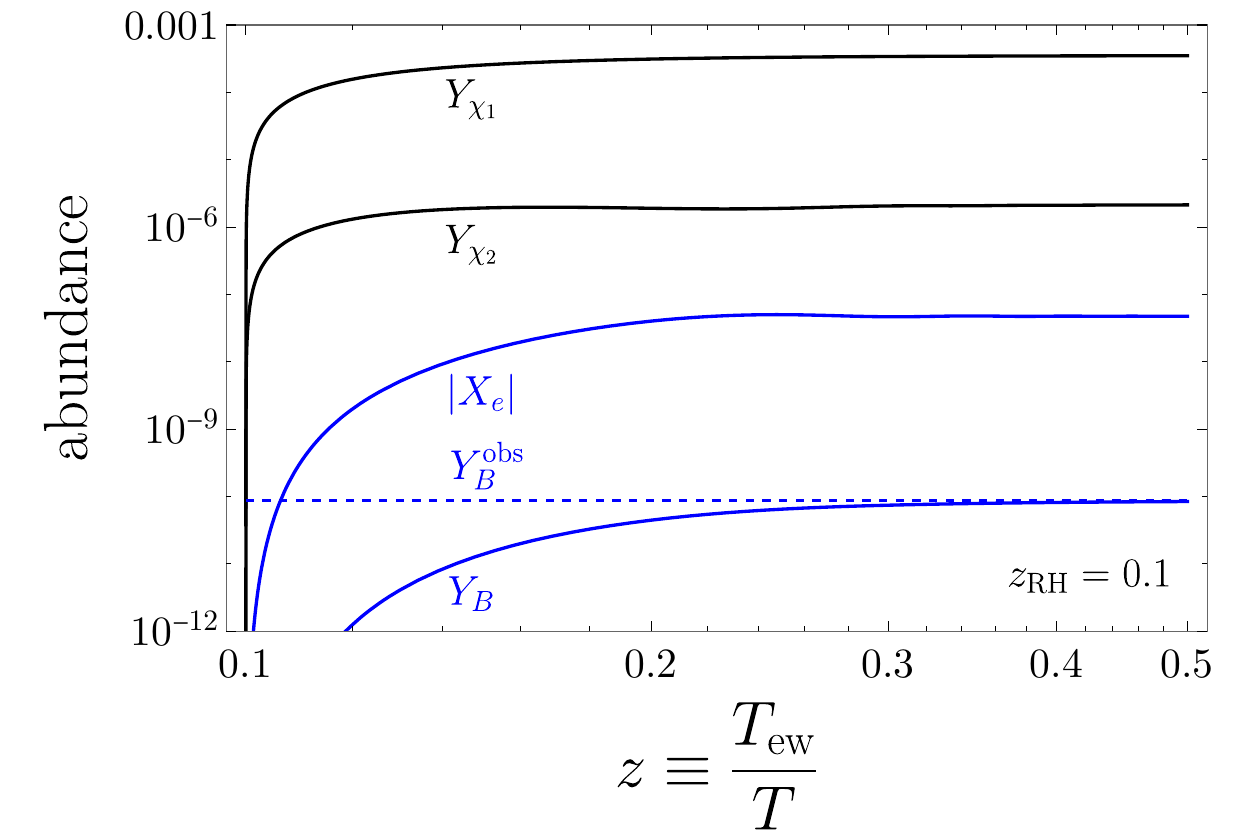}
\caption{
DM abundances, electron flavor asymmetry $X_e\equiv B/3-\Delta L_e$, and baryon asymmetry $Y_B$ as a function of dimensionless time, $z\equiv T_{\rm ew}/T$, for a benchmark point that simultaneously gives the correct DM energy density and baryon asymmetry:~$z_{\rm RH}=0.1$, $M_1=0$, $M_2=100$~keV, $\theta=0.092$, $\mathrm{Tr}D^\dagger D=8.08\times10^{-28}\,\,\mathrm{GeV}^{-4}$. The muon flavor asymmetry, $|X_\mu|$, is not shown because it is nearly indistinguishable from $|X_e|$.
}
\label{fig:DM_asym_timevol}
\end{figure}

The solutions to the QKEs for the DM abundances, electron flavor asymmetry, and baryon asymmetry are shown as a function of $z$ in Fig.~\ref{fig:DM_asym_timevol} for a benchmark point that gives the correct DM energy density. The UV nature of freeze-in is evident, with most production of DM and the baryon asymmetry occurring around $T_{\rm RH}$. The asymmetry accrues notably later than the DM yields, however, because sufficient time must pass for DM oscillations to develop.

\begin{figure*}[t]
          \includegraphics[width=3in]{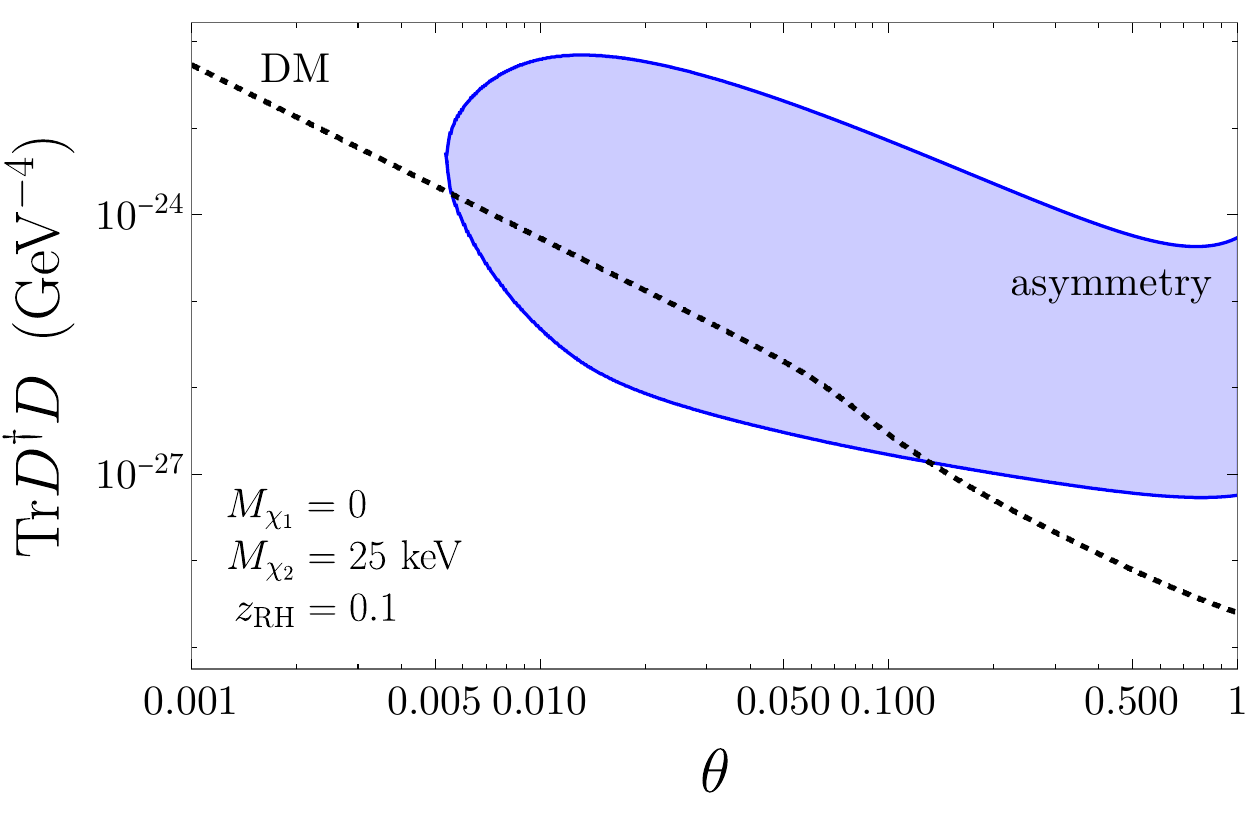}
           \quad \quad \quad
    \includegraphics[width=3in]{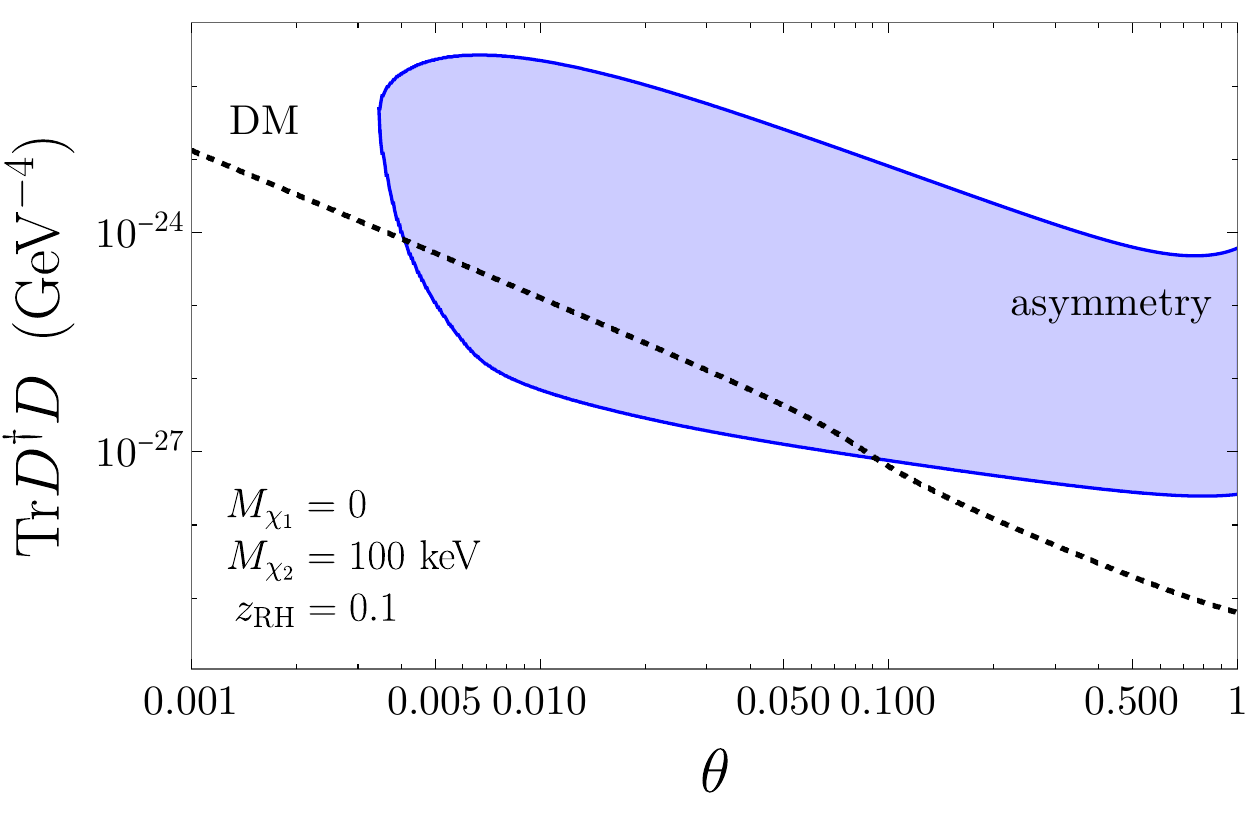}\\
    \includegraphics[width=3in]{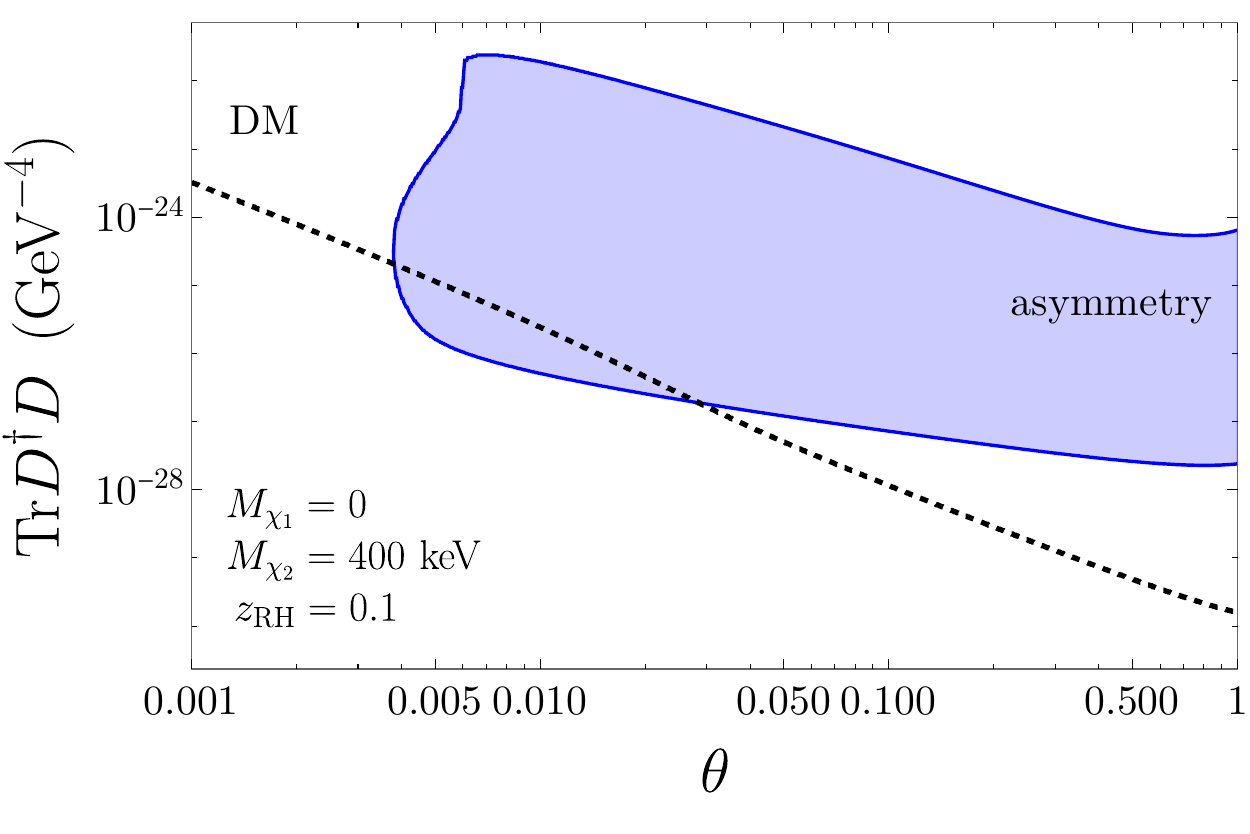}
           \quad \quad \quad
    \includegraphics[width=3in]{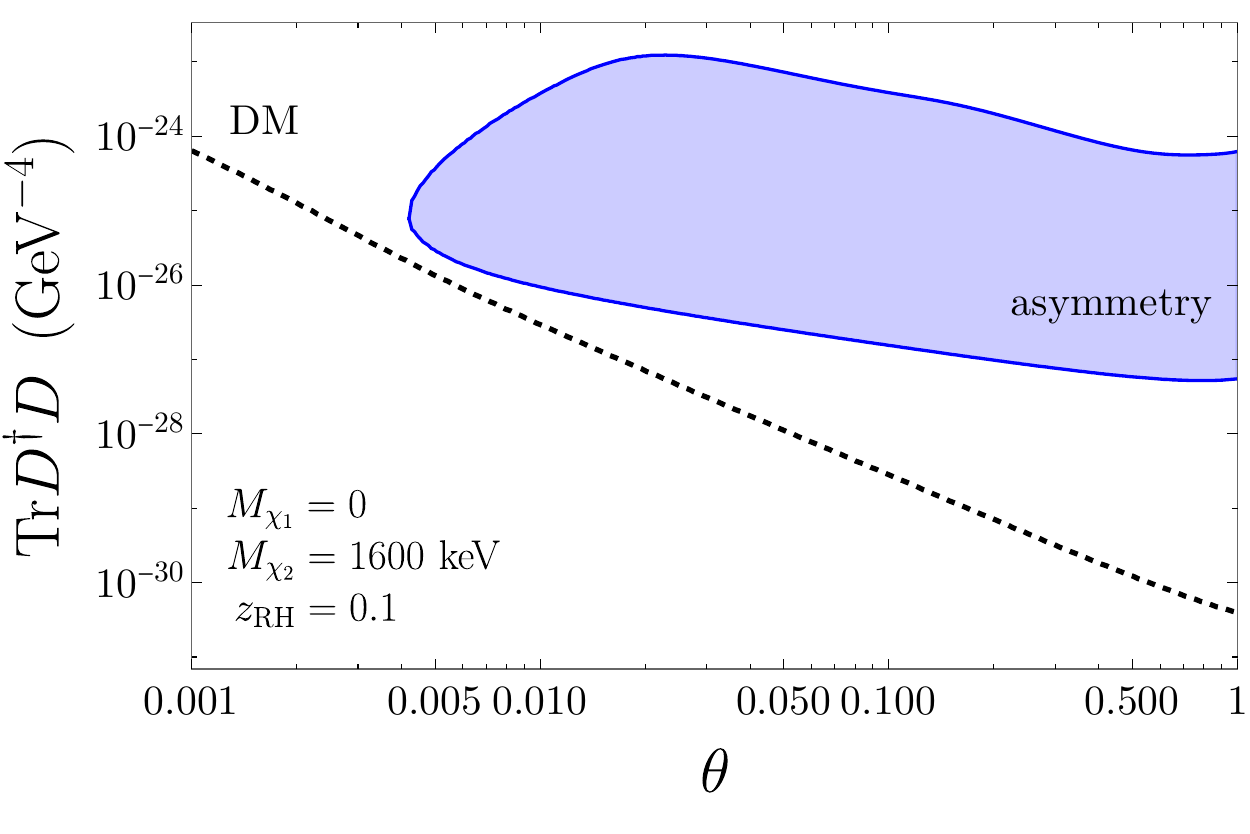}
\caption{
Coupling consistent with observed baryon asymmetry (shaded blue region) and DM abundance (dashed line) from freeze-in baryogenesis with $z_{\rm RH} \equiv T_{\rm ew}/T_{\rm RH} = 0.1$ and (top left) $M_2=25$~keV; (top right) $M_2=100$~keV; (bottom left) $M_2=400$~keV; (bottom right) $M_2=1600$~keV. We take $M_1=0$ and use the optimized semileptonic coupling benchmark from Appendix \ref{app:coupling_benchmark}; we consider a parameter point to successfully achieve the baryon asymmetry if $Y_B\ge Y_B^{\rm obs}$, since it is always possible to make the asymmetry smaller than predicted by the benchmark by modifying  $CP$-violating phases in the couplings. 
For sufficiently large couplings, the asymmetry is suppressed by washout effects, which explains why the shaded blue regions are bounded from above.
}
\label{fig:UV_masslesschi1_z0_0p1}
\end{figure*}

We now systematically explore the viable parameter space for DM and baryogenesis. We first consider the scenario with $M_1=0$. Because baryogenesis occurs at higher order in the freeze-in coupling than DM production, freeze-in baryogenesis typically overproduces the DM abundance while underproducing the asymmetry. The requirement that $M_1=0$ therefore minimizes the DM abundance without affecting baryogenesis, giving the largest possible parameter space. Because it is always possible to \emph{lower} the baryon asymmetry by changing the $CP$-violating phase or other coupling parameters away from the optimal value of our benchmark from Appendix \ref{eq:UV_benchmark}, we consider a parameter point to be viable if it gives $Y_B/Y_{\rm obs} \ge 1$ while simultaneously satisfying $\Omega_{\chi_2+\bar\chi_2}(z\to\infty)=\Omega_{\rm DM}^{\rm obs}$.

We present our results by fixing values of $(z_{\rm RH},M_2)$, and then plotting the two coupling contours that give the observed baryon asymmetry and DM abundance, respectively\footnote{We use $Y_B^{\rm obs}\approx9.7\times10^{-11}$ and $\rho_{\rm cdm}/s \equiv M_1 Y_{\chi_1+\bar\chi_1}+M_2 Y_{\chi_2+\bar\chi_2} \approx 4.32\times10^{-10}$ GeV.}. Our results for $z_{\rm RH}=0.1$ (corresponding to $T_{\rm RH}\approx1.3$~TeV) and various values of $M_2$ are shown in Fig.~\ref{fig:UV_masslesschi1_z0_0p1}. There are two competing effects of the DM mass scale. For smaller values of $M_2$,   there exists a larger number density of DM particles for fixed $\Omega_{\rm DM}$. This predicts a larger coupling, which in turn enhances the baryon asymmetry. At the same time,  asymmetry generation is most efficient when the squared mass splitting $\Delta M^2$ is such that the oscillation rate is matched to the Hubble expansion rate around the time of reheating, Eq.~\eqref{eq:osc-RH}. For $z_{\rm RH}=0.1$ and $M_1=0$, this optimal mass splitting corresponds to $M_2\approx 150$~keV; we therefore see the parameter space shrink for larger DM masses than this, with  no viable parameter space  for $M_2\gtrsim1500$~keV.

\begin{figure*}[t]
          \includegraphics[width=3in]{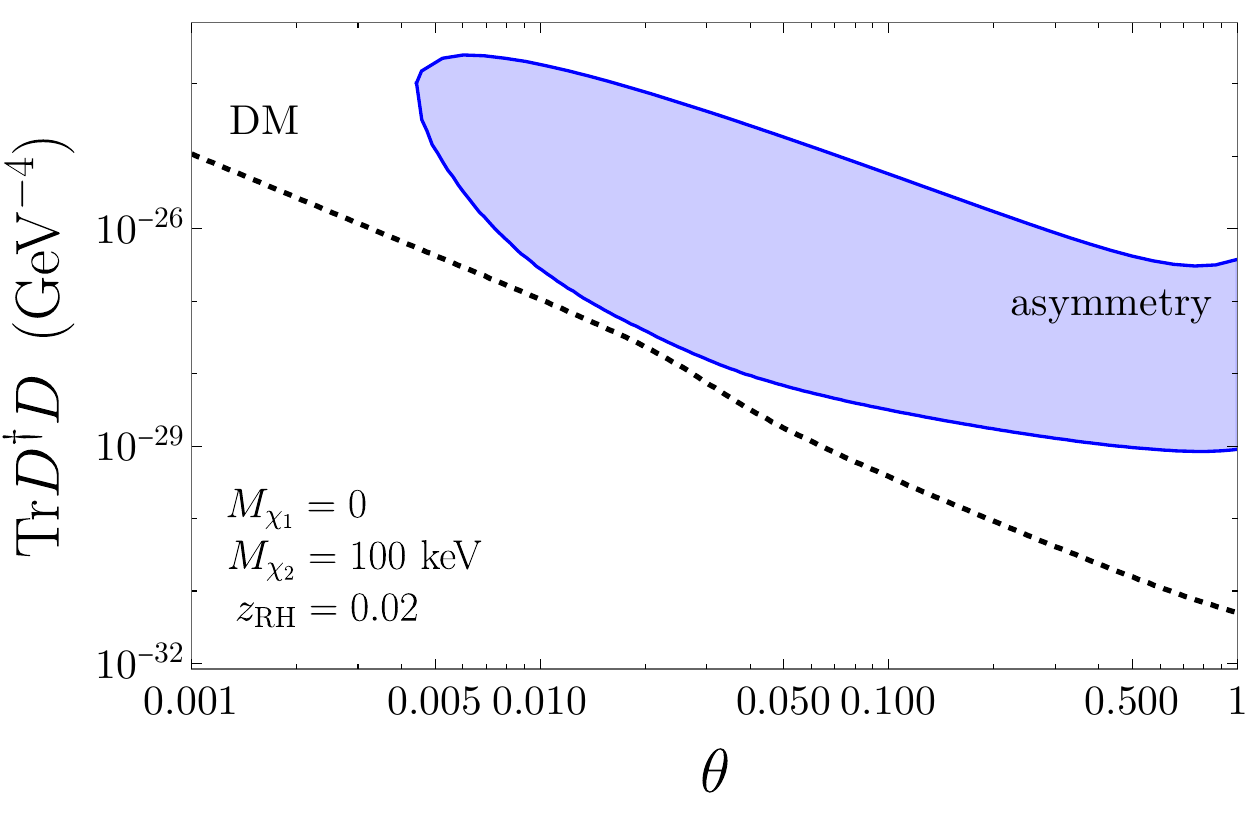}
           \quad \quad \quad
    \includegraphics[width=3in]{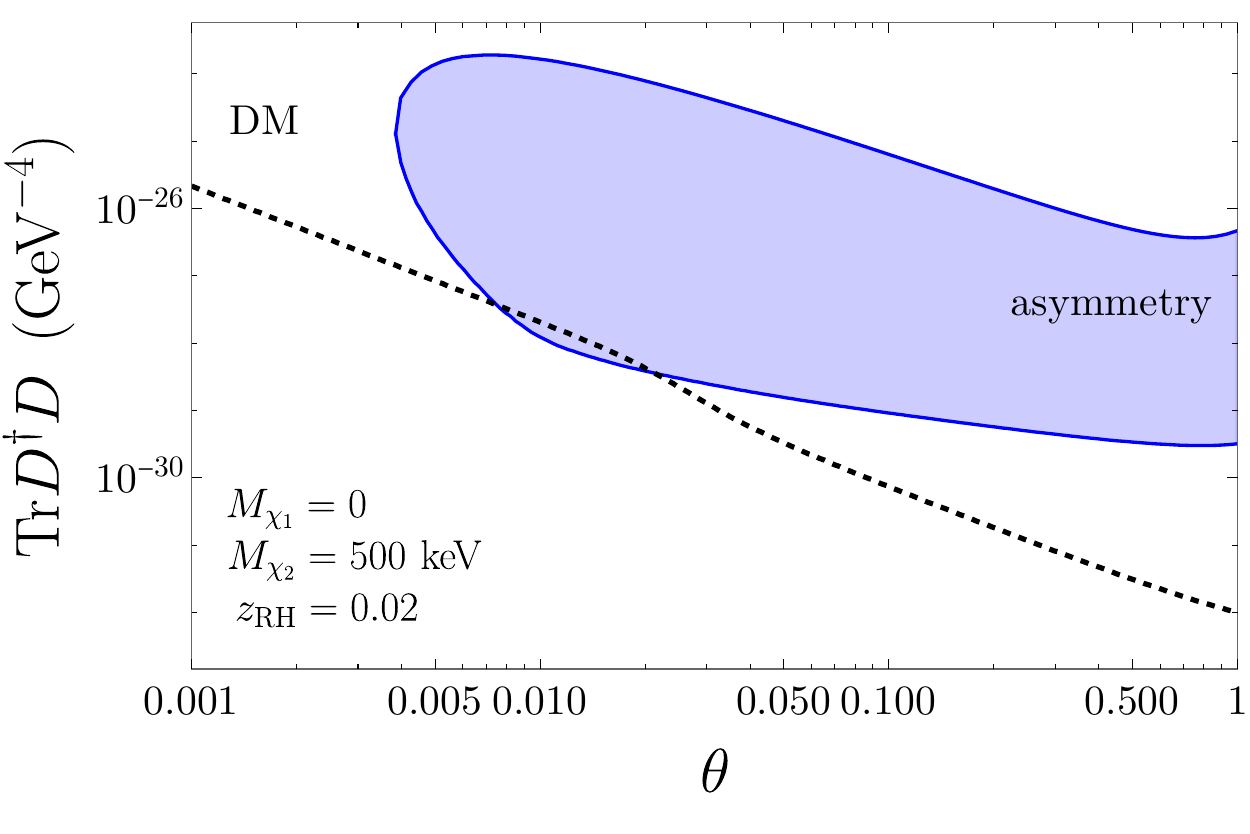}\\
    \includegraphics[width=3in]{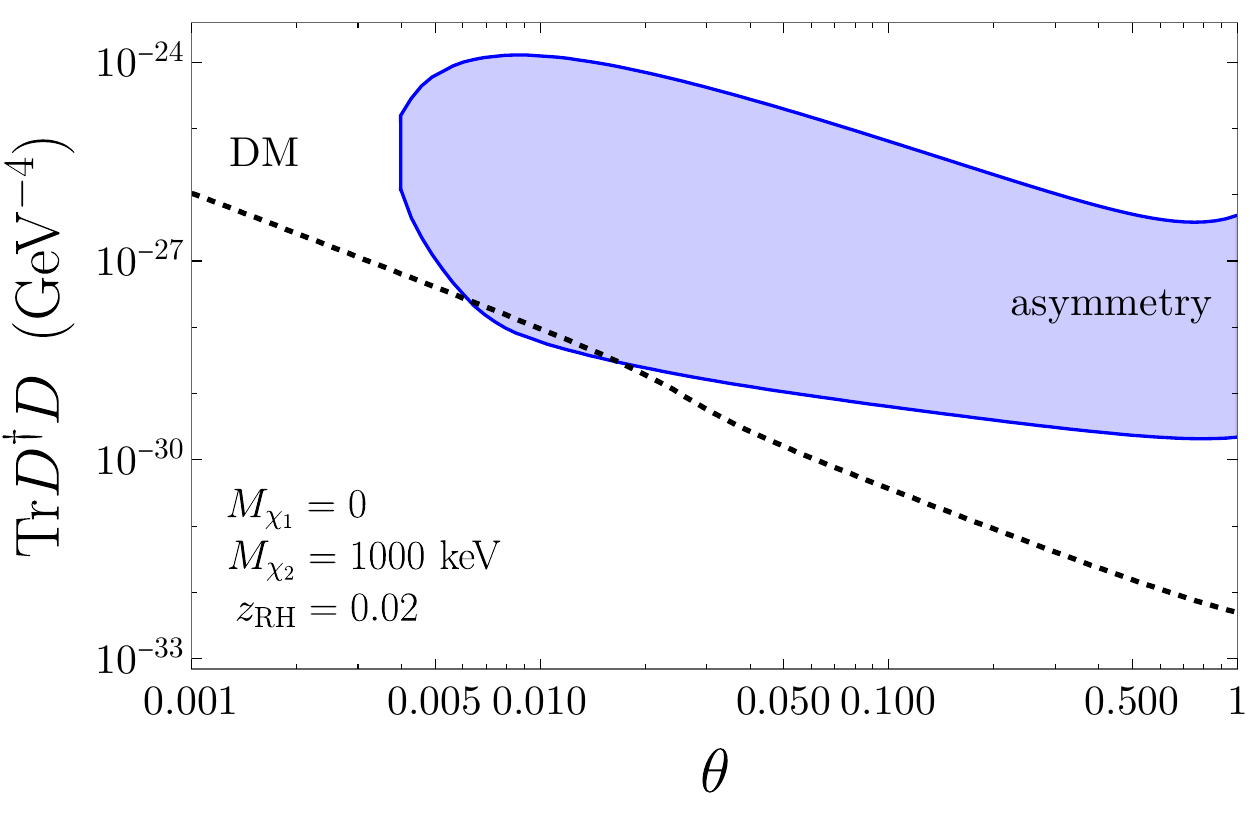}
           \quad \quad \quad
    \includegraphics[width=3in]{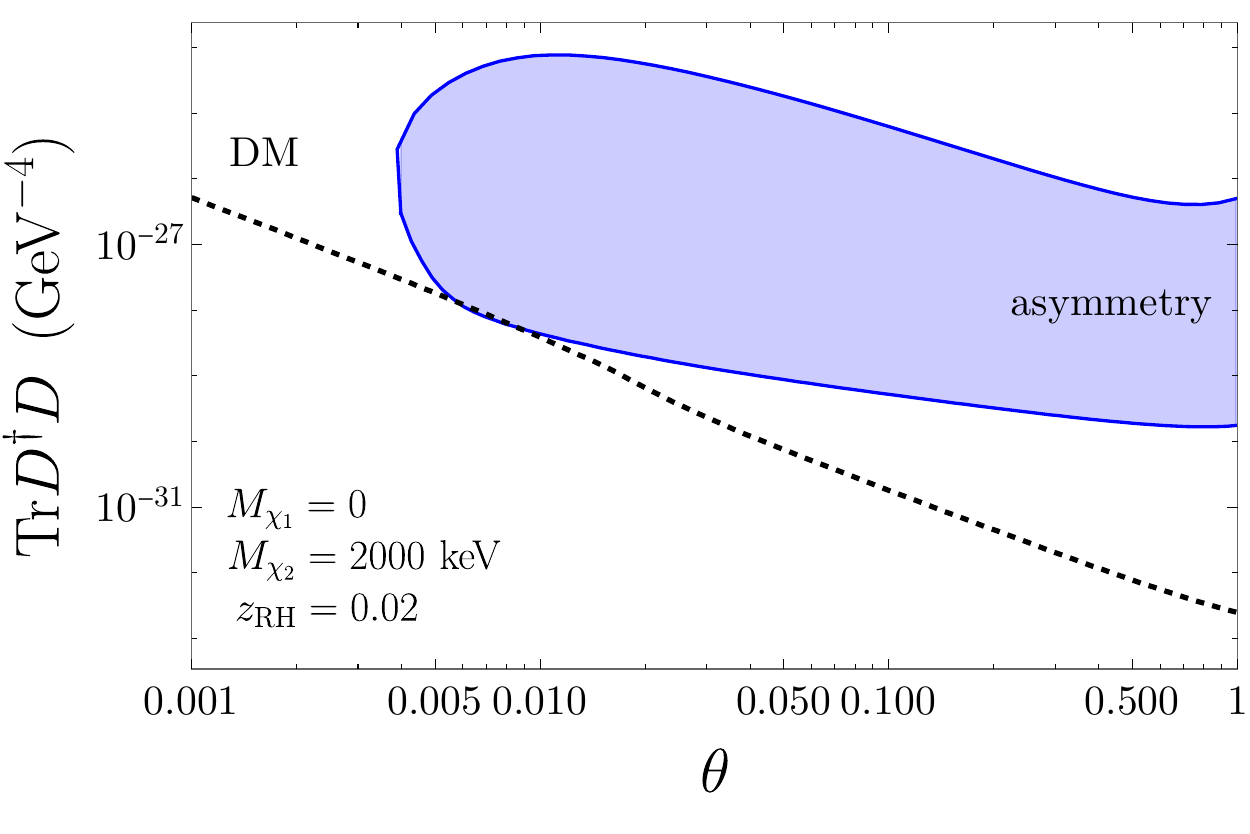}
\caption{
Coupling consistent with observed baryon asymmetry (shaded blue region) and DM abundance (dashed line) from freeze-in baryogenesis with $z_{\rm RH} \equiv T_{\rm ew}/T_{\rm RH} = 0.02$ and (top left) $M_2=100$~keV; (top right) $M_2=500$~keV; (bottom left) $M_2=1000$~keV; (bottom right) $M_2=2000$~keV. We take $M_1=0$ and other parameters the same as in Fig.~\ref{fig:UV_masslesschi1_z0_0p1}.
}
\label{fig:UV_masslesschi1_z0_0p02}
\end{figure*}

All of the viable parameter space has $\theta\ll1$. To understand this, note that the largest asymmetry results when one of the flavors of DM comes close to equilibrium (in this case, $\chi_1$), in which case there is a large $\chi_1$ abundance to participate in the second stage of scattering necessary for baryogenesis. When $M_1\approx 0$, $\chi_1$ is effectively  dark radiation that redshifts away and contributes negligibly to the Universe's energy budget today even if it comes into equilibrium. It does contribute to the effective number of neutrino species, $N_{\rm eff}$; however, current bounds  allow an equilibrium Weyl fermion contribution to $N_{\rm eff}$ provided it decouples prior to the electroweak phase transition \cite{Brust:2013ova,Antel:2023hkf}.  In contrast, $\chi_2$ must always remain far from equilibrium to avoid overclosing the Universe, and this means that a larger asymmetry is generated when $Y_{\chi_2}\ll Y_{\chi_1}$. Since $\theta$ parametrizes the relative coupling of the SM to $\chi_2$ vs.~$\chi_1$, the ratio of baryon asymmetry to DM abundance is maximized for $\theta\ll1$. Evidently, UV freeze-in for baryogenesis is only viable  in the limit where $\chi_1$ couples much more strongly to the SM than $\chi_2$.

The limit $\theta\ll1$ is stable under radiative corrections. To see this, we note that the limit $\theta\to0$ results in an additional $Z_2$ symmetry under which only $\chi_2$ is charged. Thus, small values of $\theta$ are technically natural. \\

\noindent {\bf Reheat temperature:}~we now investigate the viable parameter space for different reheat temperatures. The primary effect of raising the reheat temperature is that DM and baryogenesis occur earlier, raising the optimal value of $\Delta M^2$ according to Eq.~\eqref{eq:osc-RH}. Consequently, DM is heavier, exacerbating the tension of obtaining the observed baryon asymmetry without overproducing DM.

We show results for $z_{\rm RH}=0.02$ (corresponding to $T_{\rm RH}\approx 6.6$~TeV) in Fig.~\ref{fig:UV_masslesschi1_z0_0p02}. For oscillations to occur around the time of reheating, the optimal DM mass is $M_2\approx 1500$~keV. Consequently, we see a significantly reduced viable parameter space at small values of $M_2$; in fact, there is no parameter space for baryogenesis without overclosing the Universe for $M_2\lsim 150$~keV.  There is a small region of viable parameter space in the range $150~\mathrm{keV}\lsim M_2\lsim 2000~\mathrm{keV}$.

We find that for $z_{\rm RH}<0.0139$ (corresponding to $T_{\rm RH}\gtrsim 9.5$~TeV), it is no longer possible to simultaneously account for DM and baryogenesis. Conversely, if the reheat temperature is too low to restore electroweak symmetry, the baryon asymmetry vanishes due to the absence of sphaleron processes. The simultaneous success of UV freeze-in DM and baryogenesis thus requires reheat temperatures between $T_{\rm ew}$ and about 10 TeV. \\

\noindent {\bf Massive $\chi_1$:}~When both DM eigenstates are massive, the range of viable parameters is smaller because there is an additional contribution to the DM energy density from $\chi_1$, but with no corresponding enhancement to the baryon asymmetry. To determine the extent of the viable parameter space, we fix $z_{\rm RH}$ and $M_2$ as before, and we vary $\theta$ and $M_1$. For each parameter point, we fix the overall squared coupling magnitude $\mathrm{Tr}D^\dagger D$ so that the combined $\chi_1$ and $\chi_2$ energy densities today match the observed DM energy density. 

\begin{figure*}[t]
          \includegraphics[width=3in]{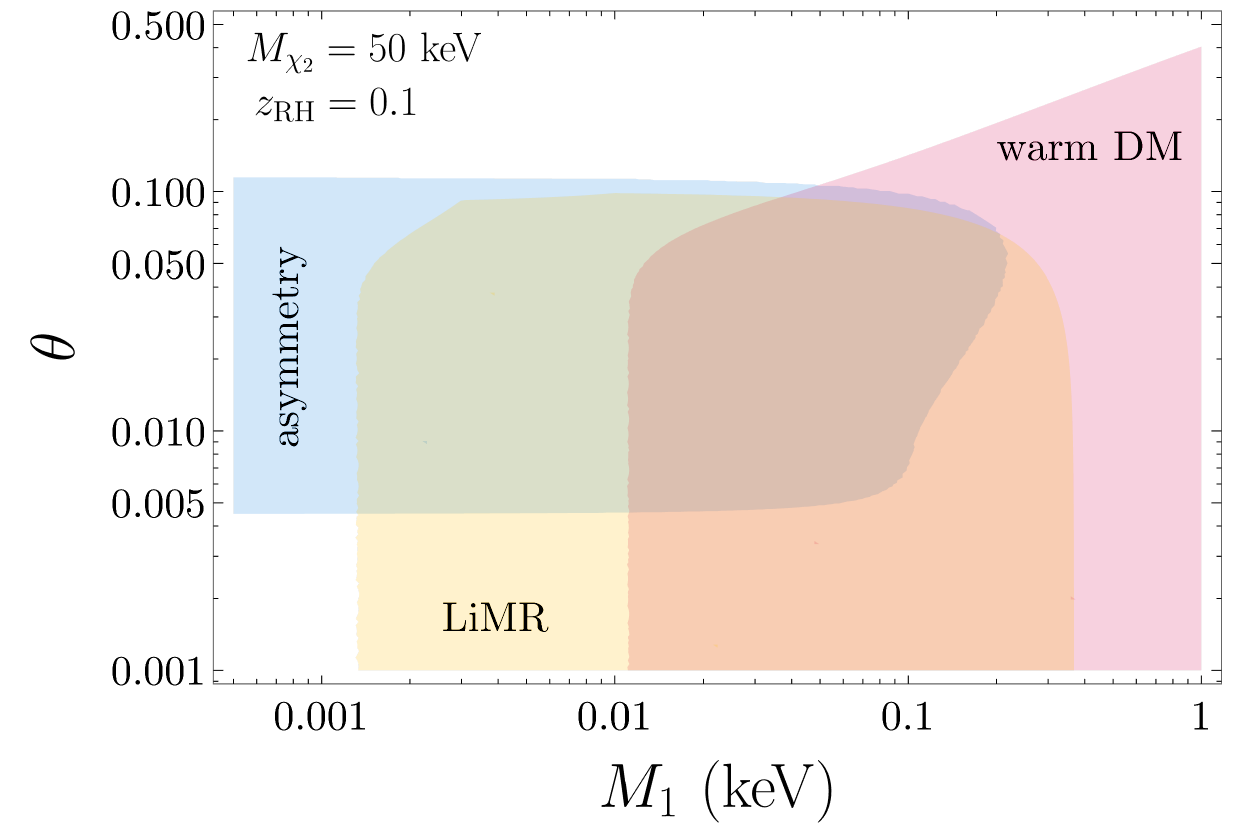}
           \quad \quad \quad
    \includegraphics[width=3in]{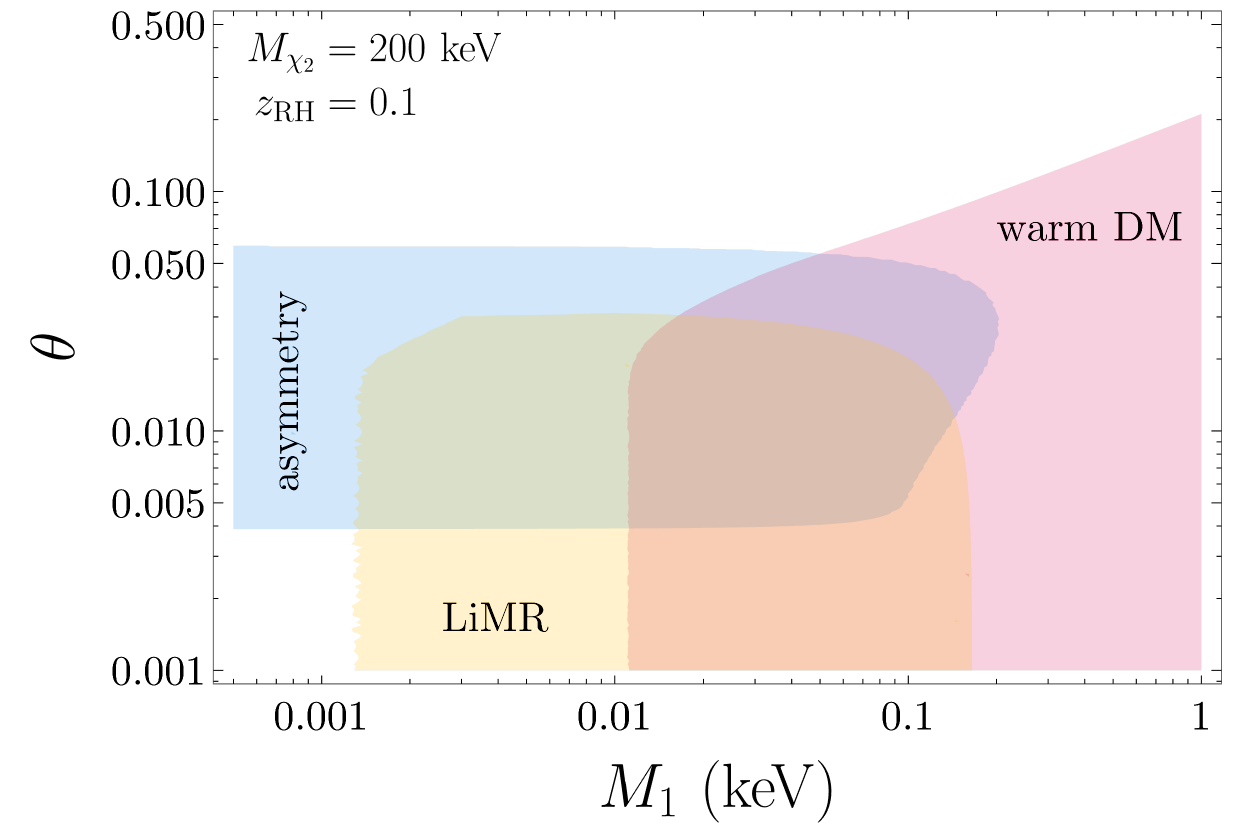}\\
    \includegraphics[width=3in]{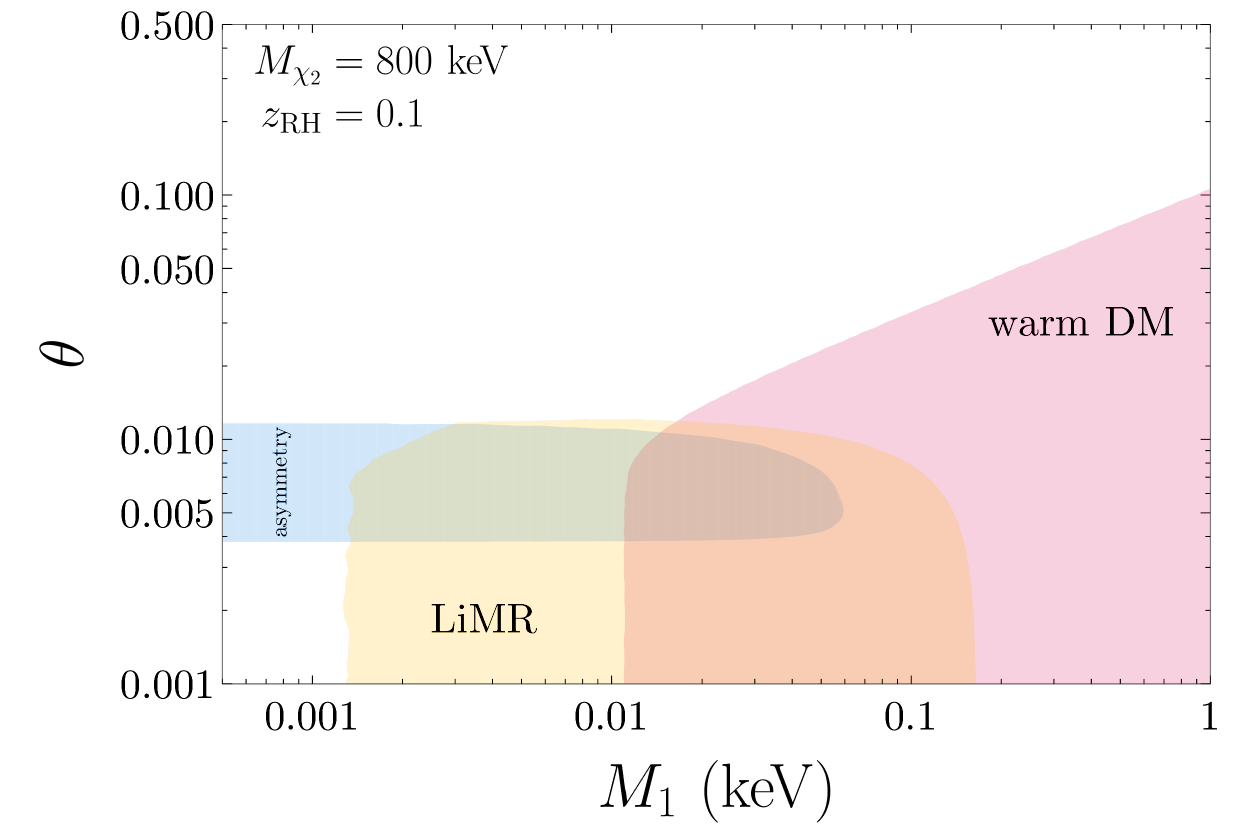}
\caption{
Parameter space for viable baryogenesis, shaded in blue, as a function of $M_1$ and coupling parameter $\theta$ for $z_{\rm RH}=0.1$ and (top left) $M_2=50$~keV; (top right) $M_2=200$~keV; (bottom) $M_2=800$~keV. The other coupling parameters are fixed to obtain the correct total DM abundance while optimizing the asymmetry.  The red and yellow shaded regions are excluded from structure formation constraints on a warm fraction of the total DM abundance ($\Omega_{\chi_1}/\Omega_{\rm DM}<0.1$) \cite{Boyarsky_2009,Kamada:2016vsc,Baur_2017}, and constraints on a light but massive relic (LiMR) from CMB, BOSS, and weak lensing \cite{Xu:2021rwg}, respectively; see Sec.~\ref{sec:structure} for more details. Note that $M_1\lesssim 0.1$~keV  for all parameters consistent with constraints giving successful leptogenesis.}
\label{fig:UV_massivechi1_z0_0p1}
\end{figure*}

We show our results in Fig.~\ref{fig:UV_massivechi1_z0_0p1} for $z_{\rm RH}=0.1$ and various values of $M_{2}$; we find similar results for other reheat temperatures. In all cases, $\chi_1$ must be significantly lighter than $\chi_2$, with $M_1\lesssim0.1$~keV to achieve successful leptogenesis and satisfy constraints. As before, we also see that the only viable parameter space for DM and leptogenesis has $\theta\ll1$:~in other words, $\chi_1$ couples more strongly than $\chi_2$ to the SM  and is brought close to equilibrium. This is why $\chi_1$ must be light enough to avoid overclosing the Universe. For $M_{1}\gtrsim$ eV, $\chi_1$ is a hot, subdominant component of DM that can affect structure formation and the CMB. In Fig.~\ref{fig:UV_massivechi1_z0_0p1}, we additionally show the excluded region from probes of structure formation. We provide details of these constraints in Sec.~\ref{sec:structure}.  \\

\noindent {\bf Validity of EFT:}~The couplings  in the EFT depend on the 
parameters of the UV-complete theory through the relation $D\sim F\lambda/M_\Phi^2$, where $F$ and $\lambda$ are $Z_2$-preserving and $Z_2$-violating couplings of $\Phi$, respectively. 
The highest value of $M_\Phi$ consistent with a perturbative UV completion of the EFT is $M_\Phi\sim1/\sqrt{D}$.
Since we assume that the reheat temperature is low enough that the scalars can never be directly created, we need $T_{\rm RH}\ll M_\Phi$, and consequently $T_{\rm RH}\ll 1/\sqrt{D}$. 
Figs.~\ref{fig:UV_masslesschi1_z0_0p1} and \ref{fig:UV_masslesschi1_z0_0p02} show 
that the largest couplings consistent with DM and baryogenesis are $|D|^2\sim 10^{-22}\,\,\mathrm{GeV}^{-4}$, leading to a constraint $T_{\rm RH}\ll 300$~TeV at its strictest. This is easily compatible with the reheat temperatures that we used in our calculations; in particular, we found that $T_{\rm RH}\lesssim10$~TeV is needed for successful baryogenesis. 

While $T_{\rm RH}\ll M_\Phi$ is a necessary condition on the validity of the EFT, it is not  sufficient. If the DM mass splitting is sufficiently small, it is possible that oscillations are inefficient for $k\sim T_{\rm RH}$ but much more efficient for $k\gg T_{\rm RH}$. The EFT is only valid if $k\ll M_\Phi$ for all momentum modes contributing to the asymmetry, and  for some parameters the enhancement of the asymmetry from the oscillations in these high-momentum modes could compensate for the Boltzmann suppression. Nevertheless, the exponential Boltzmann suppression always wins when $k$ greatly exceeds $T_{\rm RH}$. For $\mathcal{O}(1)$ couplings of the scalar to DM and SM fields, it is always possible to make $M_\Phi$ over an order of magnitude larger than $T_{\rm RH}$, and therefore the EFT can be valid for the full range of parameters that we consider. Of course, if the $F$ and $\lambda$ couplings are smaller, then $M_\Phi$ would be smaller as well; this would restrict the range of reheat temperatures and DM masses for which our approach is valid.\\

\noindent {\bf Dependence on LNV operator:}~So far, we have presented our results assuming that the $D_{\alpha I}(L_\alpha \chi_I)(Qd^{\rm c})$ operator in Eq.~\eqref{eq:EFT} is most relevant in the EFT. However, it is possible that the $E_{\alpha \beta\gamma I}(L_\alpha\chi_I)(L_\beta e_\gamma^{\rm c})$ operator in Eq.~\eqref{eq:EFT_2} is dominant instead. These two operators have different implications for X-ray constraints, which we discuss further in Sec.~\ref{sec:xray}. For the time being, however, we briefly summarize how our viable parameter space changes depending on which LNV operator is larger.

According to the discussion in Sec.~\ref{sec:xray}, there are strong X-ray constraints except when $\Phi_a$ couples exclusively to different flavors of LH and RH leptons. We consider a particular flavor structure that is safe from X-ray bounds, namely with $E_{\alpha\beta\gamma I}$   non-zero only for $\alpha\in \{1,2\}$, $\beta=2$, and $\gamma=3$ \cite{Berman:2022oht}.
Subject to this restriction, we re-optimize the components of the coupling matrix to maximize the total perturbative asymmetry (see the fully leptonic coupling benchmark in Appendix \ref{eq:UV_benchmark}).  The viable parameter space is  similar to that of the semileptonic operator in Eq.~\eqref{eq:EFT}, showing that the viability of leptogenesis does not depend significantly on the nature of the freeze-in operator provided that it explicitly violates $B-L$.

\begin{figure*}[t]
          \includegraphics[width=3in]{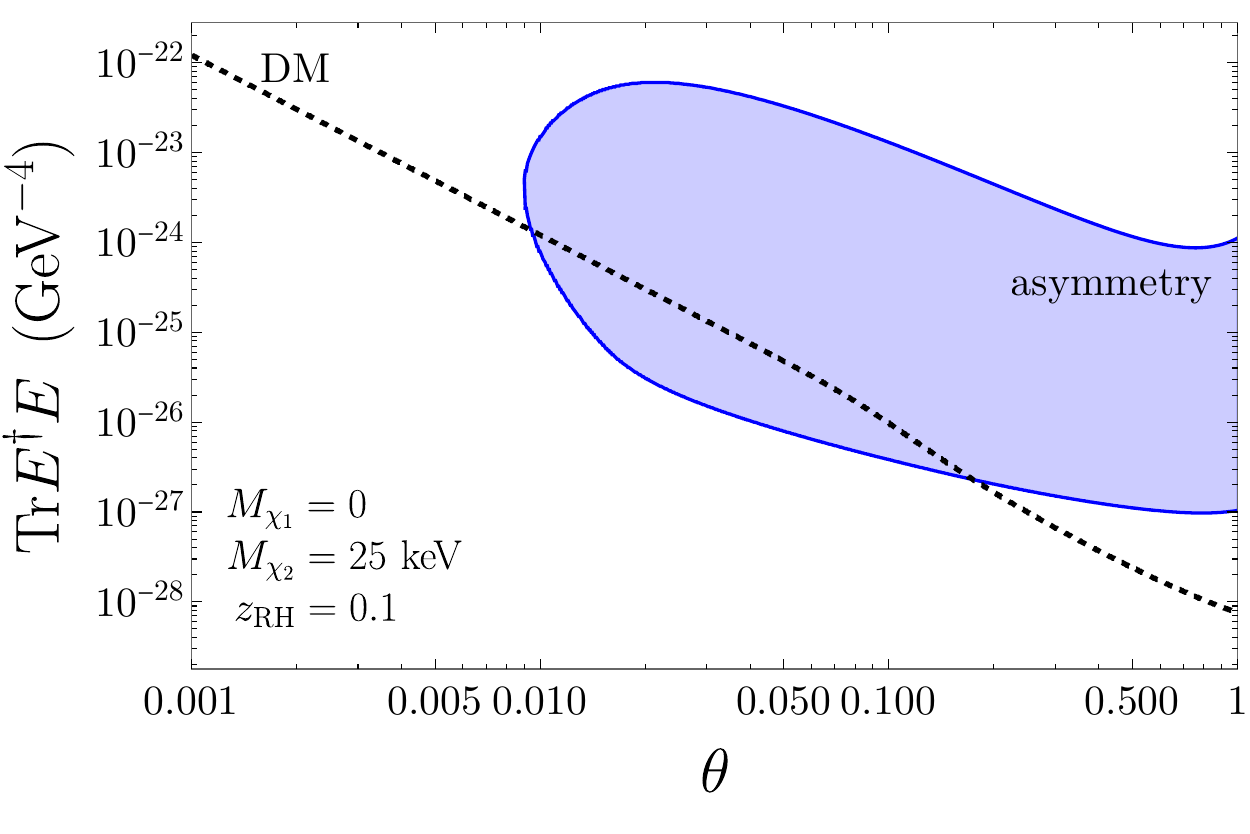}
           \quad \quad \quad
    \includegraphics[width=3in]{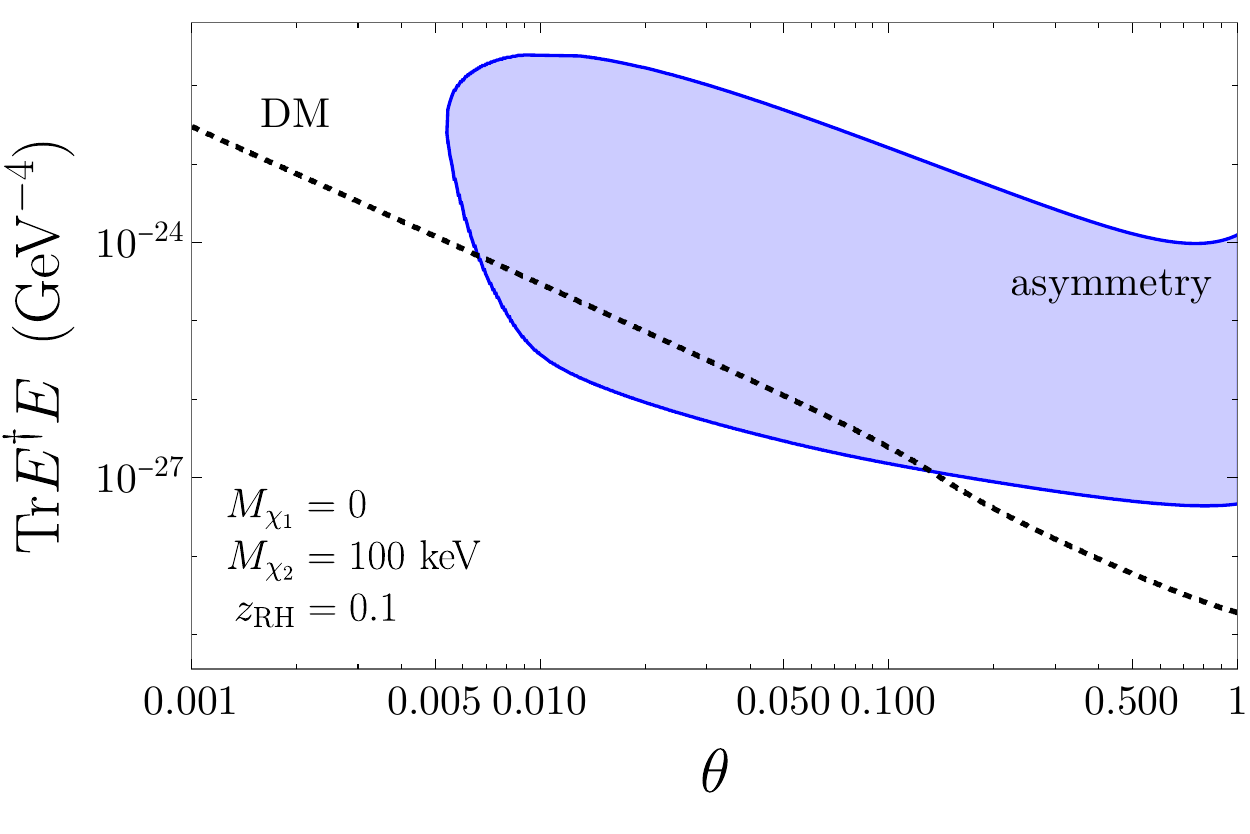}\\
    \includegraphics[width=3in]{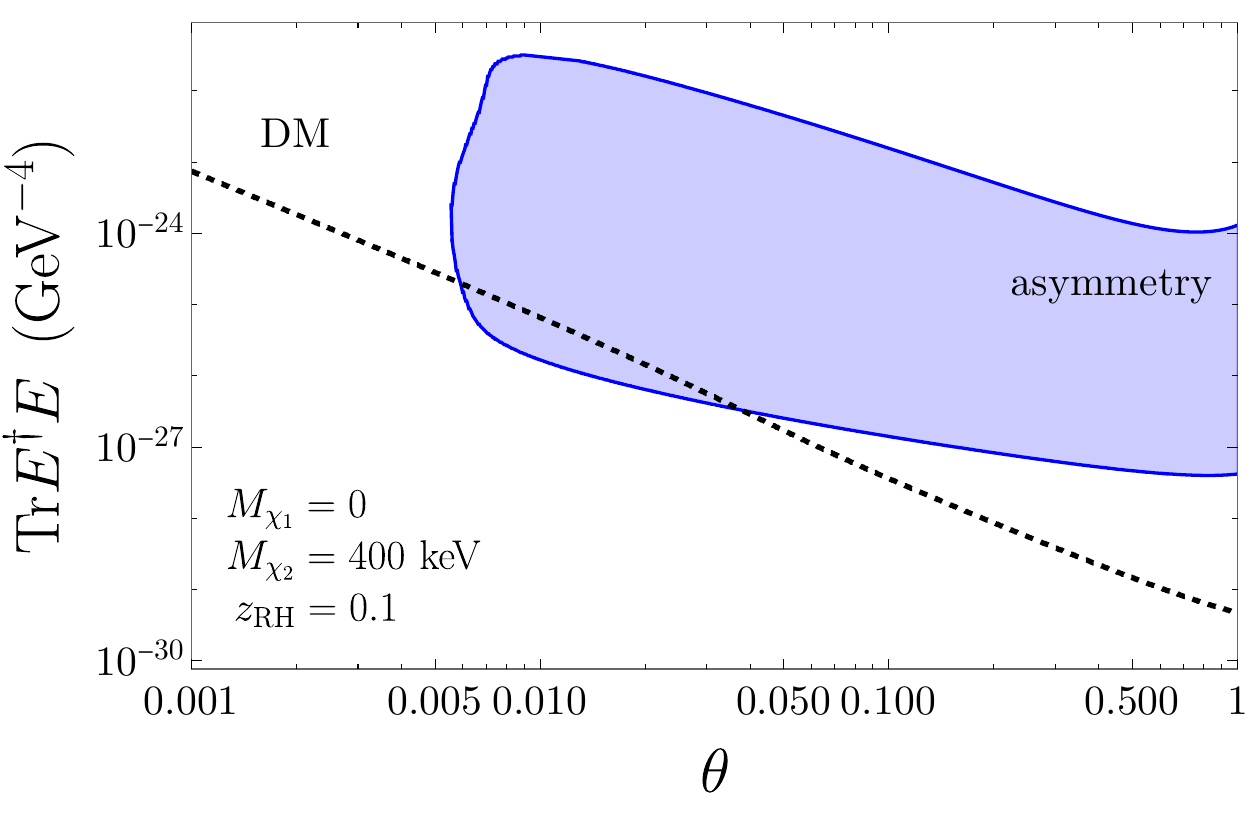}
           \quad \quad \quad
    \includegraphics[width=3in]{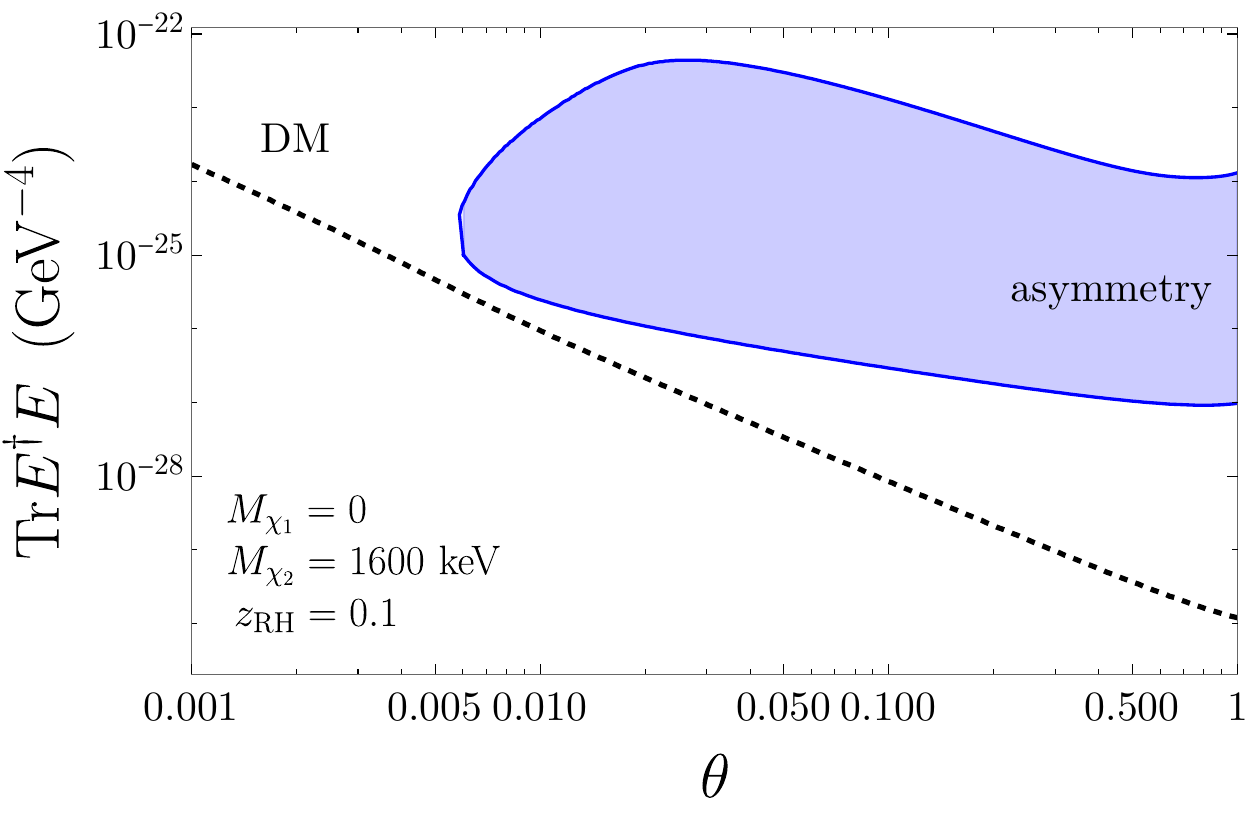}
\caption{
Coupling consistent with observed baryon asymmetry (shaded blue region) and DM abundance (dashed line) from freeze-in baryogenesis via the purely leptonic operator in Eq.~\eqref{eq:EFT_2} with $z_{\rm RH} \equiv T_{\rm ew}/T_{\rm RH} = 0.1$ and (top left) $M_2=25$~keV; (top right) $M_2=100$~keV; (bottom left) $M_2=400$~keV; (bottom right) $M_2=1600$~keV. We take $M_1=0$ and use the optimized fully leptonic coupling benchmark from Appendix 
\ref{app:coupling_fullyUV}. 
}
\label{fig:UV_masslesschi1_z0_0p1_LNValt}
\end{figure*}

\subsection{Perturbative Validation of QKE Study} \label{sec:UV_pert}

Our main results were derived using the momentum-integrated QKEs. However, these QKEs assume a thermal ansatz for the DM momentum, whereas the true DM spectrum is hotter than that of a Fermi-Dirac distribution according to Fig.~\ref{fig:DM-spectrum}. Furthermore, it is important to validate our numerical implementation of the full QKEs. Therefore, we perform a perturbative calculation of the DM abundance and baryon asymmetry, which allows for a more careful treatment of the effects of the DM spectrum on the asymmetry,  and we compare the results to those from the thermally averaged QKEs.

The approximations used in each approach are as follows:
\begin{itemize}
\item {\bf DM spectrum:}~the thermally averaged QKEs assume a thermal ansatz for the momentum distribution, while the perturbative approach  takes into account the non-thermal nature of the DM spectrum and tracks the DM oscillations separately in each momentum mode;
\item {\bf Deviation from equilibrium:}~the QKEs are valid even when one or both species are close to equilibrium, whereas the perturbative approach is only valid if all DM states are far from equilibrium;
\item {\bf Thermal masses:}~the QKEs  include  thermal corrections to $\chi$ masses, although for the purposes of comparison we can additionally solve the QKEs without thermal masses to determine their effect. The perturbative approach neglects thermal masses.
\end{itemize}
Our comparison allows us to quantify each of these effects on the DM abundance and baryon asymmetry.

We obtain the perturbative result by iteratively integrating the full, momentum-dependent QKEs from Sec.~\ref{sec:QKE}, keeping for each quantity only the terms at leading order in the freeze-in coupling, $D$. The DM abundance is thus determined at $\mathcal{O}(D^2)$, the lepton flavor asymmetries at $\mathcal{O}(D^4)$, and the total asymmetry at $\mathcal{O}(D^6)$. The full procedure is outlined in  detail in Appendix \ref{app:perturbative}.

\begin{figure*}[t]
          \includegraphics[width=3in]{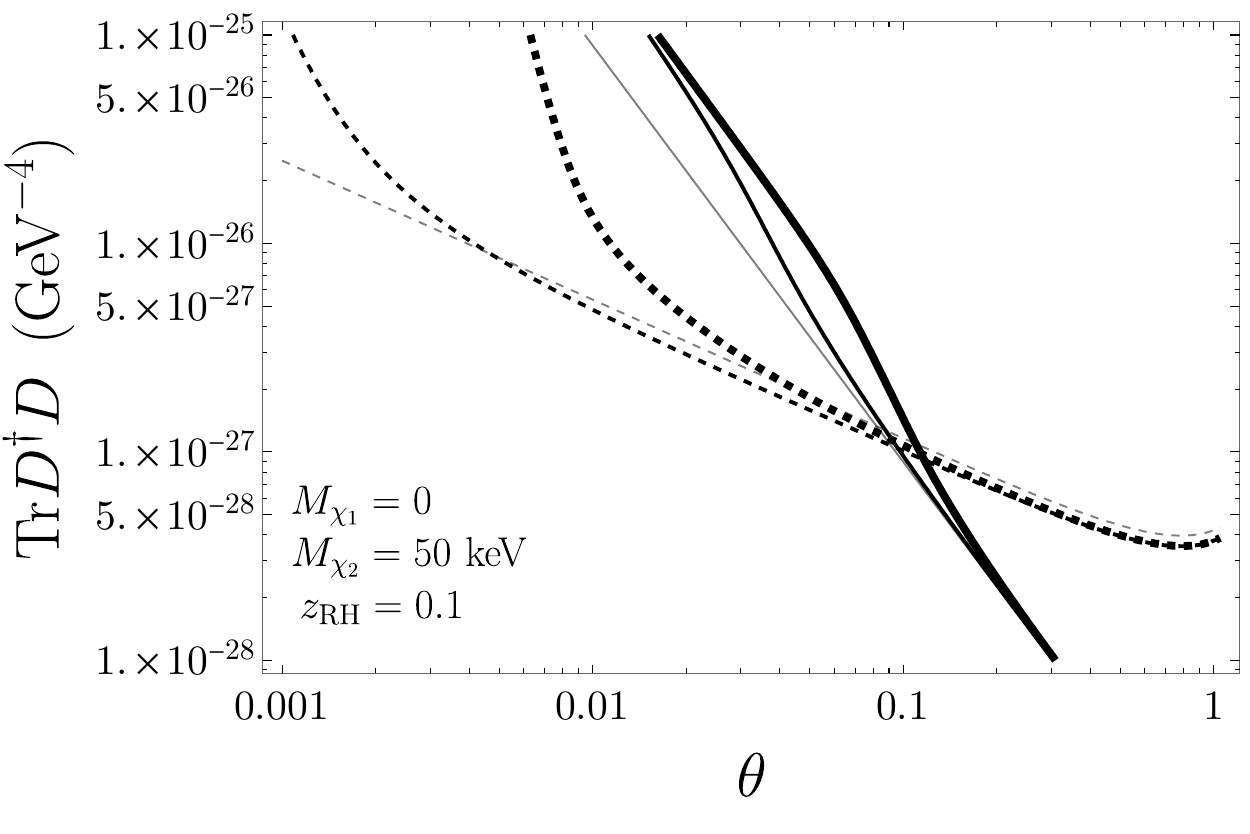}
           \quad \quad \quad
    \includegraphics[width=3in]{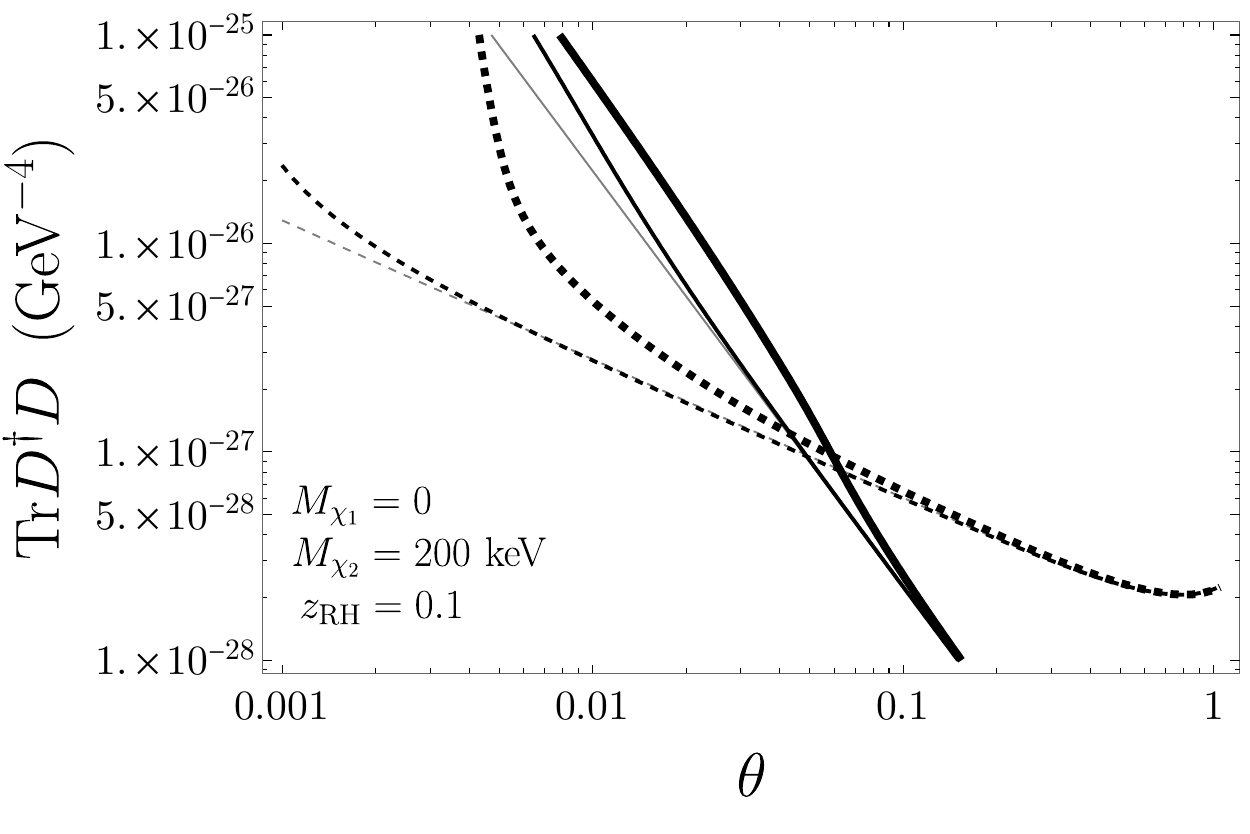}\\
    \includegraphics[width=3in]{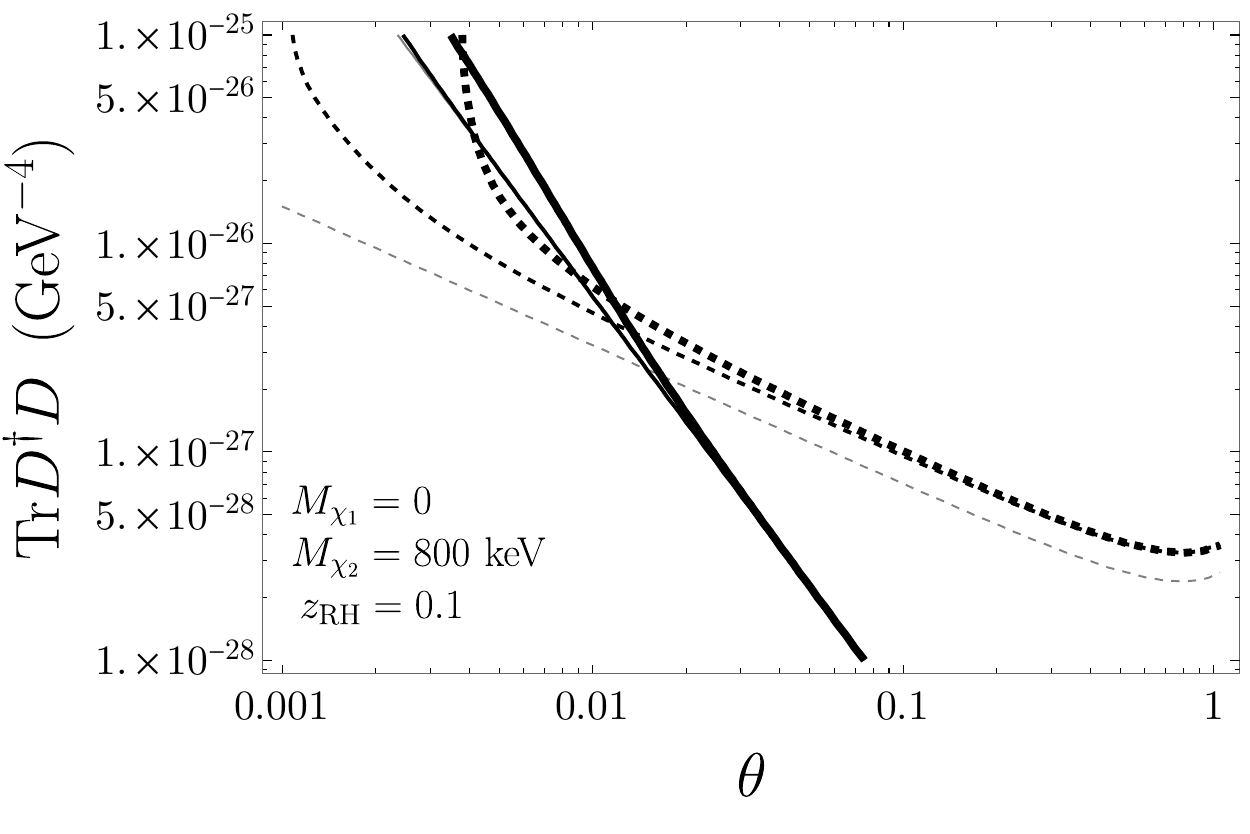}
\caption{
Parameters that give observed DM abundance (solid) and baryon abundance (dashed) for $z_{\rm RH}=0.1$ using different approximations to solve the quantum kinetic equations (QKEs):~(thin weight) perturbative solution of the full momentum-dependent QKEs; (medium weight) solution of the thermally averaged QKEs without thermal masses; (thick weight) solution of the thermally averaged QKEs including thermal masses. The $\chi_2$ mass is set to (top left) $M_2=50$~keV; (top right) $M_2=200$~keV; (bottom) $M_2=800$~keV.
}
\label{fig:pertcompare}
\end{figure*}

In our analysis, we set $z_{\rm RH}=0.1$ and we make $\chi_1$ massless; as in Sec.~\ref{sec:UV_full_results}, we vary $M_2$, $\mathrm{Tr}D^\dagger D$, and $\theta$ while fixing all other couplings to optimize the asymmetry. We show in Fig.~\ref{fig:pertcompare} the values of $\mathrm{Tr}D^\dagger D$ and $\theta$ that give the observed DM abundance and baryon asymmetry for various values of $M_2$. We calculate the results in three ways:~perturbatively solving the momentum-dependent QKEs; solving the thermally averaged QKEs neglecting thermal masses; and, solving the thermally averaged QKEs including thermal masses. 

There are several evident trends. As expected, all three calculations agree at small coupling for the DM abundance, but the perturbative calculation for the asymmetry can differ from the QKE solutions. In this limit, we expect the perturbative calculation to be more correct since it incorporates the full DM spectrum\footnote{We have checked that perturbatively solving the \emph{thermally averaged} QKEs agrees with the full solution of the thermally averaged QKEs at small coupling, which confirms that the discrepancy seen here is due to the different treatments of the DM spectrum.}. At lower masses, the thermally averaged QKEs over-predict the asymmetry (leading to a small value of $\mathrm{Tr}D^\dagger D$ to obtain the observed asymmetry), whereas at larger masses the thermally averaged QKEs under-predict the asymmetry. This can be understood from the spectra in Fig.~\ref{fig:DM-spectrum}:~because the true DM spectrum is hotter than the Fermi-Dirac distribution, the true oscillation phase is suppressed relative to that predicted for a species in equilibrium. When $M_2$ is small, oscillations don't have enough time to develop before UV freeze-in stops and a hotter spectrum further suppresses the asymmetry. Conversely, when $M_2$ is large, oscillations occur rapidly and the hotter spectrum allows a longer interval to produce the asymmetry before oscillations begin averaging to zero, which enhances the asymmetry. When the oscillation scale is matched to the reheat temperature, which for $z_{\rm RH}=0.1$ corresponds to $M_2\sim150$~keV, both methods give almost the same result.

From Fig.~\ref{fig:pertcompare}, we can estimate the error induced by the thermal ansatz in the QKEs as $\lesssim 25\%$. We expect, however, that this error could get larger for very large hierarchies between the oscillation and Hubble expansion rates around the time of reheating.

Comparing the perturbative solution and full solution without thermal masses  allows us to assess the effect on the asymmetry of higher-order processes neglected in the perturbative calculation. For $z_{\rm RH}=0.1$, the full solution diverges from the perturbative scaling when $\mathrm{Tr}D^\dagger D\gtrsim 5\times10^{-27}\,\,\mathrm{GeV}^{-4}$. The parameters giving rise to the observed DM and baryon abundances for these parameters are  within the perturbative regime, but barely:~increasing the squared coupling by an $\mathcal{O}(1)$ factor pushes into the regime where the perturbative calculation breaks down. This is consistent with our finding that $\chi_1$ has to be very light and relatively close to equilibrium in order to simultaneously account for DM and leptogenesis. 

Perhaps the most important difference comes from including or neglecting the $\chi$ thermal masses. Fig.~\ref{fig:pertcompare} shows that thermal masses are important even for small couplings where the perturbative calculation is otherwise valid. This is due to the fact that the tree-level $\chi$ mass scale is quite small.  Furthermore, at large coupling the asymmetry is greatly suppressed by thermal masses:~indeed, for $M_2=50$~keV, neglecting the thermal masses suggests that the observed asymmetry can be achieved for $\theta$ at or below $10^{-3}$, whereas including them cuts off the parameter space below $\sim5\times10^{-3}$. 

To summarize, comparing different methods of solving the QKEs has allowed us to determine that effects of the DM spectrum are largely negligible for our analysis, and that parts of the parameter space exist where the perturbative calculation gives an adequate prediction of the DM abundance and baryon asymmetry. However, nearly the entire parameter space is somewhat close to equilibrium, and a full solution of the QKEs including thermal masses is needed when the success of leptogenesis is marginal.

\section{Phenomenology}\label{sec:pheno}

\subsection{Structure Formation} \label{sec:structure}

When DM is warm, it free streams in the early Universe and can observably inhibit the formation of small-scale structure relative to cold DM. The small-scale matter power spectrum can be constrained from observations of the Lyman-$\alpha$ forest \cite{Viel_2013,Ir_i__2017},  counts of Milky Way satellite galaxies \cite{Schneider:2016uqi,Cherry:2017dwu,DES:2020fxi,Nadler:2021dft}, and strongly lensed quasars \cite{Treu:2010uj,Gilman:2019nap}. When DM is a single species that is produced from IR freeze-in (on-shell decays $\Phi\to \bar L\bar\chi$)  at $T>T_{\rm ew}$, Lyman-$\alpha$ observations require $M_{\rm DM} \gtrsim16$~keV \cite{Kamada_2019,Ballesteros:2020adh,DEramo:2020gpr} (although constraints vary among individual analyses),  while lensing and satellite counts require $M_{\rm DM}\gtrsim24$~keV \cite{Zelko:2022tgf}. For freeze-in baryogenesis where the lightest DM state is massless and acts as dark radiation, these bounds apply directly to the $\chi_2$ mass, $M_2$ \cite{Shuve:2020evk,Berman:2022oht}.

Structure formation constraints are expected to be stronger for UV freeze-in because the DM spectrum is hotter than in the IR scenario (see Fig.~\ref{fig:DM-spectrum}). We therefore reinterpret the above constraints for the spectrum resulting from UV freeze-in, with an explanation of our method provided below. We find that Lyman-$\alpha$ constraints bound $M_{\rm DM}\gtrsim24$~keV, while strong lensing and satellite counts require $M_{\rm DM}\gtrsim41$~keV. The $M_2=25$~keV benchmark in Fig.~\ref{fig:UV_masslesschi1_z0_0p1} is in tension with the latter constraint, although we include it to facilitate an understanding of how the interplay between leptogenesis and DM varies with the $\chi_2$ mass. Improvements to our understanding of the power spectrum at small scales would either allow for a discovery or, in the case of a null result, further constrain UV freeze-in.

To reinterpret existing constraints on UV freeze-in, we use existing constraints on a benchmark in which warm DM (WDM) is assumed to be thermal and constitutes the entirety of DM. Its mass is $M_{\rm WDM}$ and temperature is
\be
T_{\rm WDM} &=& \left(\Omega_{\rm DM}h^2\,\frac{93\,\,\mathrm{eV}}{M_{\rm WDM}}\right)^{1/3}T_\nu,
\ee
where $T_\nu$ is the temperature of the SM neutrinos \cite{Narayanan_2000}. The lower bounds for $M_{\rm WDM}$ are 5.3 keV  from Lyman-$\alpha$ \cite{Ir_i__2017} and 9.2 keV from strong lensing and satellite counts \cite{Zelko:2022tgf}. 

We use \texttt{CLASS} \cite{Blas:2011rf,Lesgourgues:2011rh} to compute the linear matter power spectrum for WDM, $P_{\rm WDM}(k)$, and the transfer function relative to cold DM,
\be
\mathcal{T}^2(k) &=& \frac{P_{\rm WDM}(k)}{P_{\rm CDM}(k)}.
\ee
The half mode,  $k_{1/2}$, is defined as the wavenumber at which the power in WDM is half of the power in CDM, $\mathcal{T}^2(k_{1/2}) \equiv \frac{1}{2}$ \cite{Konig:2016dzg}. $k_{1/2}$ gives an estimate of the length scale at which structure formation is suppressed by free streaming. We then compute the linear matter power spectrum and transfer function for UV freeze-in with $M_1=0$. The unnormalized spectrum is given by the function $\xi(y)$ (where $y\equiv k/T$) in Eq.~\eqref{eq:chi_spectrum} of  Appendix~\ref{app:QKE_UV}; we input tabulated values of $\xi(y)$  to \texttt{CLASS}. We then determine the DM masses $M_2$ that give the same $k_{1/2}$ scales corresponding to the excluded values of $M_{\rm WDM}$. This allows us to determine the bounds on $M_2$ of 24 keV (Lyman-$\alpha$) and 41 keV (lensing and satellites) given above.

This is a relatively simple way of comparing the constraints from UV freeze-in and thermal WDM, and as a result our mapping of the $M_{\rm WDM}$ bounds to $M_2$ is only approximate. However, we expect our results to be relatively robust since the transfer functions for UV freeze-in and thermal WDM are nearly indistinguishable. We have checked that another simple metric for mapping models to one another, $\delta A$ \cite{Murgia:2017lwo}, gives essentially the same bounds on $M_2$ as the method based on $k_{1/2}$. As an additional cross check, we repeated this procedure to re-derive the bounds on IR freeze-in DM, finding constraints on the DM mass consistent with those reported in the literature to within about 10\% \cite{Kamada_2019,Ballesteros:2020adh,DEramo:2020gpr,Zelko:2022tgf}. 

When $\chi_1$ is massive, it also contributes to warm DM given that our analysis from Sec.~\ref{sec:UV_full_results} showed that DM and leptogenesis can be simultaneously successful only if $M_1\lesssim$~keV. Depending on $M_2$, we get either mixed cold-warm or warm DM with multiple components. Refs.~\cite{Boyarsky_2009,Kamada:2016vsc,Baur_2017} constrained the mixed cold-warm scenario, putting bounds on the fraction $r$ of the DM energy density that is in warm DM. For $M_{\rm WDM}\lesssim0.7$~keV, the most stringent constraint is from analysis of Lyman-$\alpha$ data and gives $r\lesssim0.08$. This constraint assumes a thermal spectrum, whereas the $\chi_1$ spectrum will be more energetic in UV freeze-in. Additionally, the constraints are likely stronger on scenarios where both $\chi_1$ and $\chi_2$ are warm. For the present study, we take an approximate bound $r\lesssim0.1$, and we defer a full study to future work since our general results do not depend sensitively on these constraints.

We also include bounds from a recent study of light relics using observations of the matter power spectrum from the CMB,  BOSS, and weak lensing \cite{Xu:2021rwg}. Ref.~\cite{Xu:2021rwg}  sets  constraints on the temperature $T_{\chi_1}$ as a function of $M_{\chi_1}$ assuming that $\chi_1$ is fully thermalized. In our scenario, the $\chi_1$ temperature is fixed at $T_{\chi_1}\approx0.467T_\nu$ because $\chi_1$ is produced above the electroweak phase transition, but $\chi_1$ may have a sub-equilibrium abundance. Since the amplitude of the suppression of the matter power spectrum is determined by $\Omega_{\chi_1}h^2 \sim Y_{\chi_1} T_{\chi_1}^3$, we can reinterpret their constraints on $T_{\chi_1}$ as constraints on $Y_{\chi_1}$ instead. For $M_{\chi_1}\lesssim2.3$ eV, there is no constraint from the analysis of Ref.~\cite{Xu:2021rwg}, but at larger masses this analysis places constraints on otherwise viable  parameter space for DM and leptogenesis.

If the lightest DM species remains relativistic throughout the recombination era, it induces a shift in $N_{\rm eff}$ of 0.047 \cite{Xu:2021rwg}. This is  allowed by current constraints but would be within reach of CMB-S4 \cite{CMB-S4:2022ght}.

\subsection{X-ray Constraints on DM Decay}\label{sec:xray}
According to the discussion in Sec.~\ref{sec:IRtoUV}, leptogenesis in the fully UV regime requires lepton number (along with the $Z_2$ symmetry stabilizing DM) to be explicitly broken. As a result, DM can decay via $\chi_I \to \nu_\alpha \gamma,\bar\nu_\alpha\gamma$ and is thus subject to X-ray bounds on monochromatic photon emission from DM decay. The X-ray constraints depend on the specific operator(s) contributing to DM and baryogenesis. Additionally, DM can decay via $\chi_2\to \chi_1\gamma$; we will show that for every point consistent with DM and baryogenesis, it is possible to find a UV completion for which this decay is safe from observational constraints.

As noted earlier, there exist EFT operators that lead to the observed asymmetry and DM abundance while suppressing astrophysical X-ray production rates below current bounds. For example, the fully leptonic coupling $E_{\alpha\beta\gamma I}(L_\alpha \chi_I)(L_\beta e_\gamma^{\rm c})$ does not produce astrophysical X-ray lines (in the limit of vanishing neutrino masses) when the couplings $E_{\alpha \beta\gamma I}$ are only non-zero when both $\gamma\neq\alpha$ and $\gamma\neq\beta$. 
For the $(\chi d^{\rm c})({u^{\rm c}}^\dagger {e^{\rm c}}^\dagger)$ operator, DM decays to a photon and (anti)neutrino are suppressed by additional loop factors and small SM couplings. 

For operators that  induce photon-line signals, there are generally two contributions to the amplitude. One originates from a dipole operator $\chi\sigma^{\mu\nu}(LH)F_{\mu\nu}$ that directly mediates $\chi$ decay. The other comes from DM-neutrino mixing induced, for example, by a neutrino-portal coupling $h^\nu (\chi LH)$, which leads to DM decay  through a  $W$ loop. The magnitude of the ratio of the DM-neutrino mixing decay amplitude to the dipole-induced decay amplitude is proportional to $G_{\rm F}M_\Phi^2$, and so the DM-neutrino mixing contribution is expected to dominate for the UV freeze-in scenario with $M_\Phi \gg T_{\rm RH}$. We now consider each contribution in turn, starting with the neutrino-portal coupling.

We begin by considering models in which the  $\Phi$ scalars have the same quantum numbers as the SM Higgs field. We generically expect a mass mixing term in the potential, $\mu^2 H\Phi^*$, that suffers from a hierarchy problem. After electroweak symmetry breaking, this leads to a VEV $\langle\Phi\rangle \sim \mu^2 \langle H\rangle/M_\Phi^2$ for $M_\Phi\gg \mu,\langle H\rangle$. The $\Phi$~VEV then induces $\chi-\nu$ mixing (and DM decay). The $\Phi$~VEV leads to the same DM decay rate  as a direct neutrino-portal coupling $h^\nu (\chi LH)$ with $h^\nu \sim F\mu^2/M_\Phi^2$.

X-ray line bounds from NuSTAR \cite{Perez:2016tcq,Neronov:2016wdd,Ng:2019gch,Roach:2019ctw} and INTEGRAL \cite{Laha:2020ivk,Calore:2022pks} approximately require $h^\nu \lesssim10^{-15}-10^{-13}$ for $M_\chi$ ranging from 10 to 100 keV \cite{Berman:2022oht}.  In Sec.~\ref{sec:UV_full_results}, we saw that with $F\sim\lambda\sim1$, we had $M_\Phi\sim10^6-10^7$~GeV. Thus, to satisfy X-ray constraints we find $\mu\lesssim 0.1-1$~GeV. This suggests a scalar mass mixing that is tuned below the weak scale, although at a level commonly considered in many hidden-sector theories.

The neutrino-portal coupling can additionally be radiatively generated from vertex-correction diagrams featuring intermediate $\Phi$ loops. These can be present even if $\Phi$ has different quantum numbers from $H$. For example, the operator $(\chi e_\alpha^{\rm c})(L_\beta L_\gamma)$ originates from integrating out a scalar with the same quantum numbers as the RH leptons, and it induces a neutrino-portal coupling at one loop. Within the EFT, this radiative correction to the neutrino-portal coupling is quadratically divergent, suggesting that the factor of  $M_\Phi^{-2}$ in the coupling is cancelled by a factor of $M_\Phi^2$ from the cutoff, leaving a contribution of order $F\lambda'$. This na\"ive estimate agrees with the prediction for the neutrino-portal coupling, $h^\nu_{\alpha I}$, in the full UV theory  \cite{Berman:2022oht}:
\be
h^\nu_{\alpha I}\approx -\sum_\beta \frac{\lambda'_{\alpha \beta}F_{\beta I}y_\beta}{8\pi^2}\ln\frac{\Lambda}{M_\Phi},
\ee
where $y_\beta$ is the SM lepton Yukawa coupling.
This expression assumes that the neutrino-portal coupling vanishes at some scale $\Lambda$ and is generated from renormalization-group running between $\Lambda$ and $M_\Phi$. The product $F\lambda'$ must be relatively large for UV freeze-in leptogenesis to succeed:~from Fig.~\ref{fig:UV_masslesschi1_z0_0p1}, we have $F\lambda' \gtrsim 10^{-6}$ for $M_\Phi = 10$ TeV. $\Lambda$ and $M_\Phi$ are therefore tuned to be in close proximity with one another to satisfy observational constraints.  

Even if the neutrino-portal coupling to DM vanishes, loops can still induce a direct dipole coupling $\chi\sigma^{\mu\nu}(LH)F_{\mu\nu}$. This contribution appears in scenarios in which there exists a vertex-correction diagram to the neutrino-portal coupling and a photon attaches to one of the internal lines. The dipole coupling is  finite and calculable in the UV theory. For example, if we consider the UV completion of the fully leptonic operator $E_{\alpha\beta\gamma I}(L_\alpha\chi_I)(L_\beta e_\gamma^{\rm c})$, the rate of $\chi_I\to\gamma\nu+\gamma\bar\nu$ decay due to the dipole coupling is  \cite{Berman:2022oht}
\beq\label{sec:DM_decay_rate}
\Gamma_I = \frac{\alpha M_I^3}{512\pi^4}\sum_\alpha \left|\sum_\beta E_{\beta\alpha\beta I}M_\beta\left(\ln\frac{M_\Phi^2}{M_\beta^2}-1\right)\right|^2.
\eeq
This rate vanishes if the flavor structure is such that $E_{\beta\alpha\beta I}=0$ (in other words, $\Phi$ does not couple to the same flavors of LH and RH lepton). For example, if the only non-zero couplings to $\Phi$ are $F_{eI}$, $F_{\mu I}$, and $\lambda_{\mu \tau}$ (equivalent to $E_{e\mu\tau I}$ and $E_{\mu\mu\tau I}$ being non-zero), then the model is completely safe from X-ray bounds. It is still possible to generate a baryon asymmetry with this structure, and we showed numerical results for it in Fig.~\ref{fig:UV_masslesschi1_z0_0p1_LNValt}.

If, however, $\Phi$ couples to at least one common LH and RH lepton flavor, X-ray bounds can set limits on the model. If this common flavor is either $\mu$ or $\tau$, then the decay rate is enhanced because of the factor of $M_\beta^2$ in the decay rate. Using the estimates from Sec.~\ref{sec:UV_full_results} which predict the values of the couplings $E$ consistent with successful baryogenesis, we find that the $\chi_2$ lifetime is $\lesssim10^{25}$~s, which is completely excluded by X-ray constraints over the viable DM mass range \cite{Laha:2020ivk}. If instead the common flavor is exclusively $e$, then the rate is suppressed enough for a sliver of viable parameter space to exist. For example, we find that the $M_2=25$~keV benchmark in Sec.~\ref{sec:UV_full_results} predicts $\tau_{\chi_2}\lesssim2\times10^{29}$~s, which is marginally consistent with current constraints. However, larger values of $M_{\chi_2}$ are excluded because the decay rate increases as $M_{\chi_2}^3$. We therefore find that baryogenesis with the fully leptonic operator is in strong tension with X-ray bounds when there is a common coupling to the same LH and RH lepton flavor, except for a small region of parameter space with $M_{\chi_2}\lesssim25$~keV where $\Phi$ couples to LH and RH electrons.

Finally, we turn to the decay $\chi_2\to\chi_1\gamma$. This decay rate is proportional to $\sum_{\alpha,\beta}|C_{\alpha\beta 12}|^2$, where the relevant EFT Lagrangian term is $C_{\alpha\beta IJ}(L_\alpha \chi_I)(\chi_J^\dagger L_\beta^\dagger)$ and the Wilson coefficient $C_{\alpha\beta IJ}$ is defined in Eq.~\eqref{eq:coupling:C}. Both DM flavors participate in this coupling, and so it is not possible to forbid the $\chi_2\to\chi_1\gamma$ decay using flavor symmetries in the manner that was possible for $\chi_2\to\nu\gamma$. Because this operator does not contribute to leptogenesis, we cannot directly relate its role in DM decay to the parameter space for DM and leptogenesis in Sec.~\ref{sec:UV_full_results}. In terms of couplings in the UV-complete theory, the $\chi_2\to\chi_1\gamma$ decay rate is modified by a factor of $(F/\lambda)^2$ compared to $\chi_2\to\nu\gamma$. It is therefore possible to suppress $\chi_2\to\chi_1\gamma$ decays in the limit $F \ll \lambda$. We also note that the $\chi_2\to\chi_1\gamma$ decay rate scales like $M_2^5$ rather than $M_2^3 M_\beta^2$ (where $\beta$ is the lepton flavor running in the loop), which introduces an additional relative suppression in some part of the parameter space.

It is possible to realize the $F\ll \lambda$ limit while preserving perturbativity in the UV-complete theory. Both $F$ and $\lambda$ are of order unity when $M_\Phi = \left(\mathrm{Tr}D^\dagger D\right)^{-1/4}$; therefore, it is possible to make $F\ll \lambda$ if $M_\Phi \ll\left(\mathrm{Tr}D^\dagger D\right)^{-1/4} $. We further require $M_\Phi\gg T_{\rm RH}$ in order to realize UV freeze-in. We have checked that for all viable parameters giving freeze-in leptogenesis and DM in Sec.~\ref{sec:UV_full_results}, it is possible to choose parameters in the UV completion such that the $\chi_2\to\chi_1\gamma$ decay rate is below current bounds while satisfying the above constraints on $M_\Phi$; in some instances, the decay rate is safe from constraints by more than ten orders of magnitude. We therefore do not include any constraints from this decay mode in our results. At the same time, we note that for different choices of $F$, $\lambda$, and $M_\Phi$ that give rise to the same values of $D$, it is possible to make this decay rate close to the current X-ray bounds and thus potentially observable with future telescopes.

\section{Effects of DM Spectrum on Leptogenesis:~Mixed UV-IR}\label{sec:mixed}

When there exist multiple DM production mechanisms, the resulting baryon asymmetry can be significantly larger than via a single mediator \cite{Shuve:2020evk,Berman:2022oht}. Unless the two mediators have comparable masses (or the DM oscillation time is very short), the dynamics can be understood as follows:~a coherent collection of DM particles is produced by the mediator operating at higher energies. These states undergo oscillations, and the combination of relative phases from coherent superposition and oscillation leads to $CP$-violation in scattering rates through the other mediator active at lower energies. Refs.~\cite{Shuve:2020evk,Berman:2022oht} studied the scenario  with two scalars $\Phi_1$ and $\Phi_2$ (with $M_{\Phi_2}\gg M_{\Phi_1}$), in which case the coherent background of DM particles is produced at $T\sim M_{\Phi_2}$, while the asymmetry is produced predominantly by inverse decays at $T\sim M_{\Phi_1}$. The viable parameter space for this scenario is substantial, with the observed DM abundance and baryon asymmetry consistent with DM masses up to 500 keV and $\Phi_1$ masses up to 5 TeV or higher. Over much of the parameter space, all DM species are far from equilibrium and a perturbative approach to solving the QKEs is valid.

We expect that these results are largely independent of the production mechanism operating at high energies. In other words, DM could be produced through some entirely different mechanism (such as inflaton decay), and as long as the interaction eigenstate is not aligned with the $\Phi_1$-DM interaction eigenstate, an asymmetry can be produced. The magnitude of the asymmetry is most sensitive to the density of DM from the original production mechanism, and one can imagine adjusting the couplings of whatever production mechanism exists to give the same primordial DM density matrix. The resulting asymmetry would then be entirely independent of the specifics of the first stage of DM production.

There is one important caveat:~the momentum distribution of DM particles depends on the precise production mechanism. Even within the simplest scenario where the coherent collection of DM particles is produced through the interactions of a scalar $\Phi_2$, we saw in Fig.~\ref{fig:DM-spectrum} that the DM spectrum depends on whether DM is produced from on-shell $\Phi_2$ decays (with $T_{\rm RH}\gtrsim M_{\Phi_2}$) or from off-shell scattering (with $T_{\rm RH}\ll M_{\Phi_2}$). Leptogenesis in the former channel is fully IR because all DM interactions are with on-shell scalars, whereas the latter channel is mixed UV-IR because DM is produced from UV freeze-in via the heavier off-shell scalar but then undergoes inverse decays to an on-shell $\Phi_1$.

In this section, we compare baryogenesis in the fully IR and mixed UV-IR scenarios to assess the impact of the DM spectrum on baryogenesis. To do this, we use the same matrix structure for the coupling to DM in both cases to generate the same coherent superposition of background DM particles. We also adjust the magnitude of the couplings to give the same momentum-integrated DM number densities in both scenarios. For simplicity, we assume in both cases that DM couples to a single lepton flavor, which is sufficient to generate a lepton asymmetry. Together, this imposes a relationship between the on-shell $\Phi_2$ coupling $F_{IJ}^2$ in the fully IR case, and the EFT coupling $C_{IJ}$ in the mixed UV-IR case:
\be
(C^\dagger C)_{JI} &\approx& \frac{8.70}{M_{\Phi_2}T_{\rm RH}^3} (F^{2\dagger}F^2)_{IJ}, 
\ee
where the numerical pre-factor originates from differences in the integral over the DM spectrum in each scenario; see Eq.~\eqref{eq:mixed-normalization}. We then compute the ratio of the asymmetry resulting from the two spectra. Since the full effects of the DM spectrum are only captured in the perturbative method outlined in Sec.~\ref{sec:UV_pert}, we restrict our analysis to this approach and assume that the production of DM mediated by $\Phi_2$ occurs well before the inverse decays at $T\sim M_{\Phi_1}$, namely $T_{\rm RH},M_{\Phi_2}\gg M_{\Phi_1}$. For  details of the calculation, see Appendix \ref{app:mixed-pert}.

\begin{figure}[t]
          \includegraphics[width=0.48\textwidth]{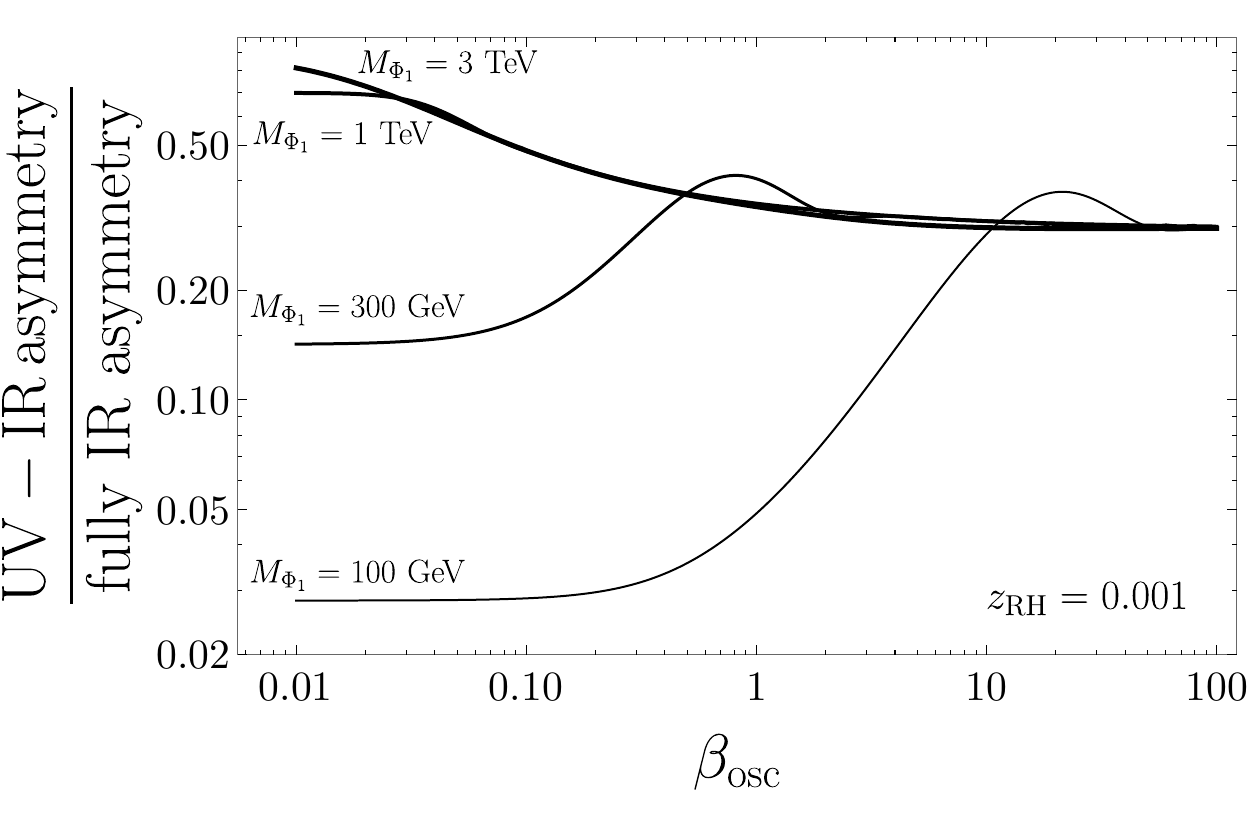}
\caption{
Ratio of asymmetries from mixed UV-IR scenario vs.~fully IR scenario assuming primordial DM abundance is normalized to be equal in both scenarios. The ratio of asymmetries is shown as a function of $\beta_{\rm osc}\equiv \Delta M^2 M_0/(6 M_{\Phi_1}^3)$, which quantifies the  oscillation phase for DM momentum mode $k=T$ when $T=M_{\Phi_1}$. The results are shown for $z_{\rm RH}=0.001$, and for the fully IR asymmetry, we take $M_{\Phi_2}=T_{\rm RH}$, although the results do not depend on these quantities provided $M_{\Phi_2},T_{\rm RH}\gg M_{\Phi_1}$. All curves converge to a common value at $\beta_{\rm osc}\gg1$, although the $M_{\Phi_1}=3$ TeV curve lies slightly below the 300 GeV and 1 TeV curves for $\beta_{\rm osc}\sim2-20$. The results indicate that the hotter spectrum predicted by UV production of DM tends to suppress the asymmetry when compared to DM production from on-shell scalar decays. 
}
\label{fig:mixeduvir-compare}
\end{figure}

The ratio of the asymmetries depends on the oscillation phase accrued by the time of leptogenesis, $T\sim M_{\Phi_1}$. This is parametrized by the dimensionless quantity
\be
\beta_{\rm osc} &\equiv& \frac{\Delta M^2 M_0}{6 M_{\Phi_1}^3}.
\ee
The rate of asymmetry generation contains the oscillatory factor $\sin\left(\frac{\beta_{\rm osc}x^3}{y}\right)$, where $x\equiv M_{\Phi_1}/T = M_{\Phi_1}z/T_{\rm ew}$. We show our results in Fig.~\ref{fig:mixeduvir-compare} for various values of $M_{\Phi_1}$. We see that the asymmetry in the mixed UV-IR scenario is always smaller than that from the fully IR scenario, and while the two primordial DM production mechanisms give comparable results over much of the parameter space, there can be a significant suppression in the UV-IR asymmetry if $\beta_{\rm osc}\ll1$ and $M_\Phi\sim T_{\rm ew}$. This indicates that the results of freeze-in baryogenesis depend in general on the spectrum of primordial DM production.  When the parameters allow rapid oscillations to develop before DM decoupling, however, the final result does not depend dramatically on the DM production mechanism. 

We can qualitatively understand the behavior of the asymmetry ratio in certain limits. When the oscillation rate is too slow for oscillations to fully develop before $\Phi_1$ decouples from the plasma ($\beta_{\rm osc}\ll1$), we can perform a series expansion on the oscillatory sine factor in calculating both asymmetries, and $\beta_{\rm osc}$ therefore cancels in the ratio of asymmetries. Thus, we see that the asymmetry ratio is flat in the limit $\beta_{\rm osc}\ll1$ for all $\Phi_1$ masses.  

In this limit, it is also possible to estimate the contribution of the asymmetry from each momentum mode, $y$. We find that each momentum mode contributes equally to the asymmetry for $y\lesssim M_{\Phi_1}^2/(4T_{\rm ew})^2$, but is suppressed for larger momenta. This follows from the fact that asymmetry production at temperature $T$ is dominated by modes with CM energy equal to the $\Phi_1$ mass, $E_{\rm CM}^2\sim 4yT^2\sim M_{\Phi_1}^2$, and the numerics work out such that the asymmetry produced per unit $y$ is effectively constant. For sufficiently large $y$, however, the condition $4yT^2\sim M_{\Phi_1}^2$ is only satisfied for $T<T_{\rm ew}$ and such modes therefore do not contribute to the asymmetry. For $M_{\Phi_1}\sim T_{\rm ew}$, only momentum modes with $y\ll1$ contribute to the asymmetry, and since the DM population for such modes is much higher for IR-dominated $1\to2$ DM production relative to UV production, the asymmetry is much larger in the former case\footnote{This calculation neglects the $\Phi_1$ thermal mass as well as $2\to2$ scattering processes involving $\Phi_1$, both of which allow a broader range of momentum modes to contribute to the asymmetry but are suppressed by additional powers of SM gauge couplings. We expect that these effects are irrelevant for the mass range we consider but could be relevant for $M_{\Phi_1}\lesssim100$~GeV. However, such parameters are largely ruled out by collider constraints \cite{Shuve:2020evk,Berman:2022oht}. }. For larger $M_{\Phi_1}$, however, a broader range of momentum modes contribute, and the asymmetries become comparable in both scenarios.

We also see universal behavior in the $\beta_{\rm osc}\gg1$ limit, which is the result of a series expansion in $1/{\beta_{\rm osc}}$. To show this, we perform a change of integration variables to $u=x^3/y$. Then, the asymmetry resulting from DM produced in a particular process can be written in the approximate form
\be
\int_0^\infty\,du\,f(u)\,\sin(\beta_{\rm osc}u),
\ee
where the function $f(u)$ encodes the time dependence of the reaction density for the DM production process; see Appendix \ref{app:mixed-pert} for complete expressions. The lower limit of integration can be well-approximated by 0 because in both scenarios, the primordial abundance of DM responsible for baryogenesis is produced well before oscillations occur. We can take the upper limit to infinity because the rapid oscillations effectively halt the asymmetry generation at late times, and hence the result is independent of the precise endpoint of asymmetry generation.

 For $\beta_{\rm osc}\gg1$, such that  oscillations occur more rapidly than any other time scale in the problem,  and $|f(u)|\to0$ as $u\to\infty$, it is possible to show that the integral  evaluates simply to  ${}-f(0)/\beta_{\rm osc}$. Since this same factor of $1/\beta_{\rm osc}$ shows up in both the $1\to2$ asymmetry and the EFT asymmetry, it cancels in the ratio, which tends to a constant value at large $\beta_{\rm osc}$.

In summary, our analysis shows that the asymmetry from freeze-in baryogenesis depends on the DM spectrum, and the asymmetry from different DM production modes can differ by many orders of magnitude in the small-oscillation regime because the asymmetry is sensitive to the low-momentum tail of the DM distribution. When oscillations are rapid, however, it appears that an $\mathcal{O}(1)$ change in the spectrum leads to an $\mathcal{O}(1)$ change in the asymmetry such that the results of freeze-in in one particular model can approximately apply to a broader range of DM production mechanisms at high temperature.

\section{Discussion and Conclusions}\label{sec:conclusion}

We have studied leptogenesis through DM oscillations when freeze-in occurs in the UV regime. We have specifically considered the case where DM couples to the SM via a dimension-6 operator, which is the EFT that results from integrating out a heavy scalar mediator. For leptogenesis to occur, the operator must explicitly break SM lepton number, and DM lacks a stabilizing $Z_2$ symmetry. Leptogenesis proceeds through an ARS-like mechanism in which asymmetries first accrue in individual SM lepton flavors and are then converted to a total lepton asymmetry through flavor-dependent washout.

It is possible to simultaneously achieve the observed DM and baryon abundances through UV freeze-in, although with restrictions on the allowed parameter space:~the reheat temperature must lie between the electroweak scale and about 10 TeV, and DM masses are viable up to about 1.5 MeV. To avoid over-production of DM, there must exist one approximately massless DM  eigenstate, leading to current bounds and future prospects for probing the model using searches for light, massive relics and dark radiation. 
The couplings needed for leptogenesis can also induce DM decay into a SM neutrino and a photon at a level inconsistent with X-ray bounds, although there exist particular operator structures that lead to a vanishing dipole moment and are thus safe from indirect detection bounds on decaying DM.

By definition, UV freeze-in depends on the physics of reheating. We have employed the instantaneous reheating approximation, which has been shown to be valid for modelling DM production for the operators we considered. Given the additional oscillation timescale relevant for leptogenesis, however, a future study is warranted which more carefully implements the dynamics of reheating to investigate the possibility of contributions to leptogenesis during the inflaton-dominated era. It would also be interesting to understand whether (p)reheating could directly produce a highly non-thermal DM population:~significant differences in the DM spectrum or abundance could sufficiently modify DM scattering and oscillation rates to meaningfully change the parameter space associated with DM and leptogenesis.

\section*{Acknowledgments}
We thank Coleman Gliddon and Rafael Porto for collaboration at an early stage of the project. We are grateful to George Fuller and Benny Weng for helpful conversations, and to the referee for  raising the question of $\chi_2\to\chi_1\gamma$ radiative decays. The work of BS is supported by Research Corporation for Science Advancement
through Cottrell Scholar Grant \#27632. The work of DTS is supported by the U.S.
National Science Foundation under Grant PHY-2310770. This work was performed in part at Aspen Center for Physics, which is supported by National Science Foundation grant PHY-2210452. BS is grateful for the hospitality of Perimeter Institute, where part of this work was carried out. Research at Perimeter Institute is supported in part by the Government of Canada through the Department of Innovation, Science and Economic Development and by the Province of Ontario through the Ministry of Colleges and Universities. 

\appendix

\section{Quantum Kinetic Equations} \label{app:QKE}

\subsection{Fully UV Scenario} \label{app:QKE_UV}

Our starting point is the full momentum-dependent quantum kinetic equation (QKE) for the $\chi$ density matrix, $f_\chi$, with comoving momentum $y\equiv k/T$, which can be derived following the methods in Refs.~\cite{Hambye:2016sby,Abada:2018oly} and references therein:
\small
\be\label{eq:QKE_general_app}
\frac{df_\chi(y)}{dt} &=& -i\left[ E_\chi(y), f_\chi(y) \right]\\
&&{}-\frac{1}{2}\sum_\alpha\left(\left\{\frac{\Gamma_\alpha^>(y)}{2E_\chi},f_\chi(y)\right\}-\left\{\frac{\Gamma_\alpha^<(y)}{2E_\chi},1-f_\chi(y)\right\}\right).\nonumber
\ee
\normalsize
The reaction rate for DM production with momentum $k$  due to scattering with a lepton of flavor $\alpha$ is quantified by
\begin{widetext}
\be
\Gamma^<_{\alpha,IJ} &=& 6D^*_{\alpha I}D^{}_{\alpha J} \int \,d\Pi_{\rm SM}\left[t^2 f_df_{\bar L_\alpha}(1-f_Q)(2\pi)^4\delta^4(p_d+p_L-k-p_Q)\right.\nonumber\\
&&{}\left.+ t^2 f_{\bar L_\alpha}f_{\bar Q}(1-f_{\bar d})(2\pi)^4\delta^4(p_L+p_Q-k-p_d)
+s^2 f_{\bar Q}f_d(1-f_{L_\alpha})(2\pi)^4\delta^4(p_Q+p_d-k-p_L)\right],
\ee\label{eq:gammaless}
\end{widetext}
where the three terms represent the scattering rates for  $d\bar{L}_\alpha\to\chi Q$, $\bar L_\alpha\bar Q\to\chi\bar{d}$, and $\bar{Q}d\to\chi L_\alpha$, respectively. For concreteness, we express results for the $(L_\alpha\chi)(Qd^{\rm c})$ effective operator from Sec.~\ref{sec:IRtoUV}, but it is straightforward to obtain the analogous reaction rates for other operators. $s$ and $t$ are the standard Mandelstam variables\footnote{These should not be confused with  entropy density and time, which are unfortunately also called $s$ and $t$. }. The integral is performed over the Lorentz-invariant phase space of the SM states (which we take to be massless),
\be
d\Pi_{\rm SM}=\frac{d^3p_L}{(2\pi)^3 2E_L}\frac{d^3p_Q}{(2\pi)^3 2E_Q}\frac{d^3p_d}{(2\pi)^3 2E_d},
\ee
and the SM statistical distributions are Fermi-Dirac distributions,
\be
f_\psi = \frac{1}{e^{(E_\psi-\mu_\psi)/T}+1}
\ee
for each species $\psi$, where $\mu_{\bar\psi}=-\mu_\psi$. The reaction rate for DM annihilation, $\Gamma_{\alpha,IJ}^>/2E_\chi$, is  almost the same as $\Gamma_{\alpha,IJ}^</2E_\chi$, but the statistical factors appearing in the initial and final states are swapped:~for example, the $\chi Q\to d\bar{L}_\alpha$ reaction density has the statistical factor $f_{Q}(1-f_d)(1-f_{\bar L_\alpha})$. The reaction densities for $f_{\bar\chi}$ are found by taking the complex conjugate of the reaction density and taking $\mu\to-\mu$ for SM chemical potentials (see Appendix \ref{app:comment-on-chibar} for more details on the choice of convention for the $\bar\chi$ density matrix).

We derive the QKEs for lepton asymmetries in terms of the asymmetries $X_\alpha =\Delta B/3-\Delta L_\alpha$ preserved by electroweak sphalerons. Since chirality-flipping DM processes are suppressed in the early Universe due to the smallness of the DM masses, there exists an approximate generalized lepton number symmetry with $L(\chi)=-1$. Consequently, any asymmetry produced in $\chi$ through interactions with a lepton flavor $\alpha$ results in an asymmetry $\Delta L_\alpha$ of equal magnitude and sign, and consequently an asymmetry $X_\alpha$ of opposite sign. This relationship relates the change in $\chi$-$\bar\chi$ number densities to the $X_\alpha$ charge normalized to the entropy density, $Y_\alpha \equiv X_\alpha/s$, giving
\be
\frac{dY_\alpha}{dt}=-\mathrm{Tr}\left(\frac{dY_\chi^\alpha}{dt}-\frac{dY_{\bar\chi}^\alpha}{dt}\right),
\ee
where the superscript $\alpha$ indicates that we only include the reaction densities corresponding to lepton flavor $\alpha$, and
\be
Y_\chi &=& \frac{1}{s}\int\,\frac{d^3k}{(2\pi)^3}\,f_\chi.
\ee
Note that the oscillation terms of the DM QKEs vanish when taking the trace of the DM density matrices. The asymmetries $X_\alpha$ are related to the SM chemical potentials through the relations given later in this Appendix.

In Appendix \ref{app:perturbative}, we provide a perturbative solution to the momentum-dependent QKEs that captures the leading order behavior when all DM species are far from equilibrium. Away from this limit, however, it is hard to solve the full momentum-dependent QKEs. We instead solve a simplified set of thermally averaged QKEs. To do this, we adopt a thermal ansatz for the DM density matrix by assuming it has a Fermi-Dirac spectrum:
\be
f_{\chi,IJ}(k,t) &=& \frac{Y_{\chi,IJ}(t)}{Y_\chi^{\rm eq}}f_{\chi0}^{\rm f.d.}(k)\\
&=& \frac{Y_{\chi,IJ}(t)}{Y_\chi^{\rm eq}}\,\frac{1}{e^{k/T}+1}.
\ee
We do not include a chemical potential for $\chi$ in the Fermi-Dirac distribution since $\chi$ is not in chemical equilibrium and its asymmetry is encoded in the difference $Y_\chi-Y_{\bar\chi}$. 

Since the thermal ansatz ensures that the only momentum-dependence is in the Fermi-Dirac distribution, it is now possible to integrate over DM momentum to obtain a differential equation for the DM density matrix normalized to entropy density, $Y_{\chi,IJ}$. Note that because DM is highly relativistic throughout its freeze-in, the equlibrium number density is effectively the same for both DM mass eigenstates. The momentum-integrated QKE for $\chi$ is
\small
\be\label{eq:QKE_momavg}
\frac{dY_\chi}{dt} &=& -i\left[ \langle E_\chi\rangle, Y_\chi \right]
+\frac{1}{s}\sum_\alpha\int\frac{d^3k}{(2\pi)^3}\frac{\Gamma_\alpha^<}{2E_\chi}\\
&&{}-\frac{1}{2s}\sum_\alpha\left\{\int\frac{d^3k}{(2\pi)^3}\frac{\Gamma_\alpha^<+\Gamma_\alpha^>}{2E_\chi}f_\chi^{\rm f.d.}(k),\frac{Y_\chi}{Y_\chi^{\rm eq}}\right\},\nonumber
\ee
\normalsize
where for later convenience we have separated out terms that are independent of $Y_\chi$, which model production of DM in the absence of any DM abundance, and the terms proportional to $Y_\chi$, which model annihilation and Pauli-blocking effects.

It is convenient to separate the reaction densities into three separate terms:~a term which is independent of $\mu$; a term that depends on $\mu$ but not $Y_\chi$; and, a term that depends on both $\mu$ and $Y_\chi$. In addition to allowing us to separate the collision integrals into more tractable components, this also turns out to be helpful in a perturbative expansion since  $\mu$ and $Y_\chi$ both vanish in the limit of zero coupling and the product of the two therefore give higher-order contributions to the QKEs. 

We now consider each of these terms in sequence.\\

\noindent {\bf (1) Reaction density independent of  $\mu$}:~the relevant reaction density we need to compute is
\be
\mathcal{C}_0 &\equiv& \frac{1}{s}\int\frac{d^3k}{(2\pi)^3}\frac{\Gamma_{\alpha0}^<}{2E_\chi}\\
&&{}-\frac{1}{2s}\left\{\int\frac{d^3k}{(2\pi)^3}\frac{\Gamma_{\alpha0}^<+\Gamma_{\alpha0}^>}{2E_\chi}f_\chi^{\rm f.d.}(k),\frac{Y_\chi}{Y_\chi^{\rm eq}}\right\},\nonumber
\ee
where $\Gamma_{\alpha0}^<$ and $\Gamma_{\alpha0}^>$ are defined as in Eq.~\eqref{eq:gammaless} but with all SM chemical potentials set to zero. Detailed balance requires that the production and annihilation rates cancel in equilibrium for each momentum mode:
\be\label{eq:KMS}
\Gamma_{\alpha0}^<(1-f_\chi^{\rm f.d.}) &=& \Gamma_{\alpha0}^> f_\chi^{\rm f.d.}.
\ee
We can then simplify $\mathcal{C}_0$ to get
\be
\mathcal{C}_0 &=& \frac{1}{2s}\left\{\int\frac{d^3k}{(2\pi)^3}\frac{\Gamma_{\alpha0}^<}{2E_\chi},1-\frac{Y_\chi}{Y_\chi^{\rm eq}}\right\}.
\ee

We define
\be
\gamma_\alpha^0(T) \equiv \int\frac{d^3k}{(2\pi)^3}\frac{\Gamma_{\alpha0}^<}{2E_\chi} = T^3\int\frac{d^3y}{(2\pi)^3}\frac{\Gamma_{\alpha0}^<}{2E_\chi}.\label{eq:gamma0_def}
\ee
$\gamma_\alpha^0$ has dimension 4, while $D^*_{\alpha I}D^{}_{\alpha J}$ has dimension $-4$. Since the only dimensional scale in the problem is $T$, we therefore know that
\be\label{eq:RD_gamma0}
\gamma_{\alpha,IJ}^0 &=&  c_0 D^*_{\alpha I}D^{}_{\alpha J} T^8,
\ee
where $c_0$ is a dimensionless number. 

To evaluate $c_0$, we compute the reaction rate
\beq\label{eq:alpha0-reaction-rate}
\frac{\Gamma_{\alpha 0,IJ}^<}{2E_\chi}(y,z)= \frac{3T_{\rm ew}^5D_{\alpha I}^*D^{}_{\alpha J} }{16\pi^3 z^5}\,\xi(y),
\eeq
where the dimensionless  factor $\xi(y)$ obtained by integrating over the SM statistical distributions is
%
%
%
\begin{widetext}
\be
\xi(y) &=& y\int_0^\infty\,dy'\,\frac{y'^3e^{y'}}{(e^{y'}+1)(e^{y+y'}-1)}\int_{-1}^1\,dc_\theta\,\frac{(1-c_\theta)^2}{q_+}\ln\left[\left(\frac{1+e^{\frac{y+y'+q_+}{2}}}{1+e^{\frac{y+y'-q_+}{2}}}\right)\left(\frac{e^{y+y'}+e^{\frac{y+y'-q_+}{2}}}{e^{y+y'}+e^{\frac{y+y'+q_+}{2}}}\right)\right]\nonumber\\
&&{}+2 y\int_0^\infty\,dy'\,\frac{y'^3}{(e^{y'}+1)(e^{y-y'}-1)}\int_{-1}^1\,dc_\theta\,\frac{(1-c_\theta)^2}{q_-}\ln\left(\frac{e^{y-y'}+e^{\frac{y-y'+q_-}{2}}}{1+e^{\frac{y-y'+q_-}{2}}}\right),\label{eq:chi_spectrum}
\ee
\end{widetext}
and
\be
q_\pm &\equiv& \sqrt{y^2+y'^2\pm 2yy'c_\theta}.
\ee
These integrals must be evaluated numerically, and the two terms come from $s$- and $t$-channel processes, respectively.

Integrating over all comoving momentum modes $y$ gives $c_0\approx0.0418$. \\

\noindent {\bf (2) Reaction density dependent on $\mu$ but not $Y_\chi$}: the collision term in the QKE is 
\be
\mathcal{C}_1&\equiv& \frac{1}{s}\int\frac{d^3k}{(2\pi)^3}\frac{\Gamma_{\alpha}^<}{2E_\chi}-\frac{1}{s}\int\frac{d^3k}{(2\pi)^3}\frac{\Gamma_{\alpha0}^<}{2E_\chi}\\
&=& \frac{1}{s}\int\,d\Pi_k\,\frac{\Gamma_{\alpha}^<}{2E_\chi} - \frac{\gamma_\alpha^0}{s}\\
&\equiv&\frac{\gamma_\alpha^1}{s},
\ee
where we have subtracted off the piece that is independent of $\mu$. Because fractional SM asymmetries are small, we linearize in SM chemical potentials,
\be
f^{\rm eq}_\psi \approx f_{\psi0}^{\rm eq} + \frac{\mu}{T}f_{\psi0}^{\rm eq}(1-f_{\psi0}^{\rm eq}),
\ee
where again the subscript 0 indicates the Fermi-Dirac distribution with zero chemical potential. 

Each of the three processes involved in DM production/annihilation has three terms associated it (one for each SM chemical potential), leading to a total of nine terms in $\mathcal{C}_1$. However, since each SM fermion has the same statistical distribution, appropriate changes of variables can be used to simplify and combine these into a smaller set of integrals to evaluate. Stripping out the coupling from the reaction density, these are of the form 
\be\label{eq:integral_form_rd}
c_{i} &=& \frac{6}{T^8}\int \,d\Pi_{\rm SM}\,d\Pi_k\, \mathcal{F}_{i}(2\pi)^4\delta^4(k+p_Q-p_L-p_d),\nonumber
\ee
where $\mathcal{F}_i$ are some combination of Mandelstam variables and statistical distribution functions. There are ultimately five independent integrals to evaluate, whose integrands are:
%
%
\be
\mathcal{F}_{1} &=& t^2\,f_{d0}f_{L0}(1-f_{Q0})(1-f_{d0}),\\ 
\mathcal{F}_{2} &=& t^2\,f_{d0}f_{L0}(1-f_{Q0})(1-f_{L0}),\\ 
\mathcal{F}_{3} &=& t^2\,f_{d0}f_{L0}f_{Q0}(1-f_{Q0}),\\
\mathcal{F}_{4} &=& s^2\,f_{d0}f_{L0}(1-f_{Q0})(1-f_{d0}),\\
\mathcal{F}_{5} &=& s^2\,f_{d0}f_{L0}f_{Q0}(1-f_{Q0}),
\ee
with
\be
\gamma_{\alpha,IJ}^1 &=& D_{\alpha I}^*D^{}_{\alpha J}T^8\left[(c_{1}+c_{3}+c_{4})\frac{\mu_d-\mu_Q}{T}\right.\nonumber\\
&&\left.{}-(2c_{2}+c_{5})\frac{\mu_{L_\alpha}}{T}\right].
\ee
Evaluating the phase space integrals, we find 

\be
c_{1} &\approx& 0.00797,\\
c_{2} &\approx& 0.00799,\\
c_{3} &\approx& 3.72\times10^{-4},\\
c_{4} &\approx& 0.0239,\\
c_{5} &\approx& 0.001118.
\ee

\noindent {\bf (3) Reaction density dependent on both $\mu$ and $Y_\chi$:}~the collision term in the QKE is
\small
\be
\mathcal{C}_2 &=& -\frac{1}{2s}\sum_\alpha\left\{\int\frac{d^3k}{(2\pi)^3}\frac{\Gamma_\alpha^<+\Gamma_\alpha^>}{2E_\chi}f_\chi^{\rm f.d.}(k),\frac{Y_\chi}{Y_\chi^{\rm eq}}\right\}\\
&&{}+ \frac{1}{2s}\sum_\alpha\left\{\int\frac{d^3k}{(2\pi)^3}\frac{\Gamma_{\alpha0}^<+\Gamma_{\alpha0}^>}{2E_\chi}f_\chi^{\rm f.d.}(k),\frac{Y_\chi}{Y_\chi^{\rm eq}}\right\}\nonumber,
\ee
\normalsize
where  we have again subtracted the term corresponding to zero chemical potential. The relevant reaction density we are interested in is
\small
\beq
\gamma_\alpha^2 \equiv \int\,d\Pi_k\left[(\Gamma_\alpha^<-\Gamma_{\alpha0}^<)+(\Gamma_\alpha^>-\Gamma_{\alpha0}^>)\right]f_\chi^{\rm f.d.}(k).
\eeq
\normalsize

For each of the production and annihilation rates, we have 9 terms (3 per process for 3 SM chemical potentials), leading to a total of 18 terms. Again, we can perform an appropriate changes of integration variables to decompose these integrals into a  set of three new integrals, along with integrals already evaluated. The new integrals have the same form as before, with integrands:
\be
\mathcal{F}_{6} &=& t^2 \,f_{L0}f_{d0}f_{\chi}^{\rm f.d.}(1-f_{d0}),\\
\mathcal{F}_{7}&=& t^2 \,f_{L0}f_{d0}f_\chi^{\rm f.d.}(1-f_{L0}),\\
\mathcal{F}_{8} &=& s^2 \,f_{Q0}f_{d0}f_\chi^{\rm f.d.}(1-f_{Q0}).
\ee
The reaction density is
\be
\gamma_{\alpha,IJ}^2 &=& D_{\alpha I}^*D^{}_{\alpha J}T^8\left[(2c_{6}+c_{8}-c_{1}-c_{3}-c_{5})\frac{\mu_d-\mu_Q}{T}\right.\nonumber\\
&&\left.{}-(2c_{7}+c_{8}-2c_{3}-c_{4})\frac{\mu_{L_\alpha}}{T}\right].
\ee
The remaining phase space integrals evaluate to
\be
c_{6} &\approx& 3.86\times10^{-4},\\
c_{7} &\approx& 4.04\times10^{-4},\\
c_{8} &\approx&0.00119. 
\ee

\noindent {\bf Relation between $X_\alpha$ asymmetries and chemical potentials:}~in our QKEs, we solve for the time evolution of the $X_\alpha$ asymmetries, denoted $Y_\alpha$. The reaction densities, however, depend on the chemical potentials $\mu_d$, $\mu_Q$, and $\mu_{L_\alpha}$. Because the only states in equilibrium are SM states, the relationships between these quantities can be derived in the conventional way using SM chemical potential relations, along with hypercharge conservation and $X_\alpha$ conservation in sphaleron interactions \cite{Harvey:1990qw}. We find
\be
\frac{\mu_d}{T} &=& \frac{38}{79}\,\frac{s}{T^3}\sum_\alpha Y_\alpha,\\
\frac{\mu_Q}{T} &=& \frac{14}{79}\,\frac{s}{T^3}\sum_\alpha Y_\alpha,\\
\frac{\mu_{L_\beta}}{T} &=& \frac{s}{T^3}\left(-2Y_\beta+\frac{32}{237}\sum_\alpha Y_\alpha\right).
\ee
The final baryon asymmetry is the usual relation,
\be
Y_B &=& \frac{28}{79}\sum_\alpha Y_\alpha.
\ee
For the final baryon asymmetry, we assume an instantaneous decoupling of sphalerons at the electroweak phase transition. Since asymmetry production in UV freeze-in  is peaked near $T_{\rm RH}$, this is a reasonable assumption provided $T_{\rm RH}$ is parametrically separated from $T_{\rm ew}$.\\

\noindent {\bf Oscillation term:}~the commutator only depends on the relative average energy $\langle E_2-E_1\rangle$, which is computed assuming a Fermi-Dirac distribution for DM. The average relative energy in vacuum in the highly relativistic limit is
\be
\langle E_2-E_1\rangle = \frac{\Delta M^2}{2}\left\langle\frac{1}{|\vec{k}|}\right\rangle = \frac{\pi^2\Delta M^2}{36\zeta(3) T}.
\ee
The vacuum energy matrix can then be chosen to be
\be
\langle E_{\rm vac}\rangle &=& \left(\begin{array}{cc} 0 & 0 \\ 0 & \langle E_2-E_1\rangle\end{array}\right).
\ee

Because the vacuum DM masses are in the keV range, thermal corrections in the early Universe can be important. We determine thermal mass corrections by computing the in-medium one-loop correction to the $\chi$ propagator and solving for the modified dispersion relation taking into account the shift in the propagator pole \cite{Weldon:1982bn}. We obtain
\be
E_{\rm thermal} &=& -\frac{7\pi^2}{45}\left(D^\dagger D \right) T^4\,k.
\ee
\\
Averaging with the thermal ansatz gives
\be
\langle E_{\rm thermal}\rangle &=& - \frac{49}{8100\zeta(3)}\left(D^\dagger D\right) T^5.
\ee
Because thermal DM masses are aligned with the interaction basis, they tend to suppress oscillations when they are comparable to or larger than the vacuum masses.\\

\noindent{\bf QKE summary:}~to numerically solve the QKEs, we switch from ordinary time to the dimensionless variable $z\equiv T_{\rm ew}/T$. The QKEs in this form are
\begin{widetext}
\be\label{eq:actualQKE_chi}
sHz\frac{dY_\chi}{dz} &=& -is\left[\langle E_{\rm vac}\rangle+\langle E_{\rm thermal}\rangle,Y_\chi\right] + \frac{1}{2}\sum_\alpha\left\{\gamma_\alpha^0,1-\frac{Y_\chi}{Y_\chi^{\rm eq}}\right\}+\sum_\alpha\gamma_\alpha^1-\frac{1}{2}\left\{\gamma_\alpha^2,\frac{Y_\chi}{Y_\chi^{\rm eq}}\right\},\\
sHz\frac{dY_{\bar\chi}}{dz} &=& -is\left[\langle E_{\rm vac}\rangle+\langle E^*_{\rm thermal}\rangle,Y_{\bar\chi}\right] + \frac{1}{2}\sum_\alpha\left\{\gamma_\alpha^{0^*},1-\frac{Y_{\bar\chi}}{Y_\chi^{\rm eq}}\right\}-\sum_\alpha\gamma_\alpha^{1*}+\frac{1}{2}\left\{\gamma_\alpha^{2*},\frac{Y_{\bar\chi}}{Y_\chi^{\rm eq}}\right\},\\
sHz\frac{dY_\alpha}{dz}&=& \mathrm{Tr}\left(\gamma_\alpha^0\frac{Y_\chi}{Y_\chi^{\rm eq}}-\gamma_\alpha^{0*}\frac{Y_{\bar\chi}}{Y_\chi^{\rm eq}}-2\gamma_\alpha^1 + \gamma_2^\alpha\frac{Y_\chi}{Y_\chi^{\rm eq}}+\gamma_2^{\alpha*}\frac{Y_{\bar\chi}}{Y_\chi^{\rm eq}}\right).\label{eq:SM_asymmetry_QKE}
\ee
\end{widetext}
When solving the QKEs, we write the chemical potentials embedded in $\gamma_\alpha^1$ and $\gamma_\alpha^2$ in terms of the comoving $X_\alpha$ asymmetries, $Y_\alpha$, using the relations given earlier.

\subsection{Mixed UV-IR Scenario} \label{app:QKE_UVIR}

\newcommand{\LPhi}{\bar{L}}
\newcommand{\LPhic}{L}
\newcommand{\mb}[1]{\mathbf{#1}}
\newcommand{\pvec}[1]{\mb{p}_{#1}}
\newcommand{\Tew}{T_{\rm ew}}

When there exist at least two mediators $\Phi_a$, it is possible that the reheat temperature of the Universe satisfies $M_{\Phi_1}<T_{\rm RH}<M_{\Phi_2}$. In this case, the production of DM proceeds through IR freeze-in mediated by $\Phi_1$ and through UV freeze-in mediated by $\Phi_2$. Because the baryon asymmetry is enhanced when freeze-in occurs through multiple mediators, the largest asymmetry will typically result through mixed UV-IR freeze-in, where a primordial DM abundance is produced through UV freeze-in, followed by inverse decay into $\Phi_1$ (\emph{e.g.}, $L\chi\to{\Phi^*_1}$).

The asymmetry in this case is determined entirely by the DM density matrix at the end of UV freeze-in. We therefore expect that the viable mass-coupling parameter space is comparable to the scenario of fully IR freeze-in through two on-shell scalars, where the primordial DM abundance is determined through ${\Phi^*_2}\to L\chi$ decay rather than through UV freeze-in. Indeed, if the DM density matrices were identical after the first stage of DM production, the resulting asymmetries from inverse decay into ${\Phi^*_1}$ would be identical too. However, the momentum spectrum of DM is different for UV and IR freeze-in (see Fig.~\ref{fig:DM-spectrum}), and so the DM density matrices are somewhat different between the two cases even if the momentum-integrated number densities of DM are identical.

In the present work, we are mainly interested in using this mixed UV-IR scenario in order to derive the effects of the DM momentum spectrum on the lepton asymmetry generated from subsequent inverse decay into an on-shell scalar.
Accordingly, we specialize to a minimal model that allows us to perform this study, namely Eq.~\eqref{eq:IR_lag},  with a single active lepton flavor coupled to DM:
\begin{equation}
    \mathcal{L}_{\text{int}} = -F_I^a\Phi_a\LPhic\chi_I + \text{h.c.}
\end{equation}
We study this in two cases:~$\Phi_1$ and $\Phi_2$ have masses below the reheat temperature (fully IR); and, $\Phi_1$ has a mass below the reheat temperature but the $\Phi_2$ mass is above the reheat temperature (mixed UV-IR).

The differences in DM spectrum between IR and UV freeze-in are most pronounced when both DM species are far from equilibrium. In this limit, we can derive a perturbative solution of the QKEs order-by-order valid in the limit of small coupling to DM. We present only the collision terms in the QKE that are relevant for the leading-order baryon asymmetry. These consist of rates for DM production and destruction from on-shell processes involving the scalars $\Phi_1,\Phi_2$ for fully IR freeze-in. For the scenario of mixed UV-IR freeze-in, we also calculate collision terms for scattering mediated by $\Phi_2$ in the limit where $\Phi_2$ is sufficiently heavy to never be on-shell. We can neglect washout in the perturbative solution, and consequently we only include terms in the QKE that persist when SM asymmetries are initially zero. All other terms proportional to SM asymmetries are higher order in the perturbative expansion.

The QKE describing the evolution of DM is the same as in Eq.~\eqref{eq:QKE_general_app}, but because DM couples to a single lepton flavor there is now only one flavor of DM production and destruction rates. We  decompose $\Gamma^<$ as
\be
\Gamma^< = \Gamma^<_{\rm IR} + \Gamma^<_{\rm UV},
\ee
where the first term corresponds to ${\Phi^*_1}\rightarrow L\chi$ decay, and the second term corresponds to  $\LPhi\LPhic\rightarrow\chi\bar{\chi}$ scattering.

 For the decay ${\Phi^*_1}\rightarrow L\chi$, we have
 \be
 \Gamma^<_{\mathrm{IR},IJ} &=& 2F^*_I F^{}_J M_{\Phi_1}^2 \int d\Pi_{\rm IR}\, (2\pi)^4 \delta^4(p_{{\Phi_1}}-p_L -k)\nonumber\\
 && {}f_{{\Phi_1}0}\left(1-f_{L0}\right),
 \ee
 where 
 \be
 d\Pi_{\rm IR} = \frac{d^3 p_{\Phi_1}}{(2\pi)^3 2E_{\Phi_1}}\frac{d^3 p_L}{(2\pi)^3 2E_L},
 \ee
 and $f_{{\Phi_1}0}$ and $f_{L0}$ are the Bose-Einstein and Fermi-Dirac distributions for $\Phi_1$ and $L$, respectively, with vanishing chemical potentials. The reaction rate is a factor of two larger than that in Ref.~\cite{Berman:2022oht} because that study considered an electroweak singlet scalar, wherease we consider a doublet. 
 
The IR DM production rate for comoving momentum $y=k/T$ is~\cite{Berman:2022oht}
\begin{align} \label{app:A:uvir:Gamma:ir}
    \frac{\Gamma^<_{\mathrm{IR},IJ}}{2E_\chi} &= \frac{M_{\Phi_1}^2}{8\pi\Tew} \frac{z}{y^2}g_0(M_{\Phi_1} z/T_{\rm ew},y) F_I^* F_J^{},
\end{align}
where 
\begin{align}
    g_0(x,y) &\equiv \frac{1}{e^y+1} \ln\frac{1+e^{-x^2/(4y)}}{1-e^{-(y+x^2/(4y))}}.
\end{align}
We neglect thermal corrections to the  $\Phi_1$ mass. As for the fully UV case, the QKE is linear in $f_\chi$, and so we can use the principle of detailed balance to conclude that
\be
f_\chi^{\rm f.d.}(y)\Gamma^>_{\rm IR} = [1-f_\chi^{\rm f.d.}(y)]\Gamma^<_{\rm IR}.
\ee
Hence we can obtain  $\Gamma^>_{\rm IR}$ from $\Gamma^<_{\rm IR}$.

For  UV scattering $L\bar{L}\rightarrow \chi\bar{\chi}$, we have
\begin{widetext}
\be
\Gamma^<_{\mathrm{UV},IJ}(k) &=& 2 \sum_{KL} C^*_{IK}C^{}_{JL} \int\,d\Pi_{\rm UV}\,(2\pi)^4\delta^4(p_L+p_{\bar L}-p_{\bar \chi}-k)\,t^2\,f_{L0}^{\rm eq}f_{\bar L 0}^{\rm eq}\left[\delta_{KL}-f_{\bar\chi\,KL}(p_{\bar\chi})\right],
\ee
\end{widetext}
where $f_{\bar\chi}$ is the density matrix for $\bar\chi$ and
\be
d\Pi_{\rm UV}=\frac{d^3p_L}{(2\pi)^3 2E_L}\frac{d^3p_{\bar L}}{(2\pi)^3 2E_{\bar L}}\frac{d^3p_{\bar\chi}}{(2\pi)^3 2E_{\bar\chi}}.
\ee

Because this process does not violate SM lepton number, it cannot directly create a lepton asymmetry. It is only effective at creating a background population of DM, which subsequently oscillates and inverse-decays into $\Phi_1$ to create an asymmetry. Thus, the leading contribution of the UV collision rate to the asymmetry is the piece that persists when $f_{\bar\chi}\to0$.

As in Eq.~\eqref{eq:gamma0_def}, we define a reaction density in the $f_{\bar\chi}\to0$ limit as
\begin{equation} \label{app:A:uvir:gamma:def}
\gamma^0_{\rm UV} \equiv T^3 \int\,\frac{d^3y}{(2\pi)^3}\,\frac{\Gamma^<_{\rm UV}}{2E_\chi}.
\end{equation}
By similar arguments to before, we use dimensional analysis to conclude that
\be
\gamma_{\mathrm{UV},IJ}^0 &=& d_0\,(CC^\dagger)_{JI}\,T^8,
\ee
where $d_0$ is a dimensionless constant.

To evaluate $d_0$, we compute the reaction rate
\be
\frac{\Gamma^<_{\mathrm{UV},IJ}}{2E_\chi}(y,z) &=& \frac{(CC^\dagger)_{JI}T_{\rm ew}^5}{16\pi^3z^5}\xi_{\rm UV}(y),
\ee
where the dimensionless  factor $\xi_{\rm UV}(y)$ obtained by integrating over the SM statistical distributions is
\begin{widetext}
\be
\xi_{\rm UV}(y) &=& y\int_0^\infty\,dy'\,\frac{y'^3}{e^{y'}+1}\int_{-1}^1\,dc_\theta\,\frac{(1-c_\theta)^2}{q_-}\ln\left(1+e^{\frac{y'-y-q_-}{2}}\right),\label{eq:chi_spectrum_mixed}
\ee
\end{widetext}
and $q_-=\sqrt{{y'}^2+y^2-2yy'c_\theta}$. These integrals must be evaluated numerically.

Integrating over all comoving momentum modes $y$ gives $d_0\approx0.00293$.

\subsection{QKEs and Antiparticle Density Matrices} \label{app:comment-on-chibar}
In the standard convention, the density matrices for  states $\chi$ and $\bar\chi$ are defined by \cite{Sigl:1993ctk}
\be
f_{\chi,IJ}(\mathbf{k}) &\equiv& \frac{1}{V}\langle a_J^\dagger(\mathbf{k})a_I(\mathbf{k})\rangle,\\
f_{\bar\chi,IJ}(\mathbf{k}) &\equiv& \frac{1}{V}\langle b_I^\dagger(\mathbf{k})b_J(\mathbf{k})\rangle,
\ee
where $a$ and $b$ are the annihilation operators for particles and antiparticles, respectively. This ensures that $f_\chi$ and $f_{\bar\chi}$ transform in the same way under unitary transformations in flavor space, given that the mode expansion of the field $\Psi_I$ is the sum of terms with $a_I$ and $b_I^\dagger$. 

Because the flavor indices are transposed for the creation and annihilation operators in $f_{\bar\chi}$ relative to $f_\chi$, the QKE for $f_{\bar\chi}$ has a flipped sign in the commutator term in the QKE, while the collision term is the same as for $f_\chi$, except taking $f_\chi \leftrightarrow f_{\bar\chi}$ and $\mu \leftrightarrow -\mu$ for SM chemical potentials (see, for example, Ref.~\cite{Cirigliano:2009yt}). 

Note, however, that this is a different prescription than what is commonly given in studies of oscillation baryogenesis such as ARS leptogenesis (see, for example, \cite{Canetti:2012kh,Drewes:2017zyw}). In the alternate prescription, the antiparticle QKE has the same sign on the oscillation term as for the particle QKE, while the couplings in the rates are the complex conjugates of those for the particle QKE (while still taking $\mu\leftrightarrow -\mu$). We can understand these QKEs as using an alternate definition for the antiparticle density matrix,
\be
\bar f_{IJ} &\equiv& \frac{1}{V}\langle b_J^\dagger(\mathbf{k})b_I(\mathbf{k})\rangle,
\ee
such that $\bar f_{IJ}^{\rm T} = f_{\bar\chi,IJ}$. Because the density matrix is Hermitian, this is equivalent to $\bar f_{IJ} = f_{\bar\chi,IJ}^*$. In most expressions of the ARS QKEs, $\bar f$  is used to quantify the antiparticle density matrix. Since in typical ARS scenarios the QKE for $f_\chi$ is independent of $\bar f$ (and vice-versa), this  redefinition of $f_{\bar\chi}\to \bar f$ is  straightforward to implement (and typically done implicitly).

In the case of the mixed UV-IR scenario, however, the collision term for $f_\chi$ depends on $f_{\bar\chi}$ (and vice-versa). Transforming $f_{\bar\chi}\to \bar f$ therefore affects the $f_\chi$ QKE as well. If this is not taken into account, the QKEs yield physically nonsensical results. For example, they predict that DM produced via the on-shell decay of $\Phi_1$ and then scattering via an off-shell heavy $\Phi_2$ produces a non-zero lepton asymmetry, even though the latter process conserves lepton number. DM oscillations should not be able to retroactively induce a lepton asymmetry at the time of $\Phi_1$ decay. When we use the correct QKE, this contribution to the asymmetry  vanishes as expected.

For the quantitative results that we need, the $f_{\bar\chi}$ term in the $f_\chi$ QKE is a higher-order effect and negligible. We therefore follow the ARS-type prescription and implicitly use $\bar f$ as our antiparticle density matrix. However, because this distinction is relevant in any model where the out-of-equilibrium state can interact directly with itself, we explicitly comment here on which density matrix we use, while neglecting to distinguish between the two definitions elsewhere because it does not affect our results.

\section{Perturbative Solutions} \label{app:perturbative}
\subsection{Fully UV Scenario} \label{app:pert_UV}
While the full momentum-dependent QKEs in Eq.~\eqref{eq:QKE_general_app} are difficult to solve in generality, they can be solved assuming all DM species are far from equilibrium. If we assume that the Universe starts with $f_\chi=0$ and no SM asymmetries, then terms involving higher powers of $f_\chi$ and/or $\mu$ are going to be parametrically smaller than those with lower powers of the same quantity.

The oscillation term in Eq.~\eqref{eq:QKE_general_app}, however, partly spoils this expansion. The reason is that oscillations become significant when $\Delta M^2/(2yT)$ becomes of order $H$, and this is true independent of the magnitude of $f_\chi$. To address this, we first move into an interaction picture that resums the oscillation contributions into exponential phases without having to solve the QKE. Note that it is challenging to incorporate thermal masses in this approach because the mass eigenstates are continually changing, so we neglect thermal masses in this calculation. Our approach closely follows that of the perturbative calculations of Ref.~\cite{Berman:2022oht}.

In the interaction picture, the density matrices have the form
\be
\tilde{f}_\chi(z) &\equiv& U^\dagger(z)f_\chi(z) U(z),
\ee
while the reaction rates are
\be \label{app:B:uv:U(z)}
\tilde\Gamma(z) &\equiv& U^\dagger(z)\Gamma(z) U(z).
\ee
The unitary matrix taking into account time evolution is
\be
U(z) &\equiv& \left(\begin{array}{cc} 1 & 0 \\ 0 & e^{-i\phi}\end{array}\right)
\ee
with
\be \label{app:B:uv:phi}
\phi \equiv \int_{t_{\rm RH}}^t\, \frac{\Delta M^2}{2k(t')}\,dt  = \frac{\Delta M^2M_0}{6yT_{\rm ew}^3}(z^3-z_{\rm RH}^3),
\ee
where $y=k/T$ is the comoving momentum and $M_0$ is defined so that $H=T^2/M_0$. We define the oscillation parameter
\be
\beta_{\rm osc} &\equiv& \frac{\Delta M^2 M_0}{6T_{\rm ew}^3},
\ee
which quantifies the oscillation phase accumulated in the $y=1$ momentum mode over the time between reheating and the electroweak phase transition.

The QKEs for the interaction-picture density matrices are the same as the ordinary QKEs, except that there is no longer an oscillation term:
\small
\be\label{eq:QKE_general_app_IP}
\frac{d\tilde f_\chi(y)}{dt} &=& -\frac{1}{2}\sum_\alpha\left(\left\{\frac{\tilde\Gamma_\alpha^>(y)}{2E_\chi},\tilde f_\chi(y)\right\}-\left\{\frac{\tilde\Gamma_\alpha^<(y)}{2E_\chi},1-\tilde f_\chi(y)\right\}\right).\nonumber
\ee
\normalsize

\noindent {\bf Leading order:~DM abundance.} At leading order, the chemical potentials vanish. We can therefore use Eq.~\eqref{eq:KMS} to simplify the QKE, and also change from $t$ to $z$:
\be\label{eq:QKE-zero-chempot}
zH\frac{d\tilde f_\chi(y)}{dz} &=& \frac{1}{2}\sum_\alpha\left\{\frac{\tilde\Gamma^<_{\alpha0}}{2E_\chi},1-\frac{\tilde f_\chi}{f_\chi^{\rm f.d.}}\right\}.
\ee
At leading order, we can also set $\tilde f_\chi=0$, giving
\be\label{eq:fchi_1order}
\tilde f^{(1)}_\chi(y,z) &=& \sum_\alpha\,\int_{z_{\rm RH}}^z\,\frac{dz_1}{z_1H(z_1)}\,\frac{\tilde\Gamma^<_{\alpha 0}(y,z_1)}{2E_\chi}.
\ee

The DM abundance at this order is determined by integrating $\tilde{f}^{(1)}_\chi$ over momentum:
\be
\tilde{Y}^{(1)}_\chi(z) &=& \frac{1}{s(z)}\int\frac{d^3k}{(2\pi)^3}\sum_\alpha \int_{z_{\rm RH}}^z\,\frac{dz_1}{z_1H(z_1)}\,\frac{\tilde \Gamma^<_{\alpha 0}}{2E_\chi}\nonumber\\
&=& \frac{T^3}{s(z)}\int\frac{d^3y}{(2\pi)^3}\sum_\alpha \int_{z_{\rm RH}}^z\,\frac{dz_1}{z_1H(z_1)}\,\frac{\tilde \Gamma^<_{\alpha 0}}{2E_\chi}\nonumber\\
&=& \frac{45}{2\pi^2 g_{*S}}\sum_\alpha \int_{z_{\rm RH}}^z\,\frac{dz_1}{z_1H(z_1)}\int\frac{d^3y}{(2\pi)^3}\frac{\tilde \Gamma^<_{\alpha 0}}{2E_\chi}\nonumber\\
&=& \frac{45}{2\pi^2 g_{*S}T_{\rm ew}^3}\sum_\alpha \int_{z_{\rm RH}}^z\,\frac{dz_1\,z_1^2}{H(z_1)}\,\tilde\gamma_\alpha^0(z_1),
\ee
where the reaction density $\gamma_\alpha^0(z)_{IJ}=c_0 D_{\alpha I}^* D^{}_{\alpha J}T_{\rm ew}^8/z^8$ was defined in Eq.~\eqref{eq:gamma0_def} and calculated in Eq.~\eqref{eq:RD_gamma0}. Note that we can interchange the order of the $y$ and $z$ integrals but not the $k$ and $z$ integrals because $k$ is not redshift-invariant.

For the DM abundance at late times, $z\to\infty$, we only care about the on-diagonal elements of $\tilde Y_\chi$ for which $\tilde\gamma_{\alpha}^0{_{II}}=\gamma_\alpha^0{_{II}}$ (in other words, oscillations are not relevant for the DM abundance). Summing over both $\chi_I$ and $\bar\chi_I$, we obtain
\be
Y_{\chi+\bar\chi,II}^{(1)}(z\to\infty) &=& \frac{15c_0(D^\dagger D)_{II}T_{\rm ew}^3M_0}{\pi^2 g_{*S}z_{\rm RH}^3}.
\ee
As expected for UV freeze-in, the DM abundance depends on the reheat temperature through $z_{\rm RH}=T_{\rm ew}/T_{\rm RH}$. Note that we would obtain the same result at this order integrating the momentum-averaged QKE, Eq.~\eqref{eq:actualQKE_chi}, since this contribution is independent of the DM abundance and therefore is also independent of any assumptions on the DM spectrum.

The momentum dependence of the DM spectrum is
\be \label{eq:DM-spectrum}
\frac{dY_\chi}{dy} &\propto& y^2\xi(y),
\ee
where $\xi(y)$ is given in Eq.~\eqref{eq:chi_spectrum}. 

Since $D^{}_{\alpha I}D^*_{\alpha I}$ is real, the number densities of $\chi$ and $\bar\chi$ are equal as a result of collisions involving lepton flavor $\alpha$, and hence the $X_\alpha$ asymmetries vanish at this order as well.\\

\noindent {\bf Second order:~flavor asymmetries.} To obtain the lepton flavor asymmetries, we substitute our expression for $\tilde f^{(1)}_\chi$ from Eq.~\eqref{eq:fchi_1order} back into the QKEs. We discard all terms independent of $\tilde f_\chi^{(1)}$ since, by the arguments in the leading-order calculation, these cannot contribute to the lepton flavor asymmetry. Likewise, since the SM chemical potentials vanish at leading order, we can set them to zero in the collision terms at this stage of the calculation as well. 

The result for the DM density matrix at this order due to scattering from lepton flavor $\beta$ is
\small
\be
\tilde{f}_{\chi,\beta}^{(2)}(y,z) &=& -\frac{1}{2f_\chi^{\rm f.d.}}\int_{z_{\rm RH}}^z\,\frac{dz_2}{z_2H(z_2)}\left\{\frac{\tilde\Gamma^<_{\beta0}}{2E_\chi}(z_2),\tilde f_\chi^{(1)}\right\}\nonumber\\
&=& -\frac{e^y+1}{2}\int_{z_{\rm RH}}^z\,\frac{dz_2}{z_2H(z_2)}\int_{z_{\rm RH}}^{z_2}\,\frac{dz_1}{z_1H(z_1)}\nonumber\\
&&{}\quad\sum_\alpha\left\{\frac{\tilde\Gamma^<_{\beta0}}{2E_\chi}(z_2),\frac{\tilde\Gamma^<_{\alpha0}}{2E_\chi}(z_1)\right\} \label{eq:density-matrix-order2}
\ee
\normalsize
For this calculation, we need the interaction-picture version of the momentum-dependent reaction density from Eq.~\eqref{eq:alpha0-reaction-rate},
\small
\beq
\frac{\tilde{\Gamma}_{\alpha 0,IJ}^<}{2E_\chi}(z)= \frac{3T_{\rm ew}^5}{16\pi^3 z^5}\,\xi(y)\, U(z)^\dagger_{II} D_{\alpha I}^*D^{}_{\alpha J} U(z)_{JJ}.
\eeq
\normalsize
The $X_\beta$ asymmetries are related to the DM asymmetries according to:
\small
\be
 Y_\beta^{(2)}(z) &=& -\frac{1}{s}\int\frac{d^3k}{(2\pi)^3}\mathrm{Tr}\left[\tilde f_{\chi,\beta}^{(2)}(z)-\tilde f_{\bar\chi,\beta}^{(2)}(z)\right]\\
&=& -\frac{e^y+1}{2s}\int_{z_{\rm RH}}^z\,\frac{dz_2}{z_2H(z_2)}\int_{z_{\rm RH}}^{z_2}\,\frac{dz_1}{z_1H(z_1)}\label{eq:leptonasym-order2}\\
&&{} \sum_\alpha\int\,\frac{d^3k}{(2\pi)^3}\,\mathrm{Tr}\left[\left\{\frac{\tilde\Gamma^<_{\beta0}}{2E_\chi}(z_2),\frac{\tilde\Gamma^<_{\alpha0}}{2E_\chi}(z_1)\right\}-\left(D\leftrightarrow D^*\right)\right].\nonumber
\ee
\normalsize
This evaluates to
\small

\begin{widetext}
\be
Y_\beta^{(2)}(z) &=& \frac{405M_0^2T_{\rm ew}^6}{256\pi^{10}g_{*S}}\mathrm{Im}\left[D_{\beta 1}^*D^{}_{\beta 2}(D^\dagger D)^{}_{21}\right]\int_0^\infty\,dy\,y^2(e^y+1)\,\xi(y)^2\int_{z_{\rm RH}}^\infty\,\frac{dz_2}{z_2^4}\int_{z_{\rm RH}}^{z_2}\,\frac{dz_1}{z_1^4}\sin\left[\frac{\beta_{\rm osc}}{y}(z_2^3-z_1^3)\right].
\ee
\end{widetext}
\normalsize
At this order, the total asymmetry vanishes, because $\sum_\beta Y_\beta^{(2)} \propto \mathrm{Im}\left[(D^\dagger D)_{12}(D^\dagger D)_{21}\right]=0 $, consistent with the arguments in Sec.~\ref{sec:IRtoUV}. Furthermore, the only states involved in DM scattering that have a non-zero chemical potential are the lepton doublets, $L_\alpha$:
\be
\frac{\mu^{(2)}_{L_\alpha}(z)}{T} = -\frac{4\pi^2 g_{*S}Y^{(2)}_\alpha(z)}{45}.
\ee
Both $\mu_d$ and $\mu_Q$ are proportional to the total asymmetry, which vanishes.\\

\noindent {\bf Third order:~total $B-L$ asymmetry.} In the perturbative expansion,  the total asymmetry arises from washout induced by $\mu_{L_\alpha}^{(2)}$, which is of $\mathcal{O}(D^4)$. The dominant term in the QKEs for the flavor asymmetries is the term that depends on $\mu$ but not on $Y_\chi$, namely the $\gamma_\alpha^1$ term in Eq.~\eqref{eq:SM_asymmetry_QKE}; the term that depends on both $\mu$ and $Y_\chi$ will necessarily be of higher order in the coupling since $Y_\chi$ is $\mathcal{O}(D^2)$. Furthermore, the third-order contribution to the DM density matrix, $\tilde{f}_\chi^{(3)}$, does not result in a total asymmetry unless there are three or more DM flavors \cite{Abada:2018oly,Shuve:2020evk}.

Using Eq.~\eqref{eq:SM_asymmetry_QKE}, we obtain the $B-L$ asymmetry by summing over all of the flavor asymmetries:
\be
Y^{(3)}_{B-L}(z) &=& 2\sum_\alpha \int_{z_{\rm RH}}^z\,\frac{dz_3}{s(z_3)H(z_3)z_3}\,\mathrm{Tr}\left(\gamma_\alpha^1\right)\\
&=& 4M_0T_{\rm ew}^3(2c_2+c_5)\int_{z_{\rm RH}}^z\,\frac{dz_3}{z_3^4}\\
&&\quad\quad\,\sum_\alpha\,(DD^\dagger)_{\alpha\alpha}\, Y_\alpha^{(2)}(z_3).\nonumber
\ee
The final baryon asymmetry is locked in at the electroweak phase transition, and so
\be
Y_B &=& \frac{28}{79}\,Y_{B-L}^{(3)}(z=1).
\ee

\subsection{Mixed UV-IR Scenario}\label{app:mixed-pert}
We follow the same perturbative method as elaborated for the fully UV scenario to calculate the leading-order DM abundance and asymmetry in the interaction picture.\\

\noindent {\bf Leading order:~DM abundance.}
The two leading contributions to DM production in the UV-IR model are $\Phi_1^*\rightarrow L\chi$ (and its CP conjugate) and $L\bar{L}\rightarrow\chi\bar{\chi}$. The DM abundance from the first process has been calculated in Ref.~\cite{Berman:2022oht}. 

For the second, we follow the same reasoning as for the fully UV scenario, except now using the reaction density $\gamma_{\rm UV}^0$:
\be
    Y^{(1)}_{\chi+\bar\chi,II}(z\to\infty) &=& \frac{15d_0(CC^\dagger)_{II}T_{\rm ew}^3 M_0}{\pi^2g_{*S}z_{\rm RH}^3}.
\ee
The momentum dependence of the spectrum for this UV-generated component of  DM is
\be
\frac{dY_\chi}{dy} &\propto& y^2 \xi_{\rm UV}(y).
\ee

As in the fully UV scenario, there are no asymmetries at this order in the solution. 
\\

\noindent {\bf Second order:~SM lepton asymmetry.} At this order, a total lepton asymmetry only results if the second DM scattering process involves a different coupling from its production \cite{Shuve:2020evk,Berman:2022oht}. As we have already argued, the UV scattering process conserves lepton number and so it cannot create a lepton asymmetry where none existed before. Therefore, the dominant contribution to the lepton asymmetry comes from DM production in $L\bar{L}\to\chi\bar\chi$ scattering, followed by $L\chi\to{\Phi^*_1}$ inverse decay. We have checked that the asymmetry indeed vanishes with the reverse ordering.

Because the lepton chemical potentials initially vanish, we can again use Eq.~\eqref{eq:QKE-zero-chempot} and substitute the appropriate UV or IR reaction rates. By analogy with Eq.~\eqref{eq:leptonasym-order2}, the SM $B-L$ asymmetry at this order is
\small
\be
 Y_{B-L}^{(2)}(z) 
&=& -\frac{e^y+1}{2s}\int_{z_{\rm RH}}^z\,\frac{dz_2}{z_2H(z_2)}\int_{z_{\rm RH}}^{z_2}\,\frac{dz_1}{z_1H(z_1)}\int\,\frac{d^3k}{(2\pi)^3}\nonumber\\
&&{} \mathrm{Tr}\left[\left\{\frac{\tilde\Gamma^<_{\rm IR}}{2E_\chi}(z_2),\frac{\tilde\Gamma^<_{\rm UV}}{2E_\chi}(z_1)\right\}-\left(C,F\leftrightarrow C^*,F^*\right)\right],\nonumber
\ee
\normalsize
where
\small
\be
\tilde{\Gamma}^<_{\mathrm{IR},IJ} &=& \frac{M_{\Phi_1}^2 z}{8\pi T_{\rm ew}y^2}g_0(M_{\Phi_1}z/T_{\rm ew},y)U(z)^\dagger_{II}F_I^* F_J^{} U(z)^{}_{JJ},\nonumber\\
\tilde{\Gamma}^<_{\mathrm{UV},IJ} &=& \frac{T_{\rm ew}^5}{16\pi^3 z^5}\xi_{\rm UV}(y)\,U(z)^\dagger_{II}(CC^\dagger)_{JI} U(z)^{}_{JJ}.\nonumber
\ee
\normalsize
This evaluates to
\small
\begin{widetext}
\be
    Y_{B-L}^{(2)}(z) &=& -\frac{45 M_0^2 M_{\Phi_1}^2}{128\pi^8 g_*}\mathrm{Im} \left[F_1^{1*}F_2^1\left(CC^\dagger\right)_{12}\right]\int_0^\infty dy\,(e^y+1)\xi_{\rm UV}(y)\int_{z_{\rm RH}}^z dz_2z_2^2 g_0(M_{\Phi_1}z_2/T_{\rm ew},y)\int_{z_{\rm RH}}^{z_2}  \frac{dz_1}{z_1^4} \sin\left[\frac{\beta_\mathrm{osc}}{y}(z_2^3-z_1^3)\right].\quad\quad
\ee
\end{widetext}
\normalsize
Finally, the portion of the SM $B-L$ asymmetry that converts to a baryon asymmetry is calculated using the methods from Refs.~\cite{Shuve:2020evk,Berman:2022oht}:
\be
Y_B^{(2)} &=& \frac{25}{79} Y_{B-L}^{(2)},
\ee
which is valid provided that $\Phi_1$ has a lifetime much longer than the age of the Universe at $T_{\rm ew}$. Otherwise, $\Phi_1$ decays will wash out the asymmetry, since $\Phi_1$ carries an asymmetry that is equal in magnitude and opposite in sign relative to the SM $B-L$ asymmetry. \\

\noindent {\bf Comparing fully IR and mixed UV-IR scenarios:} In the mixed UV-IR scenario, an initial population of DM is created through off-shell $L\bar{L}\to\chi\bar\chi$ scattering via a heavy intermediate state, $\Phi_2$. The baryon asymmetry is created through subsequent decays or inverse decays involving a lighter scalar, $\Phi_1$. We are interested in studying how the predictions of this asymmetry differ from a fully IR scenario, where the initial population of DM is produced through the decays of a heavy \emph{on-shell} state, $\Phi_2$.

To meaningfully compare the asymmetries, we normalize the couplings so that the DM population produced from off-shell $\Phi_2$-mediated scattering in the mixed UV-IR scenario is identical to the DM population produced from on-shell $\Phi_2$ decay in the fully IR scenario. The momentum-integrated $\chi$ comoving number density from UV production at $T\ll T_{\rm RH}$ is 
\be
Y_{\chi,IJ} &=& \frac{15d_0(CC^\dagger)_{JI}T_{\rm ew}^3 M_0}{2\pi^2 g_{*S}z_{\rm RH}^3}.
\ee
The abundance from on-shell $\Phi_2$ decays at $T\ll M_{\Phi_2}$ is \cite{Berman:2022oht}
\be
Y_{\chi,IJ} &=& \frac{45 d_{\rm IR} (F^{2\dagger} F^2)_{IJ} M_0}{32\pi^5 g_{*S}M_{\Phi_2}},
\ee
where
\be
d_{\rm IR} = \int_0^\infty\,dx\,x^2\,\int_0^\infty\,g_0(x,y) \approx 4.22.
\ee
Thus, we should normalize the constants so that
\be \label{eq:mixed-normalization}
(CC^\dagger)_{JI} &=& \frac{3d_{\rm IR}z_{\rm RH}^3}{16\pi^3 d_0 T_{\rm ew}^3 M_{\Phi_2}}\,(F^{2\dagger}F^2)_{IJ}\\
&\approx& \frac{8.70}{T_{\rm RH}^3 M_{\Phi_2}}\,(F^{2\dagger}F^2)_{IJ}.
\ee
Using this normalization, we can replace $C\to F^2$ in the expressions for the UV-IR asymmetry, and the coupling constants cancel when computing the ratio of the UV-IR asymmetry to the fully IR asymmetry calculated in Ref.~\cite{Berman:2022oht}.

\section{Coupling Benchmarks} \label{app:coupling_benchmark}

Our benchmarks are related to those considered in earlier works \cite{Berman:2022oht}, and we follow analogous reasoning here.

\subsection{Fully UV Scenario} \label{app:coupling_fullyUV}
The UV model we consider has many free parameters. Here, we motivate the choice of benchmarks that are optimistic in that they optimize the magnitude of the baryon asymmetry relative to the DM abundance without unnecessary complexity.  \\

\noindent {\bf Semileptonic operator:}~We consider the $D_{\alpha I}\,(L_\alpha \chi_I)(Qd^{\rm c})$ operator. When there exist two DM mass eigenstates, DM must couple to at least two flavors of lepton to generate an asymmetry. Since we do not expect qualitatively different behavior with couplings to all three flavors of leptons, we consider only the two-flavor case to simplify our analysis.
 For concreteness, we call these flavors $e$ and $\mu$, although it could in reality be any combination of lepton pairs.  At second order, the flavor asymmetries are equal and opposite, $Y_e^{(2)}=-Y_\mu^{(2)}$, but flavor-dependent washout at third order converts part of the flavor asymmetries into a total asymmetry.

We use the same Yukawa texture as for IR freeze-in, since this  optimizes the asymmetry while minimizing the number of free parameters to consider \cite{Berman:2022oht}.  The benchmark is:
\be\label{eq:UV_benchmark}
D &=& \sqrt{\mathrm{Tr}D^\dagger D}\left(\begin{array}{cc}
\cos\theta\cos\beta & \sin\theta\cos\beta \\
\cos\theta\sin\beta & i\sin\theta\sin\beta \\
0 & 0
\end{array}\right).
\ee
$\theta$ parametrizes the relative magnitudes of the coupling to $\chi_1$ vs.~$\chi_2$, while $\beta$ parametrizes the relative magnitudes of the coupling to $e$ vs.~$\mu$. We also choose a maximal relative $CP$-violating phase of $\pi/2$.

While we want the coupling to both lepton flavors to be relatively large to generate appreciable flavor asymmetries, we also do not want them to be equal:~otherwise, the washout rate is the same for each flavor and no net baryon asymmetry results. In the perturbative limit, the asymmetry is proportional to
\be
Y_{B-L} &\propto& \left[(DD^\dagger)^{}_{11}-(DD^\dagger)^{}_{22}\right]\mathrm{Im}\left[D_{11}^*D^{}_{12}(D^\dagger D)^{}_{21})\right]\nonumber\\
&\propto& \cos^2\theta\sin^2\theta\cos2\beta\sin^22\beta.
\ee
The asymmetry is therefore optimized for $\cos2\beta=1/\sqrt 3$, and we fix this value of $\beta$ throughout our work. While there is  an optimal value of $\theta$ for the asymmetry, the DM abundance also depends on $\theta$ and it is possible to enhance the asymmetry to DM abundance ratio with $\theta\ll1$. Therefore, we leave $\theta$ and $\mathrm{Tr}D^\dagger D$ as free parameters.\\

\noindent {\bf Fully leptonic operator:}~We separately consider the leptonic operator $E_{\alpha 23 I}(L_\alpha\chi_I)(L_2 e_3^{\rm c})$ for $\alpha=1,2$. At second order in the perturbative calculation, the fact that the $\chi$ and $\bar\chi$ abundances are the same ensures that the only SM flavor asymmetries that accrue are due to $CP$ asymmetries in the $L_1$ and $L_2$ scattering rates. The $Y_\alpha^{(2)}$ flavor asymmetries are furthermore identical to  the hadronic case with the replacement of $D\to E$.

The third-order calculation is now different, however. There is now washout induced by $\mu_{L_2}$ regardless of the other doublet flavor $\alpha$ involved in scattering. The washout reaction density $\gamma_{\alpha}^1$ is the same as before with the replacement $\mu_Q\to\mu_{L_2}$ and $\mu_d\to \mu_{e_3}$. Since at second order in the calculation $\mu_{L_1}=-\mu_{L_2}$, we have
\be
\mathrm{Tr}\left(\gamma_1^1\right) &\propto& (DD^\dagger)_{11}\left[-(c_1+c_3+c_4)\frac{\mu_{L_2}}{T}\right.\\
&&{}\left.-(2c_2+c_5)\frac{\mu_{L_1}}{T}\right]\\
&\approx&0.0151(DD^\dagger)_{11}\frac{\mu_{L_1}}{T}\\
\mathrm{Tr}\left(\gamma_2^1\right) &\propto& (DD^\dagger)_{22}\left[-(c_1+c_3+c_4)\frac{\mu_{L_2}}{T}\right.\\
&&{}\left.-(2c_2+c_5)\frac{\mu_{L_2}}{T}\right]\\
&\approx&0.0493(DD^\dagger)_{22}\frac{\mu_{L_2}}{T}.
\ee
The dominant contribution to the total $B-L$ asymmetry therefore arises from muon washout. We can straightforwardly compute the $B-L$ asymmetry assuming only muon washout:
\be
Y_{B-L} &\propto& (DD^\dagger)^{}_{22}\mathrm{Im}\left[D_{11}^*D^{}_{12}(D^\dagger D)^{}_{21})\right]
\\
&\propto& \cos^2\theta\sin^2\theta \sin^4\beta\cos^2\beta
\ee
where we have used the same benchmark as in Eq.~\eqref{eq:UV_benchmark}. This asymmetry is optimized for $\sin\beta=\sqrt{2/3}$, and so we use this value of $\beta$ for all studies with this operator.

\subsection{Mixed UV-IR Scenario}

In our study, we only compute the ratio of asymmetries in the mixed UV-IR scenario to the fully IR scenario. When using the coupling normalization from Sec.~\ref{app:mixed-pert}, the couplings cancel in this asymmetry ratio, and so our findings for the mixed UV-IR scenario are independent of the texture of the Yukawa couplings.

\bibliography{biblio}

\end{document}